\documentclass[12pt]{article}
\usepackage{epsfig}
\usepackage{amsfonts}
\usepackage{latexsym}
\usepackage{amsmath,amssymb}
\usepackage{verbatim}

\usepackage[textheight=9in, textwidth=6.5in, letterpaper]{geometry}

  \makeatletter
  \let\over=\@@over \let\overwithdelims=\@@overwithdelims
  \let\atop=\@@atop \let\atopwithdelims=\@@atopwithdelims
  \let\above=\@@above \let\abovewithdelims=\@@abovewithdelims

\renewcommand\section{\@startsection {section}{1}{\z@}%
                                   {-3.5ex \@plus -1ex \@minus -.2ex}%nn
                                   {2.3ex \@plus.2ex}%
                                   {\normalfont\large\bfseries}}

\renewcommand\subsection{\@startsection{subsection}{2}{\z@}%
                                     {-3.25ex\@plus -1ex \@minus -.2ex}%
                                     {1.5ex \@plus .2ex}%
                                     {\normalfont\bfseries}}

\newcommand{\Real}{\mathbb{R}\,}

\newcommand{\Integer}{\mathbb{N}}

\newcommand{\op}{\mathcal{O}\,}

\newcommand{\mL}{\mathcal{L}}
\newcommand{\mM}{\mathcal{M}}
\newcommand{\mN}{\mathcal{N}}

\newcommand{\Lie}{\pounds\hspace{-1.5pt}}
\newcommand{\ello}{\ell_{\text{o}}}

\newcommand{\hf}{\frac{1}{2}}

\newcommand{\qt}{\frac{1}{4}}

\let\a=\alpha \let\b=\beta \let\g=\gamma  \let\e=\epsilon
    
\let\l=\lambda \let\m=\mu 
 \let\r=\rho
\let\s=\sigma    \let\varf=\varphi

  \let\D=\Delta  \let\L=\Lambda

\let\del=\partial

\let\na=\nabla

\newcommand{\be}{\begin{equation}}
\newcommand{\ee}{\end{equation}}
\newcommand{\bea}{\begin{eqnarray}}
\newcommand{\eea}{\end{eqnarray}}
\def\ba{\begin{array}}
\def\ea{\end{array}}

\def\del{\partial}

\def\ie{{\it i.e.\ }}

\def\eg{{\it e.g.\ }}

\def\cov{\,^{(0)}\hspace{-1.2pt}\na}

\def\<{ \langle }
\def\>{ \rangle }

\def\({ \left( }
\def\){ \right) }

\newcommand{\Tr}{{\rm Tr} }

\newcommand{\n}{\noindent}
\newcommand{\nn}{\nonumber}

\begin{document}
\begin{titlepage}
\unitlength = 1mm
\ \\

\begin{center}

{ \LARGE {\textsc{Aspects of the zero $\mathbf{\Lambda}$ limit in the AdS/CFT correspondence}}}

\vspace{0.8cm}
R. N. Caldeira Costa

\vspace{1cm}

{\it  University of Amsterdam, Institute for Theoretical Physics,\\
 Science Park 904, 1090 GL Amsterdam, The Netherlands\\ }
{\tt Email: R.N.Caldeira-Costa@uva.nl}

\begin{abstract}
We examine the correspondence between QFT observables and bulk solutions in the context of AdS/CFT in the limit as the cosmological constant $\L \to 0$. We focus specifically on the spacetime metric and a non-backreacting scalar in the bulk, compute the one-point functions of the dual operators and determine the necessary conditions for the correspondence to admit a well-behaved zero $\L$ limit. We discuss holographic renormalization in this limit and find that it requires schemes that partially break diffeomorphism invariance of the bulk theory. In the specific case of three bulk dimensions, we compute the zero $\L$ limit of the holographic Weyl anomaly and reproduce the central charge that arises in the central extension of $\mathfrak{bms}_{3}$. We compute holographically the energy and momentum of those QFT states dual to flat cosmological solutions and to the Kerr solution and find an agreement with the bulk theory. We also compute holographically the renormalized 2-point function of a scalar operator in the zero $\L$ limit and find it to be consistent with that of a conformal operator in two dimensions less. Finally, our results can be used in a new definition of asymptotic Ricci-flatness at null infinity based on the zero $\L$ limit of asymptotically Einstein manifolds.
\end{abstract}
\vspace{0.5cm}

\vspace{1.0cm}

\end{center}

\end{titlepage}

\pagestyle{empty}
\pagestyle{plain}

\setcounter{page}{1}
\pagenumbering{arabic}
\tableofcontents

\section{Introduction}

\qquad String theory in asymptotically AdS spaces admits a dual non-perturbative formulation provided by the AdS/CFT correspondence \cite{Maldacena:1997re,Gubser:1998bc,Witten:1998qj} and several proposals have been constructed by analogy with AdS/CFT that relate string theory on spacetimes with other asymptotics to field theories formulated at the boundary. For the case of de Sitter gravity, and motivated by studies of the asymptotic symmetry group of de Sitter in a fashion similar to that of AdS \cite{Brown:1986nw}, it has been conjectured that the bulk theory can be described by an Euclidean field theory defined at the spacelike conformal boundary \cite{Hull:1998vg,Hull:1998ym,Witten:2001kn,Strominger:2001pn,Bousso:2001mw,Klemm:2001ea,
Balasubramanian:2001nb}. A further motivation lies in the fact that every solution of AdS gravity is mapped to a solution of de Sitter's by an analytic continuation, leading to a possible dS/CFT correspondence. In the context of AdS/CFT, string theory correlation functions are determined by computing QFT correlators and vice-versa, and the bulk/boundary dictionary is well established. Statements in dS/CFT can then be worked out from the AdS counterpart by analytically continuing the solutions with AdS boundary conditions to de Sitter signature.\footnote{
Note, however, that to compute correlation functions in this way one has to take into account the global properties of asymptotically de Sitter spaces \cite{Spradlin:2001nb,Bousso:2001mw}.}
In particular, the near-boundary asymptotics of AdS spaces admits an analytic continuation to dS asymptotics (see \eg \cite{Skenderis:2002wp}), leading to a well-defined mapping between asymptotic data in the bulk and boundary data in the case of a positive cosmological constant $\L$.

Despite many interesting results, a holographic description of de Sitter space remains unclear, mainly because string theory in dS is not well understood. Even though de Sitter vacua exists in string theory \cite{Kachru:2003aw}, unlike the case of flat or AdS vacua they are unstable and decay to vacua of different $\L$ signature. Another problem in a dS/CFT formulation is the fact that the conformal weights of the QFT operators are imaginary and the boundary theory is non-unitary. Nevertheless, one can still work out the details of such a correspondence and point to those ingredients that do not work.\\ 

The case of Ricci-flat gravity is substantially different. At the classical level, setting $\L$ to zero is just a fine-tuning problem and asymptotically flat spacetimes are the best controlled backgrounds in string theory to compute correlation functions. However, the mechanism in string theory by which the cosmological constant vanishes is not clear (see \eg the discussion in \cite{Witten:2000zk}). More particularly in the context of AdS/CFT, the zero $\L$ limit of the correspondence in general is not well-understood. The limit taken on boundary correlators and vacuum expectation values generically does not lead to sensible results. The conformal weigths of QFT operators dual to massive bulk fields diverge in this limit, a problem associated with the fact that the conformal boundary is null in the zero $\L$ limit. The limit taken on the near-boundary asymptotics of AdS spaces in general does not result in Ricci-flat asymptotics, unless specific constraints are imposed, and a bulk/boundary dictionary has not been established. Furthermore, and unlike the case of de Sitter gravity, holographic renormalization does not extend in a straigthforward manner to flat gravity, essentially because the asymptotics of bulk fields in this case are non-local with respect to the sources \cite{Solodukhin:1999zr,deHaro:2000wj,Skenderis:2002wp,Papadimitriou:2010as}. Nevertheless, quantum gravity in asymptotically flat spacetimes can be characterised by a unitary and analytic S-matrix and it is believed that a holographic description of the flat space S-matrix can be derived from the zero $\L$ limit of AdS/CFT. Indeed, explicit constructions for extracting S-matrix elements from boundary correlators have been proposed in \cite{Polchinski:1999ry,Gary:2009ae,Gary:2009mi,Fitzpatrick:2010zm,Okuda:2010ym,Fitzpatrick:2011jn,
Fitzpatrick:2011ia,Raju:2012zr,Raju:2012zs} (see also the discussions in \cite{Susskind:1998vk,Witten:2001kn}).\\

A different approach to flat space holography formulated as a limit of AdS/CFT is based on studies of the asymptotic symmetry group of asymptotically Minkowski spacetimes at null infinity, the BMS group. In four dimensions the symmetry algebra was originally derived in \cite{Bondi:1962px,Sachs:1962wk,Sachs:1962zza} and more recently investigated in \cite{Barnich:2006av} in general dimensions (see also \cite{Ashtekar:1996cd,Barnich:2009se}). In the three dimensional case, the $\mathfrak{bms_3}$ algebra consists of diffeomorphisms on the circle and supertranslations and is isomorphic to the two-dimensional Galilean conformal algebra (GCA) consisting of a contraction of two copies of the Virasoro algebra. The Poisson algebra of the surface charges was found to admit a central extension with central charge $c=3$ \cite{Barnich:2006av,Bagchi:2012xr},\footnote
{
The central charge $c_{LM}$ in reference \cite{Bagchi:2012xr} is related to ours as: $c_{LM}=c/12$ since we follow the convention of formula (1) in this reference.
}
representing a generalisation to the flat space case of those results originally obtained by Brown and Henneaux \cite{Brown:1986nw} for AdS$_{3}$ and which predated the AdS/CFT correspondence. In the four dimensional case, the $\mathfrak{bms_4}$ algebra is also isomorphic to a class of GCAs \cite{Bagchi:2010eg}. Based on these results, a possible connection between string theory on asymptotically flat spacetimes and non-relativistic conformal field theories defined at null infinity was proposed in \cite{Bagchi:2010eg,Barnich:2010eb,Bagchi:2012cy,Barnich:2012xq,Bagchi:2012xr}. In the same spirit, the authors in \cite{Bagchi:2012xr,Barnich:2012xq} were able to reproduce the Bekenstein-Hawking entropy of three-dimensional flat cosmological horizons by counting states in a two-dimensional Galilean conformal field theory defined at null infinity. However, these studies leave open the question of how to compute field theory correlation functions with the right properties from the bulk theory and do not establish a precise bulk/boundary dictionary. 

Similar earlier studies of flat space holography via the BMS group focused on constructing BMS-invariant field theories, see \cite{Arcioni:2003xx,Arcioni:2003td,Dappiaggi:2004kv,Dappiaggi:2005ci}. Other different approaches have investigated instead a possible dual description of flat space at spatial infinity \cite{Mann:2005yr,Mann:2006bd,Marolf:2006bk,Mann:2008ay,Park:2012bv,Mann:2009id} by analysing the variational principle for asymptotically flat spaces and determining the appropriate counterterms in a fashion similar to AdS holographic renormalization and by studying the putative boundary stress-energy tensor and correlators constructed at spatial infinity.\\

Returning to the context of AdS/CFT, let us quickly review the flat space limit in the duality \cite{Maldacena:1997re} between string theory in $AdS_5 \times S^5$ and $\mN=4$ super Yang-Mills. In the supergravity approximation, the dynamics of the massless closed string states is governed by the IIB supergravity action:
\bea
S = {1 \over 16\pi G_{10}}\, \int d^{10}x\, \sqrt{G} \( R[G] - \qt |g_s F_5 |^2 + ... \)\ ,
\eea
where $F_5 = dA_4$ is the self-dual R-R five-form and we are omitting the remaining supergravity fields. The ten-dimensional Newton constant is given in terms of the string coupling $g_s$ and the string length scale $\ell_s$ by: $G_{10} = 8\pi^6 g_s^2 \ell_s^8$. The metric solution corresponding to a stack of $N$ D3-branes that source the $A_4$ potential is given by:
\bea
ds_{10}^2 = H(r)^{-1/2} \( -dt^2 + d\vec{x}^{\, 2}_{3} \) + H(r)^{1/2} \( dr^{2} + r^{2} d\Omega_{5}^2 \)\ :\quad H(r) = 1 + \l\, {\ell_s^4 \over r^{4}}\ ,
\eea
where $\l = 4\pi g_{s} N$. The horizon of this black brane geometry is located at $r=0$. We then introduce a new radial coordinate $z$ such that $r = \ell_s^2/z$ and work in the near-horizon or decoupling limit $\ell_s \to 0$ (such that $\l / \ell_s^4 \to \infty$). In this limit, the four-dimensional worldvolume theory on the D3-branes decouples from the closed string modes and becomes $\mN = 4$ super Yang-Mills at leading order. The parameter $\l$ becomes the 't Hooft coupling of the gauge theory with $N$ the rank of the gauge group. In the bulk, the resulting near-horizon geometry is $AdS_{5} \times S^{5}$ parametrised as:
\bea
ds^{2}_{10} = \ell_s^2 \( {1 \over \sqrt{\l}}\, {-dt^2 + d\vec{x}^{\, 2}_{3} \over z^{2}} + \sqrt{\l}\, {dz^{2} \over z^{2}} + \sqrt{\l}\, d\Omega_{5}^2 \)\ .
\eea
When considering perturbations or supergravity solutions around this background, the compactification on the $S^5$ results in an effective cosmological constant $\L = -6 / (\l \ell_s^4)^{1/2}$. The flat space limit of the non-compact $AdS_5$ background can then be obtained by defining: 
\bea
&&\vec{x} = \l^{1/4} \vec{y}\ , \label{x-l-y}\\
&&t = u - \sqrt{\l}\,z\ , \label{t-u-l-z}
\eea
such that:
\bea
ds^{2}_{5} = \ell_{s}^2 \( -{1 \over \sqrt{\l}}\, {du^2 \over z^{2}} + {2 dudz \over z^{2}} + {d\vec{y}_3^{\, 2} \over z^{2}} \)\ ,
\eea
and taking the limit $\l \to \infty$ under which the near-horizon metric becomes flat.\footnote
{
One can also keep the full ten-dimensional near-horizon metric and define $\theta = \rho / \l^{1/4}$ where $d\Omega_{5}^2 = d\theta^2 + \sin^2\theta\, d\Omega^{2}_4$.
}
On the gauge theory side, observables typically diverge in this limit. A simple example is the central charge $c$ of the theory. For a CFT$_4$ with an AdS$_5$ dual, this is given at strong coupling and large $N$ by \cite{Henningson:1998gx}: $c \sim (\l \ell_s^4)^{3/4} / G_5$, where the effective five-dimensional Newton constant $G_5 = G_{10} / {\rm Vol}(S^5)$. In our case we obtain: $c \sim N^2$. Since $\l \to \infty$ requires $N \to \infty$ in string perturbation theory, we have that $c$ diverges in the flat space limit.

In the case of AdS$_3$/CFT$_2$ \cite{Maldacena:1997re}, the near-horizon geometry of the D1-D5 system is given in a similar fashion by:
\bea
ds^{2}_6 = \ell_s^2 \( {1 \over \sqrt{\l}}\, {-dt^2 + dx^{2} \over z^{2}} + \sqrt{\l}\, {dz^{2} \over z^2} + \sqrt{\l}\, d\Omega_3^2 \)\quad (\ell_s \sim 0)\ :\quad \sqrt{\l} = g_6 \sqrt{N_1 N_5}\ ,
\eea
where $g_6<1$ is the effective six-dimensional string coupling (recall that in this case we first compactify the theory on a four-manifold such as a $T^4$) and $N_{1,5}$ the number of D1,5 branes. After compactifying on the $S^3$, the flat space limit of the $AdS_3$ geometry can be taken by introducing coordinates as in \eqref{x-l-y}--\eqref{t-u-l-z} and taking the limit $\l \to \infty$ under which the geometry becomes three-dimensional flat space. In the dual CFT$_2$, the central charge at strong coupling and large charges $N_{1,5}$ is given by $c = 6 N_1 N_5$ which diverges in the limit $\l \to \infty$.\\

This type of divergences associated with the flat space limit arises in the correlation functions of the dual field theory when computed holographically and it is the main purpose of this work to study the zero $\L$ limit of these observables in AdS/CFT. Since we will not be particularising the correspondence to specific theories, in order to take the limit of a dimensionless quantity we introduce a characteristic length scale $\ello$ and rewrite the AdS radius as a multiple of $\ello$ with proportionality constant $\a$:
\bea\label{alpha}
\ell &=& \a\, \ello\ .
\eea
In the specific examples given above, $\ello$ is the string length scale and $\a$ plays the role of the effective gauge coupling constant. The zero $\L$ limit in AdS/CFT then corresponds to taking $\a \to \infty$ with $\ello$ fixed.\footnote
{
Note that the same limit has been discussed in \cite{Susskind:1998vk,Polchinski:1999ry}. In the above examples, $\a \to \infty$ corresponds to the limit of large charges $N,N_i$ with the string coupling (and length scale) fixed. Also, and as emphasized in these references, this limit involves the physics of bulk and gauge theory states with large (dimensionless) energies.
}
We will make use of the relation \eqref{alpha} throughout this work to study the limit of vacuum expectation values and specific correlatores in AdS/CFT. This will be done formally and in a fashion somewhat similar to the way vevs and boundary correlators in dS/CFT are derived from corresponding AdS results. The main difference, however, is that not every bulk solution of Einstein gravity with AdS boundary conditions is mapped to an asymptotically flat solution in the zero $\L$ limit. We will discuss this aspect in the next sections. This implies that we need to restrict the space of solutions of AdS gravity to the subspace of those that admit the limit, in the sense that they result in solutions of the bulk equations of motion with $\L=0$ once the limit $\a \to \infty$ is taken. Since gravity solutions are dual to QFT states, this corresponds to restricting the Hilbert space of the field theory to some subspace, say $\tilde{\mathcal{H}}$. Furthermore, since the limit $\a \to \infty$ is taken over solutions, on the QFT side this should correspond to some limit taken over $\tilde{\mathcal{H}}$. The objective is then to derive the correspondence between the resulting states in $\tilde{\mathcal{H}}$ and those bulk solutions of asymptotically flat gravity that result from the limit $\a \to \infty$. This will be done mainly by working out the mapping between QFT observables and the asymptotics of such solutions. We will find that well-definedness of this limit seems to be a statement about states and sources on the field theory side.\\

If the bulk field is in particular the spacetime metric, the choice of possible coordinate systems is constrained by the requirement that the solution be smooth in the zero $\L$ limit. Taking this limit on the metric must correspond to switching off the boundary lapse function so that the timelike conformal boundary of the asymptotically AdS solution becomes null as $\a \to \infty$. To some extent, it is a gauge-dependent condition the requirement that the solution be mapped to an asymptotically flat one in this limit and this fact will have an interesting implication to the holographic renormalization of the bulk theory as discussed below. This restriction to the subspace of solutions with a well-defined limit implies in particular that the standard Fefferman-Graham coordinate system used in the near-boundary analysis of asymptotically AdS and dS spaces cannot be extended to derive the asymptotics of those solutions that are smooth in $\a$.\\ 

The choice of coordinates we will then make near the asymptotic boundary are the well-known Gaussian {\it null} coordinates. This gauge is closely related to Bondi coordinates and was initially introduced by Isenberg and Moncrief \cite{Moncrief:1983} in order to prove the existence of a Killing vector field in any spacetime that contains a compact null surface with closed generators. It was further elaborated in \cite{Friedrich:1998wq} in order to generalise Isenberg and Moncrief's results, as well as Hawking's rigidity theorems, to non-analytic spacetimes (see also \cite{Racz:1999ne}) and it has been extensively used in the literature in order to study the near horizon geometry of black holes (see \cite{Reall:2002bh,Kunduri:2008rs} and references therein). This gauge choice is also motivated by those investigations of the asymptotic symmetries of asymptotically flat gravity discussed above.\footnote{
See also \cite{Barnich:2012aw} for a brief overview in three dimensions.}
In this coordinate system, the Einstein field equations decompose into a set of dynamical and constraint equations that are very tedious to solve asymptotically and increase in complexity with the spacetime dimension. For this reason we will focus specifically on the case of three and four bulk dimensions, but it is straightforward to extend the procedure to any dimension. From this analysis we will obtain in particular the unique asymptotics at null infinity of all those Ricci-flat metrics that result from the zero $\L$ limit of Einstein metrics.\\

As a final remark, it should be emphasized that, unlike the case of dS/CFT, holographic renormalization does not admit a straighforward extension to the asymptotically flat case. In general, the holographic counterterms introduce divergences in $\a$ that spoil the zero $\L$ limit of the renormalized on-shell gravity action. If one insists that the action be finite in this limit, further counterterms are needed to restore the well-definedness of the limit. The latter are finite in the holographic regulator and therefore are associated with a choice of renormalization scheme on the field theory side. These finite counterterms are covariant with respect to diffeomorphisms that preserve the spacelike foliation induced at the boundary by the bulk theory, but break invariance of the renormalized action with respect to diffeomorphisms that are not foliation-preserving. This reflects the fact that the well-definedness of the limit is a gauge-dependent requirement. We will analyse the effect of these anomalous counterterms on the holographic Ward identities of the field theory in the case of four bulk dimensions. A pathological aspect of this type of counterterms is that they introduce divergent contact terms in the two-point correlators of scalar operators. We will derive this result in section \ref{correlator}.\\

In the next section we introduce our coordinate system and determine the unique asymptotics of the bulk spacetime metric by solving the vacuum Einstein equations with a negative cosmological constant in a neighbourhood of the asymptotic boundary. We will then discuss the zero $\L$ limit of the solution and briefly compare the spacetime asymptotics in this limit with the standard definitions of asymptotic flatness at null infinity.\\

Section \ref{holog_en_tensor} contains the main results of this work. We will holographically renormalize the bulk gravity action in three and four dimensions and use the AdS/CFT prescription to compute the vacuum expectation value of the QFT energy tensor. The objective will be to analyse the correspondence between the metric asymptotics and the boundary data in the zero $\L$ limit and to address the issues associated with this limit. The three dimensional case is the best controlled setting and no major problems arise. The holographic Weyl anomaly in the zero $\L$ limit will be of particular interest in this case. The integrated anomaly is still a topological invariant and we will be able to obtain in this limit the Virasoro central charge that arises in the central extension of $\mathfrak{bms}_{3}$ as the proportionality constant between the anomaly and a geometric invariant. We will then apply our results to the zero $\L$ limit of the BTZ solution, which represents a three-dimensional flat cosmological solution, and find a matching between the energy and momentum of the QFT and those of the bulk theory.\\

In the case of four bulk dimensions we will find that the holographic renormalization spoils the zero $\L$ limit of the gravity action, as described above, by terms that are finite in the regulator and which can only be subtracted by a finite counterterm that partially breaks diffeomorphism invariance of the action. We will then compute the holographic energy tensor and address the issues associated with its zero $\L$ limit. Of particular interest will be the holographic Ward identities and the way they are affected by the anomalous counterterm. In the absence of the latter, the trace of the QFT energy tensor vanishes, but it is modified by a total derivative in the presence of the anomalous counterterm. As an application of our results, we will derive specifically the asymptotics of the Kerr solution and find a matching between the energy and momentum of this solution and those of the dual state of the field theory. At the end of this section we will address and solve the issues associated with the presence of null boundaries in the spacetime in addition to the asymptotic conformal boundary.\\

Finally, in section \ref{matter} we analyse the case of a non-backreacting massive bulk field propagating in AdS in a coordinate system appropriate to the zero $\L$ limit. We renormalize holographically the bulk action for the field, address its zero $\L$ limit and compute the vacuum expectation value and the renormalized two-point correlator of the dual scalar operator. As in the case of the spacetime metric, the objective will be to analyse the zero $\L$ limit taken on the vev and correlator. For ``large" values of the conformal weights, contact terms associated with the anomalous counterterms arise in the two-point function, but vanish away from coincident points in time. In general, the two-point functions will be consistent with that of a conformal operator in two dimensions less in this limit.\\

\section{Spacetime asymptotics}\label{spacetime_asympt}

\subsection{Choice of coordinates}\label{coordinates}

We start with the action for the spacetime metric in $d+2$ dimensions written in the form:
\bea\label{action}
16\pi G_{0}\, S &=& \int_{\mM} d^{d+2}x\, \sqrt{G} \( {d(d+1) \over \alpha^{2}\ello^{2}} + R[G] \) + 2 \int_{\del \mM} d^{d+1}x \sqrt{q}\, Q\ ,
\eea
where the cosmological constant $2\L = -d (d+1) / (\alpha\ello)^{2}$ and where $q_{ab}$ and $Q_{ab}$ are the induced metric and extrinsic curvature of the boundary. As discussed in the previous section, we have rewritten the AdS radius $\ell$ as in \eqref{alpha} so that $\L$ is switched off by taking the limit $\alpha \to \infty$ of the dimensionless parameter $\alpha$.\\

In order to solve asymptotically the Einstein field equations we introduce Gaussian {\it null} coordinates $x^{\mu} = (r,x^{a}) = (r,u,x^{i})$ near the boundary $r=\infty$ of the manifold. In such gauge, the spacetime metric has the form \cite{Moncrief:1983,Friedrich:1998wq}:
\bea\label{G_null}
ds^{2}_{d+2} &=& G_{\mu\nu}dx^{\mu}dx^{\nu} \nn\\
&=& - \phi\, du^{2} + 2 du dr + \g_{ij}(dx^{i} + \s^{i}du)(dx^{j} + \s^{j}du)\ .
\eea
where the metric components depend on all the coordinates, the spatial metric $\g_{ij}$ is positive-definite and the function $\phi$ is positive by definition. The vector $\phi^{-1/2}(\partial_{u} - \s^{i}\partial_{i})$ is future-directed timelike with unit norm. The manifold is defined to be foliated by a family of timelike hypersurfaces labelled by the coordinate $r$ and by a family of null surfaces of constant $u$. Each submanifold $\{r=constant\}$ is foliated by spacelike surfaces of constant time coordinate $u$. All the above statements hold asymptotically. In appendix \ref{appAB} we briefly deduce this coordinate system via an ADM analysis of the metric, but it all comes down to using diffeomorphisms in order to bring the metric to the desired form. In the case of asymptotically flat metrics in Gaussian null coordinates, the metric components behave asymptotically as \cite{Hollands:2003ie,Hollands:2003xp,Ishibashi:2007kb,Tanabe:2011es,Barnich:2009se}:
\bea
\g_{ij}(r,u,x) &=& r^{2} \( \g_{(0)ij}(u,x) + \op(r^{<0}) \)\ ,\\
\phi(r,u,x) &=& \phi_{(0)}(u,x) + \op(r^{<0})\ ,\\
\s^{i}(r,u,x) &=& \op(r^{<0})\ ,
\eea
with null infinity given by $r= +\infty$, so we will be interested in solving the field equations around $1/r = 0$ with $\L$ switched on and in the end analyse the limit $\alpha \to \infty$.\\

Before doing so, we introduce a new coordinate $z := \ello^{2} / r$ and also define $g_{ij} := (z/\ello)^{2} \g_{ij}$ and $\varf := (z/\ello)^{2} \phi$ such that:
\bea\label{g_null}
ds^{2}_{d+2} &=& {\ello^{2} \over z^{2}} \( - \varf\, du^{2} - 2 du dz + g_{ij}(dx^{i} + \s^{i}du)(dx^{j} + \s^{j}du) \)\ .
\eea
The decomposition of the Ricci tensor $R_{\mu\nu}[G]$ in terms of the metric components $\varf, g_{ij}$ and $\s^{i}$ is given in appendix \ref{appB}. If we solve the field equations $R_{\mu\nu} = -(d+1)/ (\alpha \ello)^{2} G_{\mu\nu}$ around $z=0$ at leading and first subleading order, we find:
\bea
\varf(z,u,x) &=& {1 \over \a^{2}} + z\, \varf_{(1)} + \op(z^{2})\ ,\\[5pt]
g_{ij}(z,u,x) &=& g_{(0)ij} + z\, g_{(1)ij} + \op(z^{2})\ ,\\[5pt]
\s^{i}(z,u,x) &=& \s_{(0)}^{i} + \op(z^{2})\ ,
\eea
where the coefficients $\varf_{(1)}(u,x)$, $g_{(0)ij}(u,x)$ and $\s^{i}_{(0)}(u,x)$ are completely arbitrary (\ie integration constants) and where $g_{(1)ij}(u,x)$ obeys the equation:
\be\label{g_1}
{1 \over \alpha^{2}}\, g_{(1)ij} = (\partial_{u} - \Lie_{\s_{(0)}}) g_{(0)ij} + \varf_{(1)} g_{(0)ij}\ ,
\ee
with $\Lie$ the Lie derivative. The asymptotic behaviour of the metric components therefore implies that the metric \eqref{g_null} is (at least $C^{2}$) conformally compact,\footnote{
See appendix \ref{appA}.}
with defining function $z/\ello$ and conformal boundary $z = 0$. For $\alpha^{-2}>0$ the boundary is timelike and it becomes null in the zero $\L$ limit. We also find from \eqref{g_1} that the leading order term $g_{(0)ij}$ becomes constrained in the case $\a^{-1}=0$. \\ 

We will now use the freedom in the choice of defining function and introduce a more judicious one. We define a new coordinate $\bar{z} := z N_{(0)}$, with $N_{(0)}(u,x)$ an arbitrary but positive smooth function of $u$ and $x^{i}$. Under this change of coordinates the spacetime metric becomes:
\be\label{bar_null}
ds^{2}_{d+2}\, =\, {\ello^{2} \over \bar{z}^{2}}\, \Big( - \bar{\varf} N_{(0)} du^{2} - 2N_{(0)} dud\bar{z} + \bar{g}_{ij} \( dx^{i} + \bar{\s}^{i}du \) \( dx^{j} + \bar{\s}^{j}du \) \Big)\ ,
\ee
where:
\bea
\bar{\varf}N_{(0)} &=& \varf N_{(0)}^{2} - 2\bar{z}\, (\partial_{u} - \Lie_{\s}) N_{(0)} + \bar{z}^{2} | \na_{g} \log N_{(0)} |^{2}\ ,\\[5pt]
\bar{\s}^{i} &=& \s^{i} + \bar{z}\, N_{(0)}^{-2} g^{ij}\partial_{j}N_{(0)}\ ,\\[5pt]
\bar{g}_{ij} &=& N_{(0)}^{2} g_{ij}\ .
\eea
The metric component $\bar{\varf}$ therefore has the asymptotics:
\bea
\bar{\varf} &=& \bar{\varf}_{(0)} + \bar{z}\, \bar{\varf}_{(1)} + \op(\bar{z}^{2})\ :\\[5pt]
\bar{\varf}_{(0)} &=& {1 \over \a^{2}}N_{(0)}\ ,\\[5pt]
\bar{\varf}_{(1)} &=& \varf_{(1)} -2 (\partial_{u} - \Lie_{\s_{(0)}}) \log N_{(0)}\ .
\eea
We then choose our function $N_{(0)}(u,x)$ such that:\footnote{
Note that if we write: $N_{(0)} := N_{(0)1}N_{(0)2}$ such that $(\partial_{u} - \Lie_{\s_{(0)}}) \log N_{(0)2} = 0$, we still have the freedom of choosing $N_{(0)2}(u,x)$ in the space orthogonal to the vector $\partial_{u} - \s_{(0)}^{i}\partial_{i}$.}
\be\label{N0}
(\partial_{u} - \Lie_{\s_{(0)}}) \log N_{(0)}^{2}\, =\, \varphi_{(1)}\ ,
\ee
which results in the asymptotics: $\bar{\varphi} = \bar{\varf}_{(0)} + \op(\bar{z}^{2})$. Recall that the coefficient $\varf_{(1)}$ was an integration constant and therefore $N_{(0)}$, or $\bar{\varf}_{(0)}$, remains arbitrary, \ie undetermined by the field equations.\\

From this choice of defining function $\bar{z}/\ello$ and the requirement that the metric components be well-defined in the limit $\a \to \infty$, it follows from equation \eqref{g_1} that:
\be
(\partial_{u} - \Lie_{\bar{\s}_{(0)}}) \bar{g}_{(0)ij}\, =\, 0\qquad (\a \to \infty)\ ,
\ee
and therefore the timelike vector $\partial_{u} - \bar{\s}^{i}\partial_{i}$ is an asymptotic Killing vector of the spatial metric $\bar{g}_{ij}$ in this limit. Furthermore, with such defining function, the normal to the boundary $m^{\mu} = \widetilde{G}^{\mu\nu}\partial_{\nu}\bar{z}$ in the conformal embedding $\widetilde{G}_{\mu\nu} = (\bar{z}/\ello)^{2} G_{\mu\nu}$ is shear, expansion and vorticity free in the zero $\L$ limit, and therefore totally geodesic: 
\be
\lim_{\a \to \infty} \widetilde{\na}_{\nu}m^{\mu} = \op(\bar{z}).
\ee
This is the standard gauge used in the study of asymptotically flat spacetimes (see \eg \cite{wald:279}). More importantly, with our choice of coordinates the boundary metric (with components $N_{(0)}, \bar{g}_{(0)ij}$ and $\bar{\s}_{(0)}^{i}$) is completely unconstrained for finite $\alpha$. In the next sections this feature will allow us to take the variations of the on-shell action with respect to all components of the boundary metric in order to derive the holographic energy tensor. The metric in the originial Gaussian null coordinates \eqref{g_null} therefore corresponds to the metric \eqref{bar_null} with the lapse function ${1 \over \alpha}N_{(0)}$ of the boundary fixed by diffeomorphisms to a constant.\\

We will drop the bar notation from now on and work with the spacetime metric in the final form:
\bea\label{metric}
ds^{2}_{d+2} &=& G_{\mu\nu} dx^{\mu}dx^{\nu} \nn\\[5pt]
&=& {\ello^{2} \over z^{2}}\, \Big( - \varf N_{(0)} du^{2} - 2N_{(0)} dudz + g_{ij} \( dx^{i} + \s^{i}du \) \( dx^{j} + \s^{j}du \) \Big)\ ,
\eea
where:
\bea
\varf &=& {1 \over \a^{2}} N_{(0)} + \op(z^{2})\ ,\\[5pt]
g_{ij} &=& g_{(0)ij} + \op(z)\ ,\\[5pt]
\s^{i} &=& \s_{(0)}^{i} + \op(z)\ .
\eea
The induced metric $q_{ab}$ of the surfaces of constant $z$ near the boundary $z=0$ is given by:
\bea\label{induced_q_leading}
ds^{2}_{d+1}\, =\, q_{ab}dx^{a}dx^{b} &=& {\ello^{2} \over z^{2}} \( - {1 \over \a^{2}}N_{(0)}^{2} du^{2} + g_{(0)ij} \big( dx^{i} + \s_{(0)}^{i}du \big) \big( dx^{j} + \s_{(0)}^{j}du \big) + \op(z) \) \nn\\
&:=& {\ello^{2} \over z^{2}}\, \Big( q_{(0)ab} + \op(z) \Big) dx^{a}dx^{b}\ ,
\eea
where $q_{(0)ab}$ represents the metric tensor of the conformal boundary and is the source for the energy tensor of the dual quantum field theory. From the determinant $\sqrt{q_{(0)}} = {1 \over \a}N_{(0)}\sqrt{g_{(0)}}$ we see clearly that the timelike boundary becomes null in the zero $\L$ limit. 

\subsection{Asymptotic solution}\label{asympt_sol}

The decomposition of the Ricci tensor $R_{\mu\nu}[G]$ in our coordinate system \eqref{metric} is given in appendix \ref{appB}. If we solve the Einstein field equations around $z=0$ with the cosmological constant switched on, we find that the asymptotics of the metric is uniquely determined:
\begin{align}
&g_{ij}\, =\, g_{(0)ij} + z\, g_{(1)ij} + z^{2}g_{(2)ij} + ... + z^{d+1}g_{(d+1)ij} + z^{d+1}\log z\, \tilde{g}_{(d+1)ij} + ...\ , \label{gij}\\[5pt]
&\varf\, =\, \varf_{(0)} + z^{2}\varf_{(2)} + z^{3}\varf_{(3)} + ... + z^{d+1}\varf_{(d+1)} + z^{d+1}\log z\, \tilde{\varf}_{(d+1)} + ...\ ,\label{varf}\\[5pt]
&\s^{i}\, =\, \s_{(0)}^{i} + z\, \s_{(1)}^{i}+ z^{2} \s_{(2)}^{i} + ... + z^{d+1} \s_{(d+1)}^{i} + z^{d+1}\log z\, \tilde{\s}_{(d+1)}^{i} + ... \label{sigma}
\end{align}
Note that the expansions in $z$ are not predetermined but uniquely fixed by the equations.\footnote{
An arbitrary term $z\, \varf_{(1)}$ in the expansion of $\varf$ can always be cancelled by a choice of $N_{(0)}$ as described above. See, however, the discussion in section \ref{ren_4}. There is also the possibility of including terms proportional to $\delta_{\L,0}$ that vanish for all finite values of $\alpha$, but we discard such terms since we are only interested in solutions for which the limit $\alpha \to \infty$ exists.\\[-5pt]} 
The coefficients $g_{(0)ij}$, $\varf_{(0)}$ and $\s_{(0)}^{i}$, which we will denote collectively by $G_{(0)\mu\nu}$, are integration constants and therefore completely arbitrary functions of $u$ and $x^{i}$. These are the standard non-normalizable modes, or sources, of asymptotically AdS metrics.\footnote{
See \cite{deHaro:2000xn,Skenderis:2002wp} for a review of the asymptotics of such metrics in Fefferman-Graham coordinates.\\[-5pt]}
The coefficients $g_{(d+1)ij}$, $\varf_{(d+1)}$ and $\s_{(d+1)}^{i}$, denoted collectively by $G_{(d+1)\mu\nu}$, are arbitrary up to specific constraints and are the standard normalizable modes. These will be associated to the different components of the holographic energy tensor and the constraints to its Ward identities. The coefficients of the logarithms, which we will denote by $\tilde{G}_{(d+1)\mu\nu}$, are only non-vanishing for odd values of $d>1$, and in such case are local functionals of the sources $G_{(0)\mu\nu}$. The remaining coefficients $G_{(n)\m\nu}$, as well as the constraints on the $G_{(d+1)\mu\nu}$, are all local functionals of the sources.

The expressions for the coefficients at first and second subleading orders are given by:
\begin{align}
&{1 \over 2\,\a^{2}}\, g_{(1)ij}\, =\, k_{(0)ij}\ ,\label{g1}\\[5pt]
&{d-1 \over \a^{2}}\, g_{(2)ij}\, =\, (d-2) \( k_{(1)ij} - {1 \over d}\, g_{(0)ij} \Tr[g_{(0)}^{-1}k_{(1)}] \) - \( R_{(0)ij} - {1 \over d} g_{(0)ij} R_{(0)} \) \nn \\
&\hspace{2em}+ {1 \over 4\alpha^{2}} \( -g_{(1)ij} \Tr[g_{(0)}^{-1}g_{(1)}] + {1 \over d} g_{(0)ij} \Tr^{2}[g_{(0)}^{-1}g_{(1)}] + 2 \big( g_{(1)}\cdot g_{(1)} \big)_{ij} + {d-3 \over d}\, g_{(0)ij} \Tr[g_{(1)}\cdot g_{(1)}] \) , \label{g2}\\
&\varf_{(0)}\, =\, {1 \over \a^{2}} N_{(0)}\ , \label{phi0}\\[5pt]
&{d(d-1) \over N_{(0)}}\, \varf_{(2)}\, =\, -2(d-1) \Tr[g_{(0)}^{-1}k_{(1)}] + R_{(0)} + {1 \over 4\a^{2}} \( \Tr^{2}[g_{(0)}^{-1}g_{(1)}] + (2d-3) \Tr[g_{(1)}\cdot g_{(1)}] \) ,\label{phi2}\\[5pt]
&\s_{(1)i}\, =\, \partial_{i}N_{(0)}\ ,\label{sigma1}\\[5pt]
&{2(d-1) \over N_{(0)}}\, \s_{(2)i}\, =\, - \cov_{j} \big( g_{(0)}^{-1}g_{(1)} \big)^{j}_{i} + \partial_{i} \Tr[g_{(0)}^{-1}g_{(1)}] - (d-1) g_{(1)ij}\cov^{j}\log N_{(0)}\ , \label{sigma2}
\end{align}
where $R_{(0)ij} := R_{ij}[g_{(0)}]$ and $\cov_{i} g_{(0)jk} := 0$, and where the indices are raised and lowered with $g_{(0)ij}$ and the inner product taken with respect to the latter. It is also useful to emphasize that in three and four bulk dimensions the coefficient $g_{(2)ij}$ simplifies as:\footnote{
For $d=1$ the coefficient $g_{(2)ij}$ is totally determined by the trace constraint: $\Tr[g_{(0)}^{-1}g_{(2)}] = \qt \Tr[g_{(0)}^{-1}g_{(1)}g_{(0)}^{-1}g_{(1)}]$ that follows from equations \eqref{Rzz} and \eqref{Einstein_eqs}. For $d=2$ we use the matrix identity \eqref{matrix_id} to simplify equation \eqref{g2}.}
\bea\label{g2_g1}
g_{(2)ij} &=& \qt \big( g_{(1)} \cdot g_{(1)} \big)_{ij} \qquad (d=1,2)\ .
\eea

In appendix \ref{appB} where the decomposition of the Ricci tensor is given we introduced the tensor $k_{ij}$ defined as:
\be
k_{ij} := {1 \over 2N_{(0)}} \( \partial_{u} - \Lie_{\s} \) g_{ij}\ .
\ee
This tensor is proportional to the extrinsic curvature of the surfaces of constant time on each submanifold $\{z = constant\}$. From the metric asymptotics, $k_{ij}$ admits the expansion:
\bea
k_{ij} &=& k_{(0)ij} + z\, k_{(1)ij} + ... + z^{d+1} k_{(d+1)ij} + z^{d+1}\log z\, \tilde{k}_{(d+1)ij} + ...
\eea
Each coefficient $k_{(n<d+1)ij}$ can be written in terms of quantities defined at the boundary. For the first and second subleading orders we find:\footnote{
In the final expression for $k_{(1)}$ we made use of equations \eqref{g1} and \eqref{sigma1} and of the standard Gauss-Codazzi identities.}
\bea
k_{(0)ij} &=& {1 \over 2N_{(0)}} \( \partial_{u} - \Lie_{\s_{(0)}} \) g_{(0)ij}\, =\, {1 \over \alpha}\, K_{(0)ij}\ ,\label{k0}\\[5pt]
k_{(1)ij} &=& {1 \over 2N_{(0)}} \Big[ \( \partial_{u} - \Lie_{\s_{(0)}} \) g_{(1)ij} - \Lie_{\s_{(1)}} g_{(0)ij} \Big]\, =\, \Lie_{n_{(0)}} K_{(0)ij} - \cov_{i} a_{(0)j} - a_{(0)i}a_{(0)j} \nn\\
&=& R_{ij}[q_{(0)}] - R_{ij}[g_{(0)}] + 2 \( K_{(0)}\cdot K_{(0)} \)_{ij} - K_{(0)}K_{(0)ij}\ ,\label{k1}
\eea
where $K_{(0)ij}$ is the extrinsic curvature of the surfaces of constant time at the boundary, $n_{(0)}^{a}\partial_{a} = \a N_{(0)}^{-1}\( \partial_{u} - \s_{(0)}^{i}\partial_{i} \)$ is the unit normal to these surfaces and $a_{(0)i} = \partial_{i} \log N_{(0)}$ the acceleration. Also, $R_{ij}[q_{(0)}]$ are the spatial components of the Ricci tensor $R_{ab}[q_{(0)}]$ of the boundary metric and we will see in section \ref{3_dims} that its trace will represent the holographic Weyl anomaly in three bulk dimensions in the zero $\L$ limit.\\

Let us start by discussing the solutions for the coefficients $G_{(n)\mu\nu}$. If the cosmological constant is non-vanishing, from the expressions \eqref{g1}--\eqref{sigma2} it follows that these coefficients are indeed local functionals of the sources $G_{(0)\mu\nu}$. On the other hand, in the case $\a^{-1}=0$ we find that the algebraic equation for a given coefficient $g_{(n)ij}$ becomes a differential equation for the coefficient $g_{(n-1)ij}$ and therefore the coefficients $G_{(n)\mu\nu}$ become non-local functionals of the sources. This feature is responsible for the fact that holographic renormalization cannot be extended in a straightforward way to Ricci-flat spacetimes (see \eg \cite{Skenderis:2002wp,deHaro:2000wj}) and we will discuss this aspect in the next sections. The asymptotic expansions \eqref{gij}--\eqref{sigma} together with the equations for the coefficients with $\a^{-1} = 0$ represent the unique asymptotics at null infinity of all Ricci-flat metrics that result from Einstein metrics in the zero $\L$ limit. 

In the case of $\a$ finite, the sources $G_{(0)\mu\nu}$ are arbitrary functions, so we may have solutions of the equations of motion with $\L$ switched on that diverge as $\a \to \infty$. The same applies to the normalisable modes $G_{(d+1)\mu\nu}$. We are interested in those Ricci-flat metrics that result from the zero $\L$ limit, so we need to restrict our space of solutions of Einstein metrics to the subspace of those that admit the limit. For this purpose we require that the coefficients in the expansions \eqref{gij}--\eqref{sigma} be non-divergent as $\a \to \infty$. For the normalisable modes, it is sufficient to restrict to those configurations that satisfy: $G_{(d+1)\mu\nu} = \op(\a^{0})$. On the other hand, since the coefficients $G_{(n)\mu\nu}$ and $\tilde{G}_{(d+1)\mu\nu}$ are all functionals of the sources up to order $z^{d+1}$, this requirement imposes specific behaviours in $\a$ of the time derivatives of $g_{(0)ij}$. From equations \eqref{g1} and \eqref{g2} for example it follows that:
\bea
&&(\partial_{u} - \Lie_{\s_{(0)}}) g_{(0)ij}\, =\, \op(\a^{-2})\ , \label{g0-alpha}\\[5pt]
&&(d-2) \( k_{(1)ij} - {1 \over d}\, g_{(0)ij} \Tr[g_{(0)}^{-1}k_{(1)}] \) - \( R_{(0)ij} - {1 \over d} g_{(0)ij} R_{(0)} \)\, =\, \op(\a^{-2})\ ,
\eea
with $k_{(1)ij}$ expressed in terms of $(\partial_{u} - \Lie_{\s_{(0)}})^{2}g_{(0)ij}$ by using equations \eqref{g1}, \eqref{sigma1} and the first identity in equation \eqref{k1}. From a holographic perspective, well-definedness of the gravity solutions in the zero $\L$ limit then translates into a statement about the sources and states on the QFT side. We will find another example of such a correspondence between the existence of the zero $\L$ limit of bulk solutions and the time behaviour of the sources when we discuss non-backreacting matter in AdS in section \ref{matter_solution}.\\

It is worth comparing the asymptotic behaviour \eqref{gij}--\eqref{sigma} of the spacetime metric in the limit $\a \to \infty$ with the standard definitions of asymptotic flatness at null infinity. For vacuum spacetimes in odd bulk dimensions higher than four, half integer powers in the asymptotics of the metric (starting at order $z^{d/2}$ in the conformal embedding $\widetilde{G}_{\mu\nu}$) are postulated in the definitions of asymptotic flatness so that linearized pertubations of the metric preserve the definition when the spacetime contains gravitational radiation \cite{Hollands:2003ie,Hollands:2003xp,Hollands:2004ac} (see also \cite{Ishibashi:2007kb,Tanabe:2011es,Godazgar:2012zq}). The absence of half integer powers in the asymptotics \eqref{gij}--\eqref{sigma} seems to indicate that vacuum, radiating spacetimes %amenable to perturbations 
cannot be obtained from the zero $\L$ limit of Einstein metrics in five or higher odd dimensions. It is also worth emphasizing the presence of the inhomogeneous logarithmic terms in the asymptotics of the metric,\footnote{ 
The fact that the logarithmic terms are non-vanishing in five or higher odd bulk dimensions is associated to the fact that the conformal anomaly of the dual field theory is non-vanishing in even ($d+1$) boundary dimensions.}
as well as the fact that the first subleading terms in the asymptotic expansions start at order $z$. The logarithmic terms are usually absent in the definitions of asymptotic flatness (see, however, the discussion in \cite{Chrusciel:1993hx}) and the first subleading terms are usually postulated to start at order $z^{d/2}$ both in even and odd bulk dimensions.\\

The above results suggest that vacuum spacetimes containing gravitational radiation in the sense of \cite{Hollands:2003ie,Hollands:2003xp,Hollands:2004ac} cannot be obtained from the zero $\L$ limit of Einstein metrics in five or higher odd dimensions. This subject will be analysed in more detail elsewhere, but we can already remark that this result is interesting in the context of AdS/CFT. One possible way to study holographically Hawking radiation from a black hole in AdS is by coupling the dual CFT to another system such as a heat bath that draws energy away from the original field theory and which is used to measure the outgoing Hawking radiation. In the bulk, this construction corresponds to letting the radiation cross the boundary and then measuring it. This type of setup can in principle be used to understand more clearly the information loss problem in AdS. One could then hope to draw lessons for the corresponding problem in flat space by scaling the coupling constant in the coupled system in a specific way and taking the zero $\L$ limit. One would then naively expect to obtain a radiating solution in the bulk after this limit is taken, but the above results suggest that this will not be the case.\footnote
{
I would like to thank the referee for raising this point and to Marika Taylor for pointing out to me this setup.\\[-5pt]
}\\

As discussed above, the integration constants of the dynamical equations of motion for the metric are the modes $G_{(0)\mu\nu}$ and $G_{(d+1)\mu\nu}$ for non-vanishing $\L$. On the other hand, we have also seen that the algebraic equation for a given $g_{(n)ij}$ is of the form:
\bea\label{g_n}
{d+1-n \over \a^{2}}\, g_{(n)ij} &=& \omega\, \big( \partial_{u} - \Lie_{\s_{(0)}} \big) g_{(n-1)ij} + ...
\eea
where the ellipsis denote lower order terms and $\omega$ is some proportionality factor. In the limit $\a \to \infty$ the algebraic equation for $g_{(n)}$ therefore results in the differential equation that defines the coefficient $g_{(n-1)}$. However, from the dynamical equation \eqref{Rij} and \eqref{E-eq} for the metric component $g_{ij}$ we find that $\omega$ is always proportional to $2(n-1)-d$.\footnote{
This fact follows from the two terms $4k_{ij}' - 2d/z\, k_{ij}$ in the last line of \eqref{Rij}.} 
This implies that the coefficient $g_{(d/2)ij}$, or more precisely $k_{(d/2)ij}$, becomes the integration constant in the limit $\a \to \infty$ instead of $g_{(d+1)ij}$. For odd values of $d$, $d/2$ is half-integer, so there is no coefficient $g_{(d/2)ij}$ in the expansion \eqref{gij}. This would be the leading mode that spoils smoothness of the metric in the definitions of asymptotic flatness at null infinity in odd dimensions as discussed above. On the other hand, the coefficient $g_{(d/2)ij}$ is non-vanishing for even bulk dimensions. Just as the integration constants $G_{(d+1)\mu\nu}$ are associated to the different components of the holographic energy tensor for the case of non-vanishing $\L$, the coefficient $g_{(d/2)ij}$, or $k_{(d/2)ij}$, will be related to the spatial components of the QFT energy tensor in even dimensions in the limit $\a \to \infty$. We will derive this result for the case $d=2$ in section \ref{vevs_4}.\\

As a final remark, we will not discuss here the asymptotic symmetry group BMS$_{d+2}$ of the metric \eqref{metric} in the limit $\a \to \infty$ and the associated asymptotic charges, but we would still like to point out that our gauge-fixed metric is not invariant under boundary diffeomorphisms (\ie transformations of the form $u \to \tilde{u}(u,x), x^{i} \to \tilde{x}^{i}(u,x)$) that do not preserve the foliation in surfaces of constant $u$. In fact, there is no gauge one can choose -- where the gauge freedom has been completely fixed -- that simultaneously admits a well-behaved zero $\L$ limit and is invariant under general boundary diffeomorphisms. This is so because the asymptotic boundary should approach a null manifold in the limit $\a \to \infty$ and therefore any gauge admitting a well-behaved zero $\L$ limit necessarily singles out the timelike direction as a preferred direction over the remaining boundary coordinates.

This observation implies in particular that the subgroup of the asymptotic symmetry group of the metric consisting of boundary diffeomorphisms must be foliation-preserving:\footnote{
See \cite{Horava:2009uw} and references therein for a review of foliation preserving diffeomorphisms.}
\bea
\begin{cases}\label{fol_preserv}
u\, \to\, \tilde{u}(u)\ ,\\
x^{i}\, \to\, \tilde{x}^{i}(u,x)\ .
\end{cases}
\eea
Furthermore, since full covariance, or gauge invariance, is weakened by the requirement that the limit $\a \to \infty$ be well-defined, the spectrum of possible holographic counterterms that we can have in the counterterm action is widened. We will see in the next sections that the canonical, fully covariant counterterms \cite{deHaro:2000xn} are sufficient to render the on-shell gravity action finite once the regulator is removed, but if we also require that the action be finite in the limit $\a \to \infty$, further counterterms are needed. The latter preserve invariance of the action under all but those diffeomorphisms that are not foliation-preserving.

Finally, it should be emphasized that the asymptotic symmetry group of the metric contains a subgroup that generates conformal transformations at the boundary. This consists of the transformation $z \to \bar{z} = z\, \Omega(u,x^{i})$ together with $x^{a} \to \bar{x}^{a} = X^{a}(u,x^{i}) + z\, Y^{a}(u,x^{i}) + \op(z^{2})$, where the functions $X^{a}$ are defined to satisfy: $q_{(0)ab}dX^{a}dX^{b} = \Omega^{2} q_{(0)ab}dx^{a}dx^{b}$, and where the functions $Y^{a}(u,x^{i})$ can be chosen so that the transformation is asymptotically a symmetry.\footnote{
See also \cite{Imbimbo:1999bj} about the relation between bulk diffeomorphisms and conformal transformations at the boundary in the context of AdS/CFT.}
%\bea
%&&q_{(0)ab}dX^{a}dX^{b} = \omega^{2} q_{(0)}dx^{a}dx^{b}\ ,\\
%&&\omega = \partial_{u}X^{u} - \omega/N_{(0)}\,q_{(0)ua}Y^{a}\ ,\\
%&&q_{(0)ab}Y^{a}Y^{b} = 2\omega\, N_{(0)}Y^{u}\ ,\\
%&&q_{(0)ia}Y^{a} = \omega N_{(0)}\partial_{i}X^{u}\ ,
%\eea
%then the metric changes as:
%\bea\label{G_new}
%G_{\mu\nu}dx^{\mu}dx^{\nu} \to {\ello^{2} \over z^{2}} \( -2\omega N_{(0)} du dz + \omega^{2}q_{(0)ab}dx^{a}dx^{b} + \op(z) \)\ .
%\eea
%The asymptotic Killing vectors associated to this transformation generate conformal transformations at the boundary.

\section{Holographic energy tensor}\label{holog_en_tensor}

\subsection{Preliminaries}\label{onshell_action}

\qquad In order to compute the vacuum expectation value of the dual QFT energy tensor via the AdS/CFT prescription, we need to evaluate the gravitational action \eqref{action} on-shell and subtract the divergences via holographic renormalization \cite{deHaro:2000xn,Skenderis:2002wp}. In the previous section we found that the coefficients in the asymptotic solution for the metric become non-local functionals of the sources in the limit $\a \to \infty$ and emphasized that this feature prevents the holographic renormalization of the action in the case of a vanishing cosmological constant. Indeed, if we attempt to renormalize the gravity action \eqref{action} with $\a^{-1} = 0$, we find that the divergent terms are functionals of the coefficients $G_{(n)\mu\nu}$. In this way, the divergences are not local functionals of the sources and therefore cannot be subtracted by local, covariant counterterms. On the other hand, it is possible to renormalize the action with the cosmological constant switched on and in the end analyse the limit $\a \to \infty$, so this is the procedure we will follow.\\

The induced metric and extrinsic curvature $q_{ab}$ and $Q_{ab}$ of the surfaces of constant $z$ are given by:
\begin{align}
&q_{ab}dx^{a}dx^{b}\, =\, {\ello^{2} \over z^{2}}\, \Big( - \varf N_{(0)} du^{2} + g_{ij} \( dx^{i} + \s^{i}du \) \( dx^{j} + \s^{j}du \) \Big) \ ,\label{induced_q}\\[5pt]
&Q_{ab}\, =\,  {1 \over 2\beta}\, \big( \partial_{z} - \Lie_{\beta n} \big) q_{ab} \nn\\[5pt]
&= n_{a}n_{b}\, {1 \over 2\beta\varf}\, \bigg[ - \varf ' + {2 \over z}\, \varf + {1 \over N_{(0)}} \( (\s_{i}\s^{i})' - {2 \over z}\,\s_{i}\s^{i} \) + (\partial_{u} - \Lie_{\s}) \log N_{(0)} - (\partial_{u} + \Lie_{\s}) \log \varf - {2 \over \varf} \s^{i}\s^{j} k_{ij} \bigg] \nn\\[5pt]
&- n_{(a} \partial_{b)}x^{i} {1 \over N_{(0)}}\, \bigg[ \s_{i}' - {2 \over z} \s_{i} - N_{(0)} \partial_{i} \log \varf - {2 \over \varf}\,N_{(0)} \s^{j} k_{ij} \bigg] \nn\\[5pt]
&+ \partial_{a}x^{i}\partial_{b}x^{j} {\b\varf \over 2N_{(0)}}\, \bigg[ g_{ij}' - {2 \over z}\, g_{ij} - {2 \over \varf}N_{(0)}k_{ij} \bigg]\ , \label{Qab}
\end{align}
where $\beta := (\ello/z)\sqrt{N_{(0)} / \varf}$ is the lapse function of the surfaces of constant $z$, $\s_{i} := g_{ij}\s^{j}$, and the prime denotes differentiation with respect to $z$. Also: $n_{a} = - \varf \beta \partial_{a}u$ , $n^{a}\partial_{a} = \varf^{-1}\beta^{-1} (\partial_{u} - \s^{i}\partial_{i})$ represents the future-directed unit normal to the surfaces of constant time on each hypersurface of constant $z$. The on-shell action is then given by:
\bea\label{on-shell}
&&16\pi G_{0}\, S^{on-shell}\, =\, \int d^{d}x\,du \int\limits^{\epsilon} dz\, {\ello^{d+2} \over z^{d+2}}\, N_{(0)} \sqrt{g} \( -2\, {d+1 \over \a^{2}\ello^{2}} \) \nn\\[5pt]
&&+ \int\limits_{z = \epsilon} d^{d}x\,du\, {\ello^{d} \over \e^{d}}\, \sqrt{g}\, \bigg( - 2(d+1)\, \e^{-1}\varphi + \partial_{\e} \varphi + \varphi\, \Tr[g^{-1}g'] + (\partial_{u} - \Lie_{\sigma}) \log{(\varphi/N_{(0)})} - 2N_{(0)} k \bigg)\ , \nn\\
\eea
where $k = g^{ij}k_{ij}$. In the above we have replaced the asymptotic boundary $\{ z=0 \}$ by a regulating surface $\{ z=\epsilon \}$ and once the vevs are computed we will remove the regulator by taking the limit $\epsilon \to 0$. Note also that the last term in \eqref{on-shell} is a total derivative and therefore can be removed from the on-shell action:\footnote{
If we also consider null boundaries $\{ u = u_{\pm}\}$ in the spacetime, such term results in a corner integral $-2\int d^{d}x \sqrt{\g}$ at $\{z=\e, u=u_{\pm}\}$, with $\g_{ij}$ the induced metric on these codimension two surfaces. Corner terms will be analysed in section \ref{corner_section}.}
\bea\label{discarded_surface_term}
-2{\ello^{d} \over \e^{d}} \int\limits_{z=\e} d^{d}x\, du\, \sqrt{g} N_{(0)} k = -2{\ello^{d} \over \e^{d}} \int\limits_{z=\e} d^{d}x\, du\, (\partial_{u} - \Lie_{\s})\sqrt{g}\ .
\eea
The next step in determining the divergences of the action is to use our asymptotic solutions \eqref{gij}--\eqref{sigma} for the fields $\varphi$, $\sigma^{i}$ and $g_{ij}$ and replace the expressions in \eqref{on-shell}. We then look for all the terms that are proportional to negative powers of $\epsilon$, as well as to factors of $\log{\epsilon}$, and rewrite the respective coefficients in terms of the sources $G_{(0)\mu\nu}$ using \eqref{g1}--\eqref{sigma2}. These terms are those that diverge if the limit $\epsilon \to 0$ is taken. Then, we invert the asymptotic expansions \eqref{gij}--\eqref{sigma} in order to express the sources $G_{(0)\mu\nu}$ order by order in $\epsilon$ in terms of the fields $\varphi$, $\sigma^{i}$ and $g_{ij}$, and then replace the inverted expansions $G_{(0)\mu\nu} = G_{(0)\mu\nu}[\varphi, \sigma^{i}, g_{ij}]$ in the coefficients of the $\epsilon^{<0}$ divergent terms (as well as the $\log \epsilon$ terms) in the on-shell action. The process results in the set of terms that contribute to the divergences of the on-shell action if the regulator $\epsilon$ is sent to zero. The divergent terms obtained in this way are written in a covariant form (except possible anomalous terms depending explicitly on the regulator via a factor of $\log{\epsilon}$) and can then be subtracted from the action by a counterterm action $S_{ct}$ consisting of minus such terms. The renormalized gravity action $S_{ren}$ will then consist of the original action \eqref{action} plus the counterterm action $S_{ct}$ derived in this way.\\ 

As the spacetime dimension increases, the number of covariant boundary counterterms increases, so we will focus separately on the cases of three and four bulk dimensions. For each case, these counterterms must nevertheless coincide with the canonical counterterms originally obtained in \cite{Henningson:1998ey,Balasubramanian:1999re,Emparan:1999pm,Kraus:1999di,deHaro:2000xn}. Although the latter were derived in a different coordinate gauge near the asymptotic boundary, these counterterms are covariant and therefore independent of the coordinate system we use. The possible exception are the anomalous counterterms in \cite{deHaro:2000xn} that depend explicitly on the regulator and therefore that are not invariant under the full diffeomorphism group.\\

Apart from the canonical counterterm action, we are always free to add finite boundary terms to the renormalized gravity action $S_{ren}$ that do not contribute with divergences in the limit $\e \to 0$ and that provide a non-vanishing contribution to the finite piece once the regulator is removed. These terms are dual to a choice of renormalization scheme in the quantum field theory. In our case, once $S_{ren}$ has been determined by the above procedure, we will have to take care of the zero $\L$ limit $\a \to \infty$. This is done by evaluating $S_{ren}$ on-shell, taking the limit $\e \to 0$ and looking for all those terms that diverge if the limit $\a \to \infty$ is taken. Such terms will always be proportional to positive powers of $\a\hspace{1pt}\e^{0}$ and the respective coefficients will always be local functionals of the sources $G_{(0)\mu\nu}$. These $\a$-divergent terms can then be subtracted by adding a finite boundary action $S_{finite}$ to $S_{ren}$ (finite in $\epsilon$) consisting of minus such terms. The subtraction of divergences associated to the zero $\L$ limit is therefore related in this way to a choice of scheme in the dual QFT. As emphasized at the end of section \ref{spacetime_asympt}, however, these finite boundary terms will be invariant under spacetime diffeomorphisms that preserve our foliation, but will break invariance of the gravity action $S_{ren} + S_{finite}$ under those diffeomorphisms that are not foliation-preserving. This fact implies that the renormalization of quantum field theories with gravity duals that admit a well-defined zero $\L$ limit must involve renormalization schemes that break invariance of the QFT under transformations that do not preserve the spacelike foliation at the boundary.

\subsection{Three bulk dimensions}\label{3_dims}

\subsubsection{Renormalization}\label{ren_3}

If we follow the procedure described above for the case $d+2=3$, we find that the counterterm action is the canonical one in standard $AdS_{3}$ holographic renormalization:
\bea\label{Sren_3}
16\pi G_{0}\, S_{ren} &=& \int d^{3}x\, \sqrt{G} \( {d(d+1) \over \alpha^{2}\ello^{2}} + R[G] \) + 2 \int\limits_{z=\epsilon} d^{2}x \sqrt{q}\, Q\ + {2\, d \over \a\,\ello} \int\limits_{z = \epsilon}d^{2}x\, \sqrt{q}\ ,\quad
\eea
Note, however, the absence of the anomalous topological invariant:
\be
\alpha\ello \int_{z = \epsilon} d^{2}x\, \sqrt{q}\, R[q] \log\epsilon\ ,
\ee
that arises in the canonical holographic counterterm action. Although such term does not contribute to the variations of the action, it plays an important role in the holographic correspondence: it represents the fact that we cannot renormalize the generating functional $Z$ of the dual QFT and preserve all its symmetries. Such term breaks invariance of the gravity action under bulk diffeomorphisms that result in a conformal transformation at the boundary and it is dual to those counterterms in the renormalization of $Z$ that do not preserve the conformal symmetry.

This term is absent in the present case because we have been careless about possible corner terms in the renormalized gravity action, \ie about integrals on the codimension two surfaces $\{ z= \epsilon, u=\pm\infty \}$. Note that in the case of a two dimensional manifold, the Ricci-scalar $R[q]$ can always be written as a total derivative (though not necessarily as an exact form). This is so because we can always imagine some hypersurface, say spacelike, in the two dimensional manifold and use the Gauss-Codazzi identities to express $R[q]$ as:
\be\label{Rq_decomp}
R[q] = R[\g] - K^{2} + K \cdot K + 2\, \mathcal{D}_{a} \( n^{a}K - a^{a} \)\ ,
\ee
where $\g_{ij}$ is the metric on the hypersurface, $K_{ij}$ its extrinsic curvature, and $n^{a}$ and $a^{a}$ the unit normal and acceleration of the surface. Also, $\mathcal{D}_{c}\,q_{ab} :=0$. Since the hypersurface is one-dimensional, then $R[\g]$ vanishes, and the terms $K^{2}$ and $K \cdot K$ cancel one another, leaving us with a total derivative. In our case, if we choose such hypersurface to be a surface of constant time $u$, we find:
\be\label{corner}
\a\ello \int_{z = \epsilon} dx\,du\, \sqrt{q}\, R[q] \log\epsilon\ = -2\a\ello \( \int_{z=\epsilon} dx \sqrt{\g}\, K \log\epsilon\)_{u=-\infty}^{u = +\infty}\ ,
\ee
which is a corner term. Such type of terms do not contribute to the computations of the vev of the QFT energy tensor and we will defer a detailed analysis of the possible corner terms until section \ref{corner_section}. There we will find that the holographic renormalization of the gravity action indeed requires the term \eqref{corner} as a counterterm.\\

Given the renormalized action \eqref{Sren_3} we now proceed as discussed at the end of section \ref{onshell_action} and analyse whether the zero $\L$ limit of the on-shell action was spoiled by the counterterm. We evaluate \eqref{Sren_3} on-shell, take the limit as the regulator $\epsilon \to 0$ and, within the set of terms that survive the limit, we look for those that are proportional to positive powers of $\a$. In three dimensions no such terms exist, which means that the canonical counterterm simultaneously renormalizes the gravity action and preserves the well-definedness of the zero $\L$ limit.

%The remarkable feature is that, both in three and four bulk dimensions, all such terms cancel and those that remain are proportional to $\ell^{\leq 0}$. This means that, in the case pure gravity, and at least in three and four dimensions, the canonical holographic counterterms are the unique counterterms that simultaneously renormalize the gravity action and preserve the well-definedness of the zero $\L$ limit. In this work we will not verify whether this feature extends to five and higher bulk dimensions due to the increasing complexity of the equations of motion for the metric, but it seems likely that this special property is a particularity of the low spacetime dimensionality. In fact, in section \ref{matter}, where we consider non-backreacting matter, we will indeed need to add specific finite counterterms to the action for the matter field in order to restore the well-definedness of the zero $\L$ limit which is spoiled by the canonical holographic counterterms.\\

\subsubsection{Vacuum expectation values}\label{vevs_3}

Now that we have guaranteed that the on-shell gravity action is free of divergences, we are in position to compute the holographic energy tensor. The variations of the renormalized on-shell action are given by:
\be
16\pi G_{0}\, \delta S_{ren}^{on-shell}\, =\, \int\limits_{z = \epsilon} d^{2}x \sqrt{q} \( Q_{ab} - q_{ab}Q \) \delta q^{ab} - {d \over \a\ello} \int\limits_{z = \epsilon} d^{2}x \sqrt{q}\, q_{ab} \delta q^{ab}\ .
\ee
The renormalized Brown-York tensor \cite{Brown:1992br} is then given by:
\bea\label{BY_3}
T_{ab} &:=& {2 \over \sqrt{q}}\, {\delta S_{ren}^{on-shell} \over \delta q^{ab}(z=\e)}\nn\\[5pt]
&=& {1 \over 8\pi G_{0}} \(Q_{ab} - q_{ab}Q - {1 \over \a\ello}\, q_{ab} \)\ .
\eea
Using the expression \eqref{induced_q} for the induced metric $q_{ab}$ we now decompose the variations $\delta q^{ab}$ in terms of the variations of the lapse, shift and spatial metric:
\be
\delta q^{ab}\, =\, \( 2 n^{a} n^{b} / N \) \delta N + \( 2\, n^{(a} {\g_{i}}^{b)} / N \) \delta \s^{i} + \g_{i}^{a}\g_{j}^{b}\, \delta \g^{ij}\ ,
\ee
where: $N=(\ello/z) \sqrt{\varf N_{(0)}}\, , \g_{ij} = (\ello/z)^2 g_{ij}$ and: $\g^{ab} = q^{ab} + n^{a}n^{b}$. Following \cite{Brown:1992br} we then define the spatial stress tensor density $s_{ij}$, and the momentum and energy densities $j_{i}$ and $\varepsilon$ as:
\begin{align}
s_{ij}\, &:=\, \g_{i}^{a} \g_{j}^{b}\, T_{ab} =\, {2 \over N\sqrt{\g}}\,{\delta S_{ren}^{on-shell} \over \delta \g^{ij}}\ ,\\[5pt]
j_{i}\, &:=\, -n^{a} \g_{i}^{b}\, T_{ab} =\, -{1 \over \sqrt{\g}}\, {\delta S_{ren}^{on-shell} \over \delta \s^{i}}\ ,\\[5pt]
\varepsilon\, &:=\, n^{a}n^{b}\, T_{ab} =\, {1 \over \sqrt{\g}}\, {\delta S_{ren}^{on-shell} \over \delta N}\ .
\end{align}
We also define the trace density $T$ as:
\bea
T &:=& q^{ab}\, T_{ab}\, =\, \( \g^{ab} - n^{a}n^{b}\) T_{ab}\, =\, \g^{ij}s_{ij} - \varepsilon\ .
\eea
Using the AdS/CFT prescription and recalling the leading order behaviour \eqref{induced_q_leading}, the expectation value of the dual field theory energy tensor is given by:
\bea
\< T_{ab} \> &=& {2 \over \sqrt{q_{(0)}}}\, {\delta S_{ren}^{on-shell} \over \delta q^{ab}_{(0)}}\, =\, \lim_{\e \to 0}\ {\ello^{d-1} \over \e^{d-1}}\, T_{ab}\ .
\eea
In terms of the above decomposition of $T_{ab}$, the spatial and time components of the holographic energy tensor are then given by:
\begin{align}
\< s_{ij} \>\, &:=\, g_{(0)i}^{~~a}g_{(0)j}^{~~b}\, \< T_{ab} \>\, =\, {2 \over {1 \over \a}N_{(0)}\sqrt{g_{(0)}}}\, {\delta S_{ren}^{on-shell} \over \delta g_{(0)}^{ij}}\, =\, \lim_{\epsilon \to 0} \( {\ello^{d-1} \over \epsilon^{d-1}}\, s_{ij} \)\ ,\label{vev_sij}\\[5pt]
\< j_{i} \>\, &:=\, -n_{(0)}^{~a}g_{(0)i}^{~~b}\, \< T_{ab} \>\, =\, - {1 \over \sqrt{g_{(0)}}}\, {\delta S_{ren}^{on-shell} \over \delta \s_{(0)}^{i}}\, =\, \lim_{\epsilon \to 0} \( {\ello^{d} \over \epsilon^{d}}\, j_{i} \)\ ,\label{vev_ji}\\[5pt]
\< \varepsilon \>\, &:=\, n_{(0)}^{~a}n_{(0)}^{~b}\, \< T_{ab} \>\, =\, {1 \over \sqrt{g_{(0)}}}\, {\delta S_{ren}^{on-shell} \over \delta ({1 \over \a}N_{(0)})}\, =\, \lim_{\epsilon \to 0} \( {\ello^{d+1} \over \epsilon^{d+1}}\, \varepsilon \)\ ,\label{vev_e}
\end{align}
where the induced metric $g_{(0)}^{ab} = q_{(0)}^{ab} + n_{(0)}^{a}n_{(0)}^{b}$. The vev of the trace of the QFT energy tensor is also given by:
\be\label{vev_t_def}
\< T \>\, :=\, q_{(0)}^{ab} \< T_{ab} \>\, =\, g_{(0)}^{ij} \< s_{ij} \> - \< \varepsilon \>\, =\, \lim_{\epsilon \to 0} \( {\ello^{d+1} \over \epsilon^{d+1}}\, T \)\ .
\ee
Now, by construction, the above vacuum expectation values cannot admit a well-behaved zero $\L$ limit because the lapse ${1 \over \a}N_{(0)}$ vanishes in this limit. For the vev of the stress tensor we have:
\be
\< s_{ij} \>\, =\, \a \( {2 \over N_{(0)} \sqrt{g_{(0)}}}\, {\delta S_{ren}^{on-shell} \over \delta g_{(0)}^{ij}} \) \to \infty\quad (\a \to \infty)\ .
\ee
Similarly, for the vev of the energy density:
\be
\< \varepsilon \>\, =\, \a \( {1 \over \sqrt{g_{(0)}}}\, {\delta S_{ren}^{on-shell} \over \delta N_{(0)}} \) \to \infty \quad (\a \to \infty)\ .
\ee
What we need to do is to work with the quantities that are well-defined in both cases $\L \neq 0$ and $\L = 0$ and these are represented by the tensor densities:
\begin{align}
&\sqrt{q_{(0)}}\, \< s_{ij} \>\, =\, 2\, {\delta S_{ren}^{on-shell} \over \delta g_{(0)}^{ij}}\ ,\label{vev_sij_2}\\[5pt]
&\sqrt{q_{(0)}}\, \< \varepsilon \>\, =\, N_{(0)}\, {\delta S_{ren}^{on-shell} \over \delta N_{(0)}}\ ,\label{vev_e_2}
\end{align}
where: $\sqrt{q_{(0)}} = {1 \over \a}N_{(0)}\sqrt{g_{(0)}}$. A straightforward computation using \eqref{BY_3} and \eqref{Qab} then leads to the following one-point functions:
\bea
\sqrt{q_{(0)}}\, \< s_{ij} \> &=& {\ello \over 8\pi G_{0}}\, N_{(0)}\sqrt{g_{(0)}} \( - {\varf_{(2)} \over 2\, N_{(0)}}\, g_{(0)ij} \)\ ,\\[5pt]
\sqrt{q_{(0)}}\, \< \varepsilon \> &=& {\ello \over 8\pi G_{0}}\, N_{(0)}\sqrt{g_{(0)}} \( {1 \over \a^{2}}\, \Tr[g_{(0)}^{-1}g_{(2)}] - {\varf_{(2)} \over 2\, N_{(0)}} - \Tr[g_{(0)}^{-1}k_{(1)}] \)\ ,\\[5pt]
\< j_{i} \> &=& -{\ello \over 8\pi G_{0}}\, N_{(0)}^{-1} \( \s_{(2)i} + \hf\, (g_{(0)}^{-1}g_{(1)})_{i}^{j}\partial_{j}N_{(0)} \)\ .
\eea
We therefore find as usual that the normalisable modes $G_{(d+1)\mu\nu}$ are directly associated to the vacuum expectation values \cite{deHaro:2000xn}. Note that these expressions admit a well-behaved limit $\a \to \infty$. 

\subsubsection{Weyl anomaly}

For the vev of the trace we find:
\be\label{vev_t}
\sqrt{q_{(0)}}\, \< T \>\, =\, {\ello \over 8\pi G_{0}}\, N_{(0)}\sqrt{g_{(0)}} \( -{1 \over \a^{2}}\, \Tr[g_{(0)}^{-1}g_{(2)}] + \Tr[g_{(0)}^{-1}k_{(1)}] \)\ .
\ee
Notice now that if we perform a decomposition of the Ricci scalar of the QFT metric as in \eqref{Rq_decomp} we obtain:
\be
R[q_{(0)}]\, =\, R[g_{(0)}] - K_{(0)}^{2} + K_{(0)} \cdot K_{(0)} + 2 ^{(0)}\hspace{-2pt}\mathcal{D}_{a} \( n_{(0)}^{a} K_{(0)} - a_{(0)}^{a} \)\ ,
\ee
where $^{(0)}\hspace{-2pt}\mathcal{D}_{c} q_{(0)ab} := 0$. A quick computation using \eqref{k0}, \eqref{g1} and \eqref{g2_g1} then reveals that:
\bea
R[q_{(0)}] &=& 2 \( - {1 \over \a^{2}} \Tr[g_{(0)}^{-1}g_{(2)}] + \Tr[g_{(0)}^{-1}k_{(1)}] \)\ .
\eea
Replacing in \eqref{vev_t} results in the standard holographic Weyl anomaly:
\bea\label{vev_t_standard}
\sqrt{q_{(0)}}\, \< T \> &=& {\ello \over 16\pi G_{0}}\, N_{(0)}\sqrt{g_{(0)}}\, R[q_{(0)}] \nn\\[5pt] 
&=&  {\a\ello \over 16\pi G_{0}}\, \sqrt{q_{(0)}}\, R[q_{(0)}] \nn\\[5pt] 
&=& {c \over 24 \pi}\, \sqrt{q_{(0)}}\, R[q_{(0)}]\ ,
\eea
where $c= 3\a\ello / (2 G_{0})$ is the standard central charge in the AdS$_{3}$/CFT$_{2}$ correspondence. Note that the anomaly admits a well-behaved zero $\L$ limit. Using equation \eqref{vev_t} we find:
\bea\label{vev_t_ZL}
\lim_{\a \to \infty}\ \sqrt{q_{(0)}}\, \< T \> &=& {\ello \over 8\pi G_{0}}\, N_{(0)}\sqrt{g_{(0)}}\ k_{(1)}\ ,
\eea
where $k_{(1)ij}$ is given in equation \eqref{k1} and therefore is non-divergent as $\a \to \infty$ (it is totally written in terms of the coefficients \eqref{gij}--\eqref{sigma} without positive powers of $\a$). Since the anomaly admits a well-defined zero $\L$ limit, a central charge can be introduced in this limit. This can be done by rewriting the right-hand side of equation \eqref{vev_t_ZL} as a geometric invariant of quantities that are well-defined at the null boundary:
\bea
\lim_{\a \to \infty}\ \sqrt{q_{(0)}}\, \< T \> &=& {\ello \over 8\pi G_0}\, N_{(0)}\sqrt{g_{(0)}}\ ^{(0)}\hspace{-2pt}\mathcal{D}_{a} \(n_{(0)}^{a} K_{(0)} - a_{(0)}^{a} \)\ .
\eea
In order to compare this result with that of \cite{Barnich:2006av,Bagchi:2012xr}, we introduce the same limit discussed in these references to obtain:
\bea
\lim_{\a \to \infty}\ \( {G_{0} \over \a \ello}\, \sqrt{q_{(0)}}\, \< T \> \) &=& {1 \over 8\pi}\, \sqrt{q_{(0)}}\ ^{(0)}\hspace{-2pt}\mathcal{D}_{a} \( n_{(0)}^{a} K_{(0)} - a_{(0)}^{a} \)\ .
\eea
The proportionality constant between the trace and the total derivative is then:
\be
{1 \over 8 \pi}\, =\, {c \over 24 \pi}\ ,
\ee
where $c=3$ is the Virasoro central charge in the central extension of the asymptotic symmetry group $\mathfrak{bms}_{3}$ of three dimensional flat gravity \cite{Barnich:2006av,Bagchi:2012xr}.\footnote
{
The central charge $c_{LM}$ in reference \cite{Bagchi:2012xr} is related to ours as: $c_{LM} = c/12$ since we follow the convention of formula (1) in this reference as can be seen by comparing the central charge in this formula with that in the $AdS$ case \eqref{vev_t_standard}.
}

\subsubsection{Improved energy tensor}\label{Improved_T}

If we return to the full Weyl anomaly \eqref{vev_t} or \eqref{vev_t_standard} for generic $\L$ and use equations \eqref{g2_g1}, \eqref{k1} and \eqref{sigma1}, we can rewrite it in terms of the coefficient $g_{(1)ij}$ as:
\bea
\sqrt{q_{(0)}}\ \< T \> &=& {\ello \over 8\pi G_{0}}\, N_{(0)} \sqrt{g_{(0)}} \( -{1 \over 4\a^{2}}\, (g_{(1)}\cdot g_{(1)}) + {1 \over 2N_{(0)}}\, g_{(0)}^{ij} (\partial_{u} - \Lie_{\s_{(0)}})g_{(1)ij} - {1 \over N_{(0)}}\, ^{(0)}\square N_{(0)} \) \ . \nn\\
\eea
Notice now from equation \eqref{g1} that a non-vanishing coefficient $g_{(1)ij}$ represents the fact that the QFT metric is time dependent. The boundary shift $\s_{(0)}^{i}$ can always be fixed to any configuration by boundary diffeomorphisms; in particular, we can fix $\s_{(0)}^{i}$ to zero by the transformation $x^{i} \to x^{i} - \int du\, \s_{(0)}^{i}$. In such coordinates, equation \eqref{g1} becomes: ${N_{(0)}/\a^{2}}\, g_{(1)ij} = \partial_{u}g_{(0)ij}$. Therefore, if the QFT metric \eqref{induced_q_leading} is static the Weyl anomaly becomes:
\bea
\sqrt{q_{(0)}}\ \< T \> &=& - {\ello \over 8\pi G_{0}}\, \sqrt{g_{(0)}}\ ^{(0)}\square N_{(0)} \nn\\[5pt]
&=& - {\a\ello \over 8\pi G_{0}}\, \sqrt{q_{(0)}}\ ^{(0)}\hspace{-2pt}\mathcal{D}_{b}\, a_{(0)}^{b}\ ,
\eea
where the acceleration $a_{(0)}^{a} = g_{(0)}^{ab}\partial_{b}\log{N_{(0)}}$ as before. Using the definition of $g_{(0)}^{ab}$ this can be rewritten as:
\bea
\sqrt{q_{(0)}}\ \< T \> &=& - {\a\ello \over 8\pi G_{0}}\, \sqrt{q_{(0)}} \( ^{(0)}\hspace{-2pt}{\mathcal{D}_{a}}  ^{(0)}\hspace{-2pt}\mathcal{D}^{a} \log{N_{(0)}} + {^{(0)}\hspace{-2pt}\mathcal{D}_{a}} \( n_{(0)}^{a} n_{(0)}^{b}\partial_{b}\log{N_{(0)}} \) \)\ .
\eea
Since the boundary metric is static, the second total derivative vanishes: $n_{(0)}^{b}\partial_{b} \log{N_{(0)}} = \a/N_{(0)} (\partial_{u} - \s_{(0)}^{i}\partial_{i}) \log{N_{(0)}} = 0$. The first total derivative that remains is unphysical in the sense that it can be absorbed in an improved energy tensor $\Theta^{ab}$ defined in terms of the QFT energy tensor $T^{ab}$ and covariant derivatives of the acceleration $a_{(0)}^{a}$, or of the lapse $\log{N_{(0)}}$ (see \eg \cite{Nakayama:2013is}). The conformal Ward identity then becomes:
\bea
\sqrt{q_{(0)}}\ \< \Theta^{a}_{~a} \> &=& 0\ ,
\eea
for a static metric $q_{(0)ab}$. In section \ref{vevs_4} we will find another example where an improved energy tensor can be defined such that staticity of the boundary metric restores conformal invariance of the field theory.

\subsubsection{Diffeomorphism Ward identity}

In order to verify that the holographic energy tensor is conserved we need to solve the constraint equations \eqref{Rui}--\eqref{Ruu} using \eqref{Einstein_eqs} for the normalisable modes. At first subleading order with $d=1$ these two equations result in the constraints:
\bea
&& {1 \over \sqrt{g_{(0)}}} \(\partial_{u} - \Lie_{\s_{(0)}}\) \( \sqrt{g_{(0)}}\ N_{(0)}^{-1} \( \s_{(2)i} + \hf\, (g_{(0)}^{-1}g_{(1)})_{i} ^{~j} \partial_{j}N_{(0)} \) \) \nn\\[5pt]
&&\qquad = - \hf\, \partial_{i}\varf_{(2)} + \( {1 \over \a^{2}} \Tr[g_{(0)}^{-1}g_{(2)}] - {\varf_{(2)} \over 2N_{(0)}} - \Tr[g_{(0)}^{-1}k_{(1)}] \) \partial_{i}N_{(0)}\ ,\\[10pt]
&& {1 \over \sqrt{g_{(0)}}} \(\partial_{u} - \Lie_{\s_{(0)}}\) \( \sqrt{g_{(0)}} \( -{1 \over \a^{2}}\, \Tr[g_{(0)}^{-1}g_{(2)}] + {\varf_{(2)} \over 2N_{(0)}} + \Tr[g_{(0)}^{-1}k_{(1)}] \) \) \nn\\[5pt]
&&\qquad = \hf\, \varf_{(2)}\, k_{(0)} - {1 \over \a^{2}N_{(0)}}\, \cov_{i} \( N_{(0)} \( \s_{(2)}^{i} + \hf g_{(1)}^{ij}\partial_{j}N_{(0)} \) \) \ .
\eea
These constraints result in the conservation equations for the QFT energy tensor:
\bea
0 &=& ^{(0)}\hspace{-2pt}\mathcal{D}_{a} \( \sqrt{q_{(0)}}\, \< T^{a}_{~i} \> \) \nn\\[5pt]
&=& \( \partial_{u} - \Lie_{\s_{(0)}} \) \( \sqrt{g_{(0)}}\, \< j_{i} \> \) + \cov_{j} \( \sqrt{q_{(0)}}\, \< s^{j}_{~i} \> \) + \sqrt{q_{(0)}}\, \< \varepsilon \>\, \partial_{i} \log{N_{(0)}}\ ,\label{conserv_1}\\[10pt]
0 &=& {1 \over \a}\, n_{(0)}^{b}\, ^{(0)}\hspace{-2pt}\mathcal{D}_{a} \( \sqrt{q_{(0)}}\, \< T^{a}_{~b} \> \) \nn\\[5pt]
&=& -\(\partial_{u} - \Lie_{\s_{(0)}}\) \( N_{(0)}^{-1}\sqrt{q_{(0)}}\, \< \varepsilon \> \) - N_{(0)}^{-1}\, \cov_{i} \( (N_{(0)}/\a)^{2} \sqrt{g_{(0)}}\, \< j^{i} \> \) + \sqrt{q_{(0)}}\, \< s^{ij} \>\, k_{(0)ij}\ . \label{conserv_2}\nn\\
\eea

\subsubsection{BTZ and 3-dimensional cosmology}

In this section we would like to make a brief application of the results obtained so far to a particular bulk metric and the spacetime we are interested in is the BTZ black hole and its zero $\L$ limit, which represents a cosmological solution \cite{Cornalba:2002fi}. In Eddington-Finkelstein coordinates the BTZ metric is given by:
\be
ds^{2}\, =\, - \( -8 M G_{0} + {r^{2} \over \ell^{2}} + \({4 a G_{0} \over r}\)^{2} \) du^{2} + 2dudr + r^{2} \( d\theta - {4 a G_{0} \over r^{2}}\, du \)^{2}\ ,
\ee
where the cosmological constant $\L = - \ell^{-2}$, $M$ is the mass of the spacetime and $a$ the angular momentum. Also, the angular coordinate $\theta \in [0,2\pi[$. In order to bring the metric to the form \eqref{metric} we introduce a coordinate $z := \ello^{2} / r$:
\bea
ds^{2}\, =\, {\ello^{2} \over z^{2}} \( - \( {1 \over \a^2} - {8 M G_{0} \over \ello^2}\, z^{2} + \( {4 a G_{0} \over \ello^3} \)^{2} z^{4} \) du^{2} - 2 dudz + \ello^{2} \( d\theta - {4 a G_{0} \over \ello^{4}}\, z^{2} du^{2} \)^{2} \)\ . \nn\\
\eea
Note that the metric is well-defined in the limit $\a \to \infty$. For this solution the holographic energy tensor reads:
\bea
\sqrt{q_{(0)}}\, \< s_{ij} \> &=& \sqrt{g_{(0)}} \( {M \over 2\pi \ello}\, g_{(0)ij} \)\ ,\\[5pt]
\sqrt{q_{(0)}}\, \< \varepsilon \> &=& \sqrt{g_{(0)}} \( {M \over 2 \pi \ello} \)\ ,\\[5pt]
\< j_{i} \> &=& {a \over 2 \pi \ello}\ ,
\eea
where the spatial metric $g_{(0)ij}dx^{i}dx^{j} = \ello^{2} d\theta^{2}$. In this case, the characteristic length $\ello$ represents the radius of the boundary cylinder. If we then introduce the average energy $\< E \>$ over a time interval $2T$ we obtain:
\be\label{En_average}
\< E \>\, :=\, {1 \over 2T}\, \int\limits_{-T}^{T}du\int d^{d}x \sqrt{q_{(0)}}\, \< \varepsilon \>\, =\, M\ .
\ee
Also, for the angular momentum we find:
\be
\< J_{i} \> :=\, \int d^{d}x \sqrt{g_{(0)}}\, \< j_{i} \>\, =\, a\ .
\ee
Note that these results can be extended to the zero $\L$ limit of the solution and coincide with those obtained in \cite{Barnich:2012xq,Bagchi:2012xr} via a thermodynamics analysis of the respective three-dimensional cosmological solution.

\subsection{Four bulk dimensions}\label{four_dims}

\subsubsection{Renormalization}\label{ren_4}

In the case of $d+2=4$ dimensions the renormalized action is given by:
\bea\label{Sren_4}
16\pi G_{0}\, S_{ren} &=& \int d^{4}x\, \sqrt{G} \( {d(d+1) \over \alpha^{2}\ello^{2}} + R[G] \) + 2 \int\limits_{z=\epsilon} d^{3}x \sqrt{q}\, Q \nn\\[5pt] 
&+& {2\, d \over \a\ello} \int\limits_{z = \epsilon}d^{3}x\, \sqrt{q} + {\a\ello \over d-1} \int\limits_{z=\e} d^{3}x \sqrt{q}\, R[q]\ ,
\eea
where the counterterms again coincide with the canonical ones in four bulk dimensions. The next step is to determine whether the renormalization spoils the zero $\L$ limit of the action. In the present case, if we evaluate $S_{ren}$ on-shell and take the limit $\e \to 0$ as described in section \ref{onshell_action}, we find again that no terms proportional to positive powers of $\a$ survive and therefore that the limit $\a \to \infty$ is still wel-defined in the presence of the counterterm action. However, this feature is peculiar to our particular choice of boundary lapse function $N_{(0)}$. In section \ref{coordinates} and \ref{asympt_sol} we found that, in general, the metric component $\varf$ admits an arbitrary term $z\varf_{(1)}$ in the asymptotic expansion. We then argued that it is always possible to redefine the coordinate $z$ and choose some new function $N_{(0)}$ as in \eqref{N0} such that the term $\varf_{(1)}$ is removed from the asymptotics. On the other hand, if we choose to decouple $N_{(0)}$ from $\varf_{(1)}$ by requiring that equation \eqref{N0} for $N_{(0)}$ does not hold, then the asymptotic solution \eqref{varf} for $\varf$ will admit a term $z\varf_{(1)}$, with $\varf_{(1)}(u,x)$ an arbitrary function. In such case, the solution \eqref{g1} is modified to:
\be\label{g1_varf1}
{1 \over 2\a^{2}}\, g_{(1)ij} = k_{(0)ij} + {\varf_{(1)} \over 2N_{(0)}}\, g_{(0)ij}\ .
\ee
Although the renormalization in the three dimensional case analysed in the previous section remains unaffected, if we switch on the coefficient $\varf_{(1)}$ by allowing the lapse $N_{(0)}$ to be independent, the canonical counterterm action in the four dimensional case will spoil the zero $\L$ limit via the term:
\bea
\lim_{\e \to 0}\ 16\pi G_{0}\, S_{ren}^{on-shell} &=& {\a^{2}\ello^{2} \over 2} \int\limits_{z=0} d^{3}x\, \sqrt{g_{(0)}}\, \varf_{(1)} R[g_{(0)}] + \op(\a^{\leq 0})\ ,
\eea
where $\op(\a^{\leq 0})$ denotes terms proportional to non-positive powers of $\a$. From equation \eqref{g1_varf1} it follows that the (finite) counterterm that restores the well-definedness of the limit is given by:
\bea\label{Sren2_4}
16\pi G_{0}\, S_{ren} &=& \int d^{4}x\, \sqrt{G} \( {d(d+1) \over \alpha^{2}\ello^{2}} + R[G] \) + 2 \int\limits_{z=\epsilon} d^{3}x \sqrt{q}\, Q \nn\\[5pt] 
&+& {2\, d \over \a\ello} \int\limits_{z = \epsilon}d^{3}x\, \sqrt{q} + {\a\ello \over d-1} \int\limits_{z=\e} d^{3}x \sqrt{q}\, R[q] \nn\\[5pt]
&+& {\a^{2}\ello^{2} \over 2} \int\limits_{z=\e} d^{3}x\, \sqrt{q}\, K R[\g]\ ,
\eea
with $\g_{ij}$ and $K_{ij}$ the induced metric and extrinsic curvature of the surfaces of constant time at the boundary as defined in section \ref{ren_3}. This last counterterm is covariant with respect to diffeomorphisms that preserve our foliation of the spacetime, but breaks invariance of the action under those transformations that are not foliation-preserving, as discussed in section \ref{onshell_action}. The latter include those bulk diffeomorphisms that result in a conformal transformation at the boundary and therefore the trace Ward identity will be affected by such term as discussed in the next section. This counterterm is also finite in the regulator $\e$ and therefore must be related to a choice of renormalization scheme in the dual field theory. In particular, it signals the fact that the scheme cannot preserve the invariance of the QFT under those transformations that are not foliation-preserving at the boundary if the gravity dual has a well-defined zero $\L$ limit. We will find more examples of counterterms of this type in section \ref{matter}. With our choice of $N_{(0)}$, however, the canonical action \eqref{Sren_4} is well-defined, so we will ignore for now this extra counterterm and discuss its necessity and implications in the next section.

\subsubsection{Vacuum expectation values and the Ward identities}\label{vevs_4}
The variations of the renormalized on-shell action \eqref{Sren_4} are given by:
\bea
16\pi G_{0}\, \delta S_{ren}^{on-shell} &=& \int\limits_{z = \epsilon} d^{3}x \sqrt{q} \( Q_{ab} - q_{ab}Q \) \delta q^{ab} \nn\\[5pt] 
&-& {d \over \a\ello} \int\limits_{z = \epsilon} d^{3}x \sqrt{q}\, q_{ab} \delta q^{ab} + {\a\ello \over d-1} \int\limits_{z=\e} d^{3}x\, \sqrt{q} \( R_{ab}[q] - \hf\, q_{ab}\, R[q] \) \delta q^{ab}\ . \nn\\
\eea
The spatial and time components of the Brown-York tensor as defined in section \ref{vevs_3} are then given by:
\bea
s_{ij} &=& \g_{i}^{a} \g_{j}^{b} \({2 \over \sqrt{q}}\, {\delta S_{ren}^{on-shell} \over \delta q^{ab}(z=\e)}\) \nn\\[5pt]
&=& {1 \over 8\pi G_{0}} \( Q_{ij} - \g_{ij} Q - {d \over \a\ello}\, \g_{ij} + \a\ello \( R_{ij}[q] - \hf\, \g_{ij} R[q] \) \)\ ,\\[10pt]
j_{i} &=& -n^{a} \g_{i}^{b} \({2 \over \sqrt{q}}\, {\delta S_{ren}^{on-shell} \over \delta q^{ab}(z=\e)}\) \nn\\[5pt]
&=& {1 \over 8\pi G_{0}} \( -n^{a}Q_{ai} - \a\ello\, n^{a}R_{ai}[q] \)\ ,\\[10pt]
\varepsilon &=& n^{a}n^{b} \({2 \over \sqrt{q}}\, {\delta S_{ren}^{on-shell} \over \delta q^{ab}(z=\e)}\) \nn\\[5pt]
&=& {1 \over 8\pi G_{0}} \( \g^{ij}Q_{ij} + {d \over \a\ello} + {\a\ello \over 2} \( n^{a}n^{b}R_{ab}[q] + \g^{ij}R_{ij}[q] \) \)\ .
\eea
A lengthy computation using the prescriptions \eqref{vev_sij_2}--\eqref{vev_e_2} and \eqref{vev_ji} for the vacuum expectation values of the components of the dual QFT energy tensor in $d+2=4$ dimensions results in the following one-point functions:
\bea
\sqrt{q_{(0)}}\, \< s_{ij} \> &=& {\ello^{2} \over 8\pi G_{0}}\, N_{(0)}\sqrt{g_{(0)}}\ \Bigg[ {3 \over 2\a^{2}}\, g_{(3)ij} - {\varf_{(3)} \over 2N_{(0)}}\, g_{(0)ij} \nn\\[5pt]
&+& \a^{2} \( \big( \partial_{u} - \Lie_{\s_{(0)}} \big) k_{(1)ij} - \hf\, g_{(0)ij} \Tr\Big[g_{(0)}^{-1} \big( \partial_{u} - \Lie_{\s_{(0)}} \big) k_{(1)}\Big] \) \nn\\[5pt]
&+& \partial_{(i}\log N_{(0)}\partial_{j)}\Tr[g_{(0)}^{-1}g_{(1)}] - \cov_{k} (g_{(0)}^{-1}g_{(1)})^{k}_{~(i}\, \partial_{j)}\log N_{(0)} \nn\\[5pt]
&+& X_{ij} - \hf\, g_{(0)ij} \Tr[g_{(0)}^{-1}X]\, \Bigg]\ ,\label{sij_4}\\[10pt]
\sqrt{q_{(0)}}\, \< \varepsilon \> &=& {\ello^{2} \over 8\pi G_{0}}\, N_{(0)}\sqrt{g_{(0)}}\ \Bigg[ -{\varf_{(3)} \over N_{(0)}} \nn\\
&+& \partial_{i}\log N_{(0)} \cov^{i} \Tr[g_{(0)}^{-1}g_{(1)}] - \cov^{i} g_{(1)ij} \cov^{j} \log N_{(0)} \Bigg] \ ,\\[10pt]
\< j_{i} \> &=& {\ello^{2} \over 8\pi G_{0}}\, \Bigg[ -{3 \over 2N_{(0)}}\, \s_{(3)i} \nn\\
&-& \a^{2} \cov^{j} \( k_{(1)ij} - \hf\, g_{(0)ij} \Tr[g_{(0)}^{-1}k_{(1)}] \) + {\a^{2} \over 4}\, \partial_{i}R[g_{(0)}] + X_{i} \Bigg]\ .\label{ji_4}
\eea
From the trace constraint equation \eqref{Rzz} using \eqref{Einstein_eqs} it follows that the normalisable mode $g_{(3)ij}$ is traceless: $\Tr[g_{(0)}^{-1}g_{(3)}] = 0$. The trace \eqref{vev_t_def} of the holographic energy tensor is then given by:
\bea\label{Weyl_4}
\< T \> &=& g_{(0)}^{ij} \< s_{ij} \> - \< \varepsilon \> =\, 0\ .
\eea
This is the expected result for a conformal field theory in three dimensions. From the above one-point functions for finite $\a$, we find that the normalisable modes $G_{(d+1)\mu\nu}$ are again mapped to the vacuum expectation values. The expressions for the terms $X_{ij}$ and $X_{i}$ are given in appendix \ref{Xij} and consist in a set of terms in $g_{(1)ij}$ proportional to non-positive powers of $\a$. These terms are scheme dependent in the sense that they can be subtracted by a choice of finite counterterms of the form:
\bea\label{anomalous_cts}
\a^{2}\ello^{2} \int_{z=\e} d^{3}x \sqrt{q}\, \Big( a_{1} K^{3} + a_{2} K (K \cdot K) + a_{3} (K\cdot K\cdot K) + a_{4} K R[\g] + a_{5} \square_{\g}K + ... \Big)\ . \nn\\
\eea
As discussed in section \ref{Improved_T}, a non-vanishing coefficient $g_{(1)ij}$ represents the fact that the QFT metric is time dependent. It follows that the terms $X_{ij}$ and $X_{i}$ are possibly non-vanishing only if the boundary metric is not static.\\

Let us then discuss the terms in the second line of \eqref{sij_4} and \eqref{ji_4} that depend on $\a^{2}$. The last of these, $(\a^{2}/4)\, \partial_{i} R[g_{(0)}]$, diverges in the limit $\a \to \infty$. Note, however, that if we preserve the counterterm introduced in \eqref{Sren2_4}, it will contribute to the variations of the on-shell action as:
\bea\label{var_anomalous_ct}
\delta\, \bigg({\a^{2}\ello^{2} \over 2} \int\limits_{z = \epsilon} d^{3}x \sqrt{q}\, K R[\g] \bigg) &=& {\a^{2}\ello^{2} \over 2} \int\limits_{z=\e} d^{3}x\, \sqrt{g}\, \partial_{i}R[g]\, \delta\s^{i} \nn\\[5pt]
&+& {\a^{2}\ello^{2} \over 2} \int\limits_{z=\e} d^{3}x\, \sqrt{g} \( g_{ij} \na^{k}\na^{l} \( N_{(0)} k_{kl} \) - \na_{i}\na_{j} \( N_{(0)} k \) \) \delta g^{ij}\ , \nn\\
\eea
where $\na_{i}g_{jk} := 0$. This result implies that the one-point functions will be modified to:\footnote{
As a technical point, if the coefficient $\varf_{(1)} \neq 0$ then the last integral in \eqref{var_anomalous_ct} will contribute with terms $\a^{2}\varf_{(1)}$ to $\<s_{ij}\>_{new}$. However, the previous spatial stress $\<s_{ij}\>$ will contain the symmetric of such terms if $\varf_{(1)} \neq 0$ such that they cancel overall.}
\bea
\< j_{i} \> &\to& \< j_{i} \>_{new}\, =\, \< j_{i} \> - {\ello^{2} \over 8\pi G_{0}} \( {\a^{2} \over 4}\, \partial_{i}R[g_{(0)}] \)\ ,\label{ji_4_new}\\[10pt]
\sqrt{q_{(0)}}\, \< \varepsilon \> &\to& \sqrt{q_{(0)}}\, \< \varepsilon \>_{new}\, =\, \sqrt{q_{(0)}}\, \< \varepsilon \>\ ,\\[15pt]
\sqrt{q_{(0)}}\, \< s_{ij} \> &\to& \sqrt{q_{(0)}}\, \< s_{ij} \>_{new}\, =\, \sqrt{q_{(0)}}\, \< s_{ij} \>\nn\\
&+& {\ello^{2} \over 32\pi G_{0}} \sqrt{g_{(0)}} \( g_{(0)ij} \cov^{k}\cov^{l} \( N_{(0)}g_{(1)kl} \) - \cov_{i}\cov_{j} \( N_{(0)} \Tr[g_{(0)}^{-1}g_{(1)}] \) \)\ . \nn\\ \label{sij_4_new}
\eea
The anomalous counterterm therefore provides a contribution to $\<j_{i} \>$ that cancels the $\a$-divergence proportional to the gradient of the Ricci scalar without introducing further divergences. This is done, however, at the expense of modifying the conformal Ward identity \eqref{Weyl_4} by a total derivative:
\bea\label{trace_Ward_mod}
\sqrt{q_{(0)}}\, \< T \>\, \to\, \sqrt{q_{(0)}}\, \< T \>_{new} &=& 0 + {\ello^{2} \over 16 \pi G_{0}}\, \sqrt{g_{(0)}}\, \cov^{i}\cov^{j} \bigg[ N_{(0)} \( g_{(1)ij} - \hf\, g_{(0)ij} \Tr[g_{(0)}^{-1}g_{(1)}] \) \bigg] \nn\\[5pt]
&=& {\ello^{2} \over 8 \pi G_{0}}\, \sqrt{g_{(0)}}\ \cov^{i}\cov^{j} \bigg[ \a N_{(0)} \( K_{(0)ij} - \hf\, g_{(0)ij} K_{(0)} \) \bigg]\ , 
\eea
which is finite in the limit $\a \to \infty$. Note that if we define:
\bea
v_{ij} &:=& \a N_{(0)} \( K_{(0)ij} - \hf\, g_{(0)ij}K_{(0)} \)\ , \\[5pt]
v^{ab} &:=& g_{(0)}^{ai}g_{(0)}^{bj}v_{ij}\ ,
\eea
and use the standard identities from the theory of embedded hypersurfaces, we obtain that:
\bea
&& \a\sqrt{q_{(0)}}\, ^{(0)}\hspace{-2pt}\mathcal{D}_{a} \( N_{(0)}^{-1} \cov_{b} v^{ab} \)\, =\, \sqrt{g_{(0)}}\, \cov^{i} \cov^{j} v_{ij}\ , \nn\\[5pt]
&& N_{(0)}^{-1} \cov_{b} v^{ab}\, =\, ^{(0)}\hspace{-2pt}\mathcal{D}_{b}L^{ab} - {1 \over \a}\, n_{(0)}^{a} \( L_{ij}L^{ij} \)\ ,
\eea
with $\cov_{a}$ the covariant derivative induced on the surfaces of constant time at the boundary manifold, associated to the induced metric $g_{(0)ab} = q_{(0)ab} + n_{(0)a}n_{(0)b}$, and where:
\bea
L^{ab} &:=& N_{(0)}^{-1} v^{ab}\, =\, \a\, g_{(0)}^{ai}g_{(0)}^{bj} \( K_{(0)ij} - \hf\, g_{(0)ij} K_{(0)} \)\ .
\eea
The modified trace Ward identity can then be rewritten as:
\bea\label{trace_Ward_mod2}
\sqrt{q_{(0)}}\ \< T \>_{new} &=& {\ello^{2} \over 8\pi G_{0}}\, \sqrt{q_{(0)}}\, \bigg[\, \a\, {^{(0)}}\hspace{-2pt}{\mathcal{D}_{a}} ^{(0)}\hspace{-2pt}\mathcal{D}_{b}\, L^{ab} -  ^{(0)}\hspace{-2pt}\mathcal{D}_{a} \( n_{(0)}^{a} \big(L \cdot L\big) \) \bigg]\ .
\eea
The first total derivative is unphysical because it can be absorbed in an improved energy tensor $\Theta^{ab}$ defined in terms of the QFT energy tensor $T_{ab}$ and covariant derivatives of $L_{ab}$ \cite{Nakayama:2013is}, but the second term remains. The Ward identity in such case becomes:
\bea
\sqrt{q_{(0)}}\ \< \Theta^{a}_{~a} \> &=& {\ello^{2} \over 8\pi G_{0}}\, \sqrt{q_{(0)}}\ ^{(0)}\hspace{-2pt}\mathcal{D}_{a} v^{a}\ ,
\eea
where:
\bea
\sqrt{q_{(0)}}\ ^{(0)}\hspace{-2pt}\mathcal{D}_{a} v^{a} &=& - \sqrt{q_{(0)}}\ ^{(0)}\hspace{-2pt}\mathcal{D}_{a} \( n_{(0)}^{a} \big(L \cdot L\big) \) \nn\\[5pt]
&=& -\partial_{u} \( \sqrt{g_{(0)}}\ \qt\, \Big( g_{(1)}\cdot g_{(1)} - \hf\, \Tr[g_{(0)}^{-1}g_{(1)}]^{2} \Big) \) \nn\\[5pt] 
&+& \partial_{i} \( \sqrt{g_{(0)}}\ \s_{(0)}^{i}\, \qt\, \Big( g_{(1)}\cdot g_{(1)} - \hf\, \Tr[g_{(0)}^{-1}g_{(1)}]^{2} \Big) \)\ ,
\eea
which is finite in the limit $\a \to \infty$. This result is expected because the anomalous counterterm in \eqref{Sren2_4} breaks, in particular, invariance of the renormalized gravity action under bulk diffeomorphisms that result in a conformal transformation at the boundary. The generating functional of the dual QFT therefore will not be conformally invariant unless the QFT metric is static (which requires $g_{(1)ij}=0$). As in section \ref{Improved_T}, we find here another relation between metric staticity and conformal invariance. Scale invariance of the dual field theory is, however, preserved because the anomaly is a total derivative. Recall that the breaking of conformal symmetry follows from the requirement that the renormalized gravity action be finite in the zero $\L$ limit. Below we will still discuss the implications of the anomalous counterterm to the diffeomorphism Ward identity.\\

With the divergent term $(\a^{2}/4)\partial_{i}R[g_{(0)}]$ subtracted in this way, the terms proportional to $\a^{2}$ that remain in the expressions for the vacuum expectation values represent derivatives of the traceless part of the coefficient $k_{(1)ij}$:\footnote{
Note that: $\a^{2}g_{(0)ij} \Tr\Big[g_{(0)}^{-1} \big( \partial_{u} - \Lie_{\s_{(0)}} \big) k_{(1)}\Big]\, =\, \a^{2} \big( \partial_{u} - \Lie_{\s_{(0)}} \big) \( g_{(0)ij} \Tr[g_{(0)}^{-1}k_{(1)}] \) + \op(\a^{0})$.} 
\bea
\begin{cases}
\a^{2} \big( \partial_{u} - \Lie_{\s_{(0)}} \big) \( k_{(1)ij} - \hf\, g_{(0)ij} \Tr[g_{(0)}^{-1}k_{(1)}] \)\ ,\\[15pt]
\a^{2} \cov^{j} \( k_{(1)ij} - \hf\, g_{(0)ij} \Tr[g_{(0)}^{-1}k_{(1)}] \)\ .
\end{cases}
\eea
These terms cannot be subtracted by covariant counterterms, nor by counterterms of the form \eqref{anomalous_cts}. This fact implies that the traceless part of $k_{(1)ij}$ needs to admit an expansion in $\a$ of the form:
\bea\label{k1_alpha}
k_{(1)ij} - \hf\, g_{(0)ij} \Tr[g_{(0)}^{-1}k_{(1)}] &=& {1 \over \a^{2}} \( \kappa_{[0]ij} + \op(\a^{< 0}) \)\ ,
\eea
with $\kappa_{[0]ij}$ independent of $\a$. In other words, in three boundary dimensions, only field theory states dual to bulk solutions that admit the behaviour \eqref{k1_alpha} in $\a$ result in finite vacuum expectation values in the limit $\a \to \infty$. The expression for $\kappa_{[0]ij}$ is given by the vev of the QFT stress tensor in the zero $\L$ limit. As discussed at the end of section \ref{asympt_sol}, in this limit the coefficient $k_{(1)ij}$ replaces the normalisable mode $g_{(3)ij}$ as the integration constant of the equations of motion for the case $d=2$. Notice then that the coefficient $g_{(3)ij}$ drops out of equation \eqref{sij_4} for the expectation value of the spatial stress $s_{ij}$ in the limit $\a \to \infty$ and the latter is mapped to the Lie derivative of $\kappa_{[0]ij}$ along $n_{(0)}^{a}$ in this limit. In this way, $\kappa_{[0]ij}$ is part of the asymptotic bulk data that is mapped to boundary data in the zero $\L$ limit.\\
%Note: the asymptotic data seems to consist of the boundary metric and some subleading order of the extrinsic curvature.
%Discuss this: We find that solutions of the bulk theory with a vanishing cosmological constant cannot be reconstructed holographically from the boundary theory unless the correspondence between bulk and boundary data is known for generic $\L$.

Finally, we will not compute here the diffeomorphism Ward identity for the general case in $d+2=4$ dimensions because the constraint equations for the metric are very tedious to solve at second subleading order, but we will verify it explicitly for the Kerr solution discussed below. However, we would still like to emphasize that the terms in the holographic energy tensor that arise from the anomalous counterterm should not contribute to the spatial component of the Ward identity. Indeed, if we use the second identity in equation \eqref{conserv_1} and the expressions for the components of $\< T_{ab} \>_{new}$ given in equations \eqref{ji_4_new}--\eqref{sij_4_new}, we find:
\bea
^{(0)}\hspace{-2pt}\mathcal{D}_{a} \( \sqrt{q_{(0)}}\ \< T^{a}_{~i} \>_{new} \) &=& ^{(0)}\hspace{-2pt}\mathcal{D}_{a} \( \sqrt{q_{(0)}}\ \< T^{a}_{~i} \>\)\ .
\eea
This is the statement that the anomalous counterterm does not break invariance under boundary diffeomorphisms \eqref{fol_preserv} that are foliation preserving.\footnote{
These are essentially spatial diffeomorphisms plus a possible redefinition of the time coordinate.}
On the other hand, if we compute the time component of the divergence of $\< T_{ab} \>_{new}$ using the second identity in \eqref{conserv_2}, we find in general that it is not equal to that of $\< T_{ab} \>$. This must necessarily be the case because the anomalous counterterm is not invariant under those boundary diffeomorphisms in which the time coordinate transforms as $u \to \tilde{u}(u,x^{i})$, and therefore break the spatial foliation of the boundary.

\subsubsection{Kerr solution}

As an application of the results of the previous section, we would like to compute the expectation value of the QFT energy tensor evaluated on those states dual to the asymptotically flat Schwarzschild and Kerr spacetimes. For the case of Schwarzschild-AdS$_4$, the metric in the coordinate system \eqref{metric} reads:
\bea
ds^{2} &=& {\ello^{2} \over z^{2}} \( - \( {1 \over \a^{2}} + {z^{2} \over \ello^{2}} - {2MG_{0} \over \ello^{4}}z^{3} \) du^{2} - 2dudz + \ello^{2}\,d\Omega^{2} \)\ ,
\eea
with $d\Omega^{2} = d\theta^{2} + \sin^{2}\theta d\phi^{2}$ the metric on the $S^{2}$ and where the cosmological constant $\L = - 3 / (\a^{2}\ello^{2})$.  In the limit $\a \to \infty$ the metric tends to four dimensional Schwarzschild. The expectation values of the components of the holographic energy tensor in this case become:
\bea
\sqrt{q_{(0)}}\ \< s_{ij} \> &=& \sqrt{g_{(0)}} \( {M \over 8\pi \ello^{2}}\, g_{(0)ij} \)\ , \\[5pt]
\sqrt{q_{(0)}}\ \< \varepsilon \> &=& \sqrt{g_{(0)}} \( {M \over 4\pi \ello^{2}} \)\ , \\[5pt]
\< j_{i} \> &=& 0\ ,
\eea
where the spatial metric $g_{(0)ij}dx^{i}dx^{j} = \ello^{2}\, d\Omega^{2}$. These expressions still hold in the zero $\L$ limit. The average energy $\<E\>$ as defined in \eqref{En_average} is then equal to $M$.\\

In the case of Kerr spacetime, the metric in Gaussian null coordinates is very complicated,\footnote{
See \cite{Fletcher:2003,Venter:2005cs} for specific examples. Note that Bondi-Sachs coordinates are related to the Gaussian null gauge by a simple redefinition of the radial coordinate.}
but we can deduce its asymptotics up to the desired order from the following considerations. The Kerr metric follows from the zero $\L$ limit of Kerr-AdS$_4$ and the latter is asymptotically exactly AdS$_4$ -- with the cross section of the asymptotic boundary with a spacelike hypersurface topologically an $S^{2}$. In our coordinate system, Kerr-AdS$_4$ must therefore be of the form:
\bea
ds^{2} &=& {\ello^{2} \over z^{2}} \( - \( {1 \over \a^{2}} + \op(z^{2}) \) du^{2} - 2dudz + \( g_{(0)ij} + \op(z) \) \( dx^{i} + \op(z)du\) \( dx^{j} + \op(z)du \) \)\ , \nn\\
\eea
where $g_{(0)ij}dx^{i}dx^{j} = \ello^{2}\,d\Omega^{2}$. Since the lapse $N_{(0)}=1$, from equation \eqref{sigma1} we have that $\s_{(1)}^{i} = 0$. Furthermore, since $\s_{(0)}^{i} = 0 = \partial_{u}g_{(0)ij}$, we find from equation \eqref{g1} that $g_{(1)ij} = 0$. From equations \eqref{g2_g1} and \eqref{sigma2} we then find that $g_{(2)ij} = 0 = \s_{(2)}^{i}$. Also, the spatial Ricci scalar $R[g_{(0)}] = 2/\ello^{2}$, so from \eqref{phi2} we have $\varphi_{(2)} = \ello^{-2}$. In this way, Kerr-AdS$_4$ must be asymptotically of the form:
\bea
ds^{2} &=& {\ello^{2} \over z^{2}}\, \bigg[ - \( {1 \over \a^{2}} + {z^{2} \over \ello^{2}} + \varphi_{(3)}z^{3} + \op(z^{>3}) \) du^{2} - 2dudz  \nn\\[5pt]
&+& \Big( g_{(0)ij} + z^{3}g_{(3)ij} + \op(z^{>3}) \Big) \( dx^{i} + \( z^{3}\s_{(3)}^{i} + \op(z^{>3}) \)du \) \( dx^{j} + \( z^{3}\s_{(3)}^{j} + \op(z^{>3}) \)du \) \bigg]\ . \nn\\
\eea
The coefficients $\varphi_{(3)}, g_{(3)ij}$ and $\s_{(3)}^{i}$ are the normalisable modes $G_{(d+1)\mu\nu}$ and from the constraint equations \eqref{Rzz}--\eqref{Ruu}, supplemented by \eqref{Einstein_eqs}, it follows that they satisfy:
\bea
&&\Tr[g_{(0)}^{-1}g_{(3)}] = 0 \ ,\\[5pt]
&&{1 \over \a^{2}} \cov_{j} (g_{(0)}^{-1}g_{(3)})^{j}_{~i} = \partial_{u}\s_{(3)i} + {1 \over 3}\, \partial_{i}\varphi_{(3)}\ ,\\[5pt]
&&{3 \over 2 \a^{2}} \cov_{i} \s_{(3)}^{i} = -\partial_{u}\varphi_{(3)}\ .
\eea
The holographic energy tensor so far reads:
\bea
\sqrt{q_{(0)}}\ \< s_{ij} \> &=& {\ello^{2} \over 8\pi G_{0}} \sqrt{g_{(0)}} \( {3 \over 2\a^{2}}\, g_{(3)ij} - \hf\, \varphi_{(3)}\, g_{(0)ij} \) \ ,\\[5pt]
\sqrt{q_{(0)}}\ \< \varepsilon \> &=& {\ello^{2} \over 8\pi G_{0}} \sqrt{g_{(0)}} \( -\varphi_{(3)} \) \ ,\label{en_dens_asympt_flat}\\[5pt]
\< j_{i} \> &=& {\ello^{2} \over 8\pi G_{0}} \( -{3 \over 2}\, \s_{(3)i} \) \ .
\eea
By using the second identity in equations \eqref{conserv_1} and \eqref{conserv_2} it then follows from the above constraints that the energy tensor is covariantly conserved:
\bea
^{(0)}\hspace{-2pt}\mathcal{D}_{a} \( \sqrt{q_{(0)}}\, \< T^{a}_{~i} \> \) &=& 0\, =\, n_{(0)}^{b}\, ^{(0)}\hspace{-2pt}\mathcal{D}_{a} \( \sqrt{q_{(0)}}\, \< T^{a}_{~b} \> \)\ .
\eea
Note that, apart from the constraints, the normalisable modes are so far arbitrary. We then require that the solution be stationary and axi-symmetric, which results in the constraints:
\bea
&&\Tr[g_{(0)}^{-1}g_{(3)}] = 0 \ ,\\[5pt]
&&{1 \over \a^{2}} \cov_{j} (g_{(0)}^{-1}g_{(3)})^{j}_{~i} = {1 \over 3}\, \partial_{i}\varphi_{(3)}\ , \label{g3_constraint}\\[5pt]
&&{1 \over \a^{2}} \cov_{i} \s_{(3)}^{i} = 0\ , \label{s3_constraint}
\eea
where the modes now depend only on the boundary coordinate $\theta$. These are the necessary conditions for Kerr-AdS$_4$. In the zero $\L$ limit, however, there will be a further constraint. Recall that the equation for a given coefficient $g_{(n)ij}$ is of the form \eqref{g_n} and, therefore, that it becomes a differential equation for $g_{(n-1)ij}$ in the limit $\a \to \infty$. For the particular case of $n=4$ in $d+2=4$ bulk dimensions, the equation for $g_{(4)ij}$ turns into a differential equation for the normalisable mode $g_{(3)ij}$ in the zero $\L$ limit. Therefore, if we solve the dynamical equation \eqref{gij}, together with \eqref{Einstein_eqs}, at order $z^{2}$ we find in the limit $\a \to \infty$:
\bea\label{k3}
4\, k_{(3)ij} - g_{(0)ij}\, \Tr[g_{(0)}^{-1}k_{(3)}] + \varf_{(4)}\, g_{(0)ij} + 3 \cov_{(i}\s_{j)}^{(3)} &=& 0\ ,
\eea
where we have used the fact that $g_{(1)ij} = g_{(2)ij} = \s_{(1)i} = \s_{(2)i} = 0$ in our case. The equation for the coefficient $\varf_{(4)}$ follows from the dynamical equation \eqref{Rzu} and \eqref{Einstein_eqs} for $\varf$:
\bea
\varf_{(4)} - 2 \Tr[g_{(0)}^{-1}k_{(1)}] - {3 \over 2}\, \cov_{i}\s_{(3)}^{i} &=& 0\ .
\eea
Replacing in \eqref{k3}, we find:
\bea\label{k3_2}
4\, k_{(3)ij} + g_{(0)ij}\, \Tr[g_{(0)}^{-1}k_{(3)}] + 3 \( \cov_{(i}\s_{j)}^{(3)} + \hf g_{(0)ij} \cov_{i}\s_{(3)}^{i} \) &=& 0\ .
\eea
Now, in our case we have:
\bea
k_{(3)ij} &=& {1 \over 2N_{(0)}} \( (\partial_{u} - \Lie_{\s_{(0)}}) g_{(3)ij} - \Lie_{\s_{(1)}}g_{(2)ij} - \Lie_{\s_{(2)}}g_{(1)ij} - \Lie_{\s_{(3)}}g_{(0)ij} \) \nn\\
&=& - \cov_{(i}\s_{j)}^{(3)}\ .
\eea
Replacing in equation \eqref{k3_2} results in the following constraint for $\s_{(3)i}$:
\bea
\cov_{(i}\s_{j)}^{(3)} - \hf\, g_{(0)ij} \cov_{k}\s_{(3)}^{k} &=& 0\ .
\eea
Now, the constraint \eqref{s3_constraint} for $\s_{(3)}^{i}$ holds for all values of $\a \in \Real$, so we extend this to the limit $\a \to \infty$ so that the metric is continuous in $\a$. If this were not the case, then $\s_{(3)}^{i}$ would contain terms proportional to $\delta_{\L,0}$ and therefore Kerr would not follow from the zero $\L$ limit of Kerr-AdS$_4$. The same argument applies to the $\phi$-component of the constraint \eqref{g3_constraint}. The constraint equations for the normalisable modes in the limit $\a \to \infty$ therefore become:
\bea
&&\Tr[g_{(0)}^{-1}g_{(3)}] = 0 = \cov_{j}(g_{(0)}^{-1}g_{(3)})^{j}_{~i=\phi}\ ,\\[5pt]
&&\partial_{i}\varphi_{(3)} = 0\ ,\\[5pt]
&&\cov_{i}\s_{(3)j} + \cov_{j}\s_{(3)i} = 0\ ,
\eea
where the modes depend only on $\theta$. The coefficient $\s_{(3)}^{i}$ is therefore a Killing vector of the spatial metric $g_{(0)ij}$ on the $S^{2}$ and hence we choose: $\s_{(3)}^{i}\partial_{i} := a/\ello^{4}\, \partial_{\phi}$, with $a$ some dimensionless constant. Furthermore, $\varf_{(3)}$ is constant, so we define: $\varf_{(3)} := -2MG_{0}/\ello^{4}$. Note also that in the limit $a \to 0$ we must recover the Schwarzschild metric, so $g_{(3)ij}$ must be proportional to the parameter $a$. The average energy and angular momentum of those states dual to asymptotically flat Kerr are then given by:
\bea
\< E \> &=& {1 \over 2T}\, \int\limits_{-T}^{T}du\int d^{2}x \sqrt{q_{(0)}}\, \< \varepsilon \>\, =\, M\ ,\\[5pt]
\< J^{i} \> \partial_{i} &=& \int d^{2}x \sqrt{g_{(0)}}\, \< j^{\,i} \> \partial_{i}\, =\, - {3 \over 4 G_{0}}\, a\, \partial_{\phi}\ .
\eea
More generally, for an asymptotically Minkowski spacetime we have that $g_{(1)ij}=0$, so the energy density will be of the form \eqref{en_dens_asympt_flat}. The average energy will then be given by:
\bea
\< E \> &=& - {1 \over 2T}\, {\ello^{2} \over 8\pi G_{0}}\int d^{2}x \sqrt{g_{(0)}}\int\limits_{-T}^{T}du\, \varf_{(3)} \nn\\
&=& {1 \over 2T} \int\limits_{-T}^{T}du\, M(u)\ ,
\eea
where $M(u)$ is the Bondi mass (see \eg \cite{Tanabe:2011es}).\\

\subsection{Null boundaries and corner terms}\label{corner_section}

So far we considered a single timelike boundary $\{z=\e\}$ for the spacetime and neglected all possible corner integrals evaluated on the codimension two surfaces $\{z=\e, u=\pm \infty\}$ that may arise in the gravitational action. If one also considers null boundaries $\{u=u_{\pm}\}$ in the spacetime, where these surfaces can be at infinity, the original bare action \eqref{action} is not the appropriate one in the sense that the variational problem is not well-defined, and a further surface term is needed. Furthermore, the renormalized gravity action in each dimension will require corner counterterms at $\{z=\e, u=u_{\pm}\}$ that ensure that the action is finite once the regulator $\e$ is removed. In order to derive the correct bare action in general, we start by performing an ADM decomposition of the spacetime metric with respect to timelike hypersurfaces of constant $z$ as:
\bea
ds^{2}_{d+2} &=& G_{\mu\nu}dx^{\mu}dx^{\nu} \nn\\[5pt]
&=& \b^{2}dz^{2} + q_{ab} \( dx^{a} + \b^{a} dz \) ( dx^{b} + \b^{b} dz )\ .
\eea
The inverse and determinant of the metric are given by:
\bea\label{inverse_det}
G^{\mu\nu} &=&
\begin{pmatrix} 
{1 \over \b^{2}} & -{1 \over \b^{2}}\, \b^{a}\\
-{1 \over \b^{2}}\, \b^{a}\ & q^{ab} + {1 \over \b^{2}}\, \b^{a} \b^{b}
\end{pmatrix}\ ,\label{inverse_G}\\[5pt]
\sqrt{G} &=& \b\sqrt{q}\ .
\eea
The unit normal $m^{\mu}$ to the surfaces of constant $z$ is given by:
\bea
m_{\mu} &=& \b \partial_{\mu}z\ ,\\
m^{\mu}\partial_{\mu} &=& {1 \over \b}\, \( \partial_{z} - \b^{a}\partial_{a} \)\ ,\label{m_up}\\
m^{\mu}m^{\nu}G_{\mu\nu} &=& 1\ .
\eea
The metric $q_{ab}$ represents the induced metric of the hypersurfaces of constant $z$ and we can extend it to a tensor in the whole spacetime by defining: $q^{\mu\nu} := G^{\mu\nu} - m^{\mu}m^{\nu}$. Next we perform an ADM decomposition of $q_{ab}$ with respect to surfaces of constant $u$. In each submanifold $\{z=constant\}$, we define these surfaces to be spacelike:
\bea
ds^{2}_{d+1} &=& q_{ab}dx^{a}dx^{b} \nn\\[5pt]
&=& -N^{2}du^{2} + \g_{ij} (dx^{i} + \s^{i} du ) (dx^{j} + \s^{j} du)\ .
\eea
The determinant of this metric is given by: $\sqrt{q} = N\sqrt{\g}\,$, so that: $\sqrt{G} = \beta N \sqrt{\g}$. In each submanifold $\{z=constant\}$, the future-directed unit normal $n^{a}$ to the surfaces of constant $u$ is given by:
\bea
n_{a} &=& - N \partial_{a}u\ ,\\
n^{a}\partial_{a} &=& {1 \over N} \( \partial_{u} - \s^{i}\partial_{i} \)\ ,\\
n^{a}n^{b}q_{ab} &=& -1\ .
\eea
We can extend this unit normal to a vector in the whole spacetime by defining: 
\bea
n^{\mu} := q^{\mu\nu} \( -N \partial_{\nu}u \)\ .
\eea 
We then find: $n^{\mu}n^{\nu}G_{\mu\nu} = -1$ and: $m^{\mu}n^{\nu}G_{\mu\nu} = 0$. Finally, with the two unit normals $m^{\mu}$ and $n^{\mu}$ we construct two null vectors $n_{\pm}^{\mu}$ defined as:
\bea
n_{\pm}^{\mu} := n^{\mu} \pm m^{\mu}\ .
\eea
We find that: $n_{\pm}^{\mu}n_{\pm}^{\nu}G_{\mu\nu} = 0$ and: $n_{\pm}^{\mu}m^{\nu}G_{\mu\nu} = \pm 1$. Given this general construction, we will now show that, if the surfaces $\{u=u_{\pm}\}$ are null in the spacetime, the bare gravitational action for which the variational problem is well-posed is given by:
\bea\label{correct_bare_action}
16\pi G_{0}\, S &=& \int dzdud^{d}x\, \sqrt{G} \( {d(d+1) \over \a^{2}\ello^{2}} + R[G] \) \nn\\[5pt]
&+& 2\int\limits_{z=\e} du d^{d}x\, \sqrt{q}\, Q - 2 \int\limits^{u=u_{+}}_{u=u_{-}} dz d^{d}x\, \beta \sqrt{\g}\ \na_{\mu}n_{+}^{\mu}\ ,
\eea
with $\na_{\mu}G_{\nu\a} := 0$, and where $Q$ is the extrinsic curvature of the hypersurfaces of constant $z$ as before, such that: $Q = \na_{\mu}m^{\mu}$. Also, the last integral represents the difference: $\int\limits^{u=u_{+}}_{u=u_{-}} := \int\limits_{u=u_{+}} - \int\limits_{u=u_{-}}$. In order to show that the variational problem is well-defined, we perform a Gauss-Codazzi decomposition of the Ricci scalar $R[G]$:
\bea
R[G] &=& R[q] + Q^{2} - Q\cdot Q - 2 \na_{\mu} \( m^{\mu}\, \na\cdot m - m\cdot\na m^{\mu} \)\ .
\eea
Replacing in \eqref{correct_bare_action} and integrating the total derivatives results in the action:
\bea\label{correct_bare_action_2}
16\pi G_{0}\, S &=& \int dzdud^{d}x\, \beta\sqrt{q} \( {d(d+1) \over \a^{2}\ello^{2}} + R[q] + Q^{2} - Q\cdot Q \) \nn\\[5pt]
&-&2 \int\limits^{u=u_{+}}_{u=u_{-}} dzd^{d}x\, \beta\sqrt{\g}\, \bigg( K + \Big( 1 + N m^{\mu}\partial_{\mu}u \Big) \na\cdot m \bigg)\ ,
\eea
where $K = q^{ab}D_{a}n_{b} = q^{\mu\nu} \na_{\mu} \( q_{\nu}^{~\a}n_{\a} \) = q^{\mu\nu}\na_{\mu}n_{\nu}$ is the extrinsic curvature of the surfaces of constant $u$ in each submanifold $\{ z = constant \}$, with $D_{a}q_{bc} :=0$. Now, from the decomposition \eqref{inverse_G} we find in particular that:
\bea
\partial_{\mu}u\partial_{\nu}u\, G^{\mu\nu} &=& q^{uu} + \(\b^{u}/\beta\)^{2}\, =\, - N^{-2} + \(\b^{u}/\beta\)^{2}\ .
\eea
If the surfaces $u=u_{\pm}$ are null in the spacetime, the left-hand side vanishes at $u=u_{\pm}$ and we find up to a sign: $\b^{u} = \beta/N$ at $u=u_{\pm}$. If we choose the opposite sign, then we should replace the null vector $n_{+}$ in \eqref{correct_bare_action} by its dual $n_{-}$. Replacing this condition for $\beta^{u}$ in equation \eqref{m_up} results in: 
\bea\label{m_u}
1 + N\,m^{\mu}\partial_{\mu}u = 0\qquad (u=u_{\pm})\ .
\eea
Note that this holds everywhere if the surfaces of constant $u$ are everywhere null, and in such case the null vector $n_{+}$ is given by: $n_{+\mu} = -N \partial_{\mu}u$. Finally, using equation \eqref{m_u} in the action \eqref{correct_bare_action_2} yields our final result:\footnote{
Note that the Gibbons-Hawking surface term takes a minus sign because we have defined the unit normal $n^{a}$ to be future-directed.}
\bea
16\pi G_{0}\, S &=& \int dz \( \int dud^{d}x\, \beta\sqrt{q} \( {d(d+1) \over \a^{2}\ello^{2}} + R[q] + Q^{2} - Q\cdot Q \) -2 \int\limits^{u=u_{+}}_{u=u_{-}} d^{d}x\, \beta\sqrt{\g}\, K \)\ . \nn\\
\eea
This is the correct action for which the variational problem is well-posed \cite{Brown:1992br}. Taking variations with respect to the lapse, shift, and induced metric $\beta, \beta^{a}$ and $q_{ab}$, and requiring only that the boundary configurations of the fields are fixed, results in the ADM equations of motion.\\

If the spacetime contains null boundaries, the holographic renormalization of the gravitational action \eqref{correct_bare_action} will result in corner counterterms as emphasized above. We will exemplify this for the particular case of $d+2=3$ dimensions and derive the anomalous counterterm \eqref{corner} discussed in section \ref{ren_3}. Returning to our gauge-fixed metric \eqref{metric} for generic $d$, if we evaluate on-shell the last integral in the action \eqref{correct_bare_action}, we obtain:
\bea
- 2 \int\limits^{u=u_{+}}_{u=u_{-}} dz d^{d}x\, \beta \sqrt{\g}\ \na_{\mu}n_{+}^{\mu} &=& -2 \int\limits^{u=u_{+}}_{\substack{u=u_{-}\\ \hspace{-7pt} z=\e}} d^{d}x\, \sqrt{g} \( { \ello \over \e} \)^{d} +2 \ello^{d} \int\limits^{u=u_{+}}_{u=u_{-}} dz d^{d}x\, \sqrt{g} \( z^{-(d+1)} - \hf\, z^{-d} \partial_{z}\log \varf \)\ . \nn\\
\eea
Using our asymptotic solutions \eqref{gij}--\eqref{sigma} we find that, for $d=1$, the divergences of this term are given by:
\bea
- 2 \int\limits^{u=u_{+}}_{u=u_{-}} dz dx\, \beta \sqrt{\g}\ \na_{\mu}n_{+}^{\mu} &=& -4 \int\limits^{u=u_{+}}_{\substack{u=u_{-}\\ \hspace{-7pt} z=\e}} dx\, \sqrt{g_{(0)}} \({\ello \over \e}\) + 2\a^{2} \ello \int\limits^{u=u_{+}}_{\substack{u=u_{-}\\ \hspace{-7pt} z=\e}} dx\, \sqrt{g_{(0)}}\, k_{(0)} \log\e\ + \op(\e^{0})\ .\nn\\
\eea
The counterterm that subtracts these divergences is given by:
\bea
4 \int\limits^{u=u_{+}}_{\substack{u=u_{-}\\ \hspace{-7pt} z=\e}} dx\, \sqrt{\g} - 2\, \a\ello \int\limits^{u=u_{+}}_{\substack{u=u_{-}\\ \hspace{-7pt} z=\e}}dx\, \sqrt{\g}\, K \log\e\ .
\eea
If we also take into account the surface term \eqref{discarded_surface_term} that we discarded and use the result we found in \eqref{Sren_3}, we find that the renormalized gravitational action in $d+2=3$ spacetime dimensions in the presence of null boundaries $u=u_{\pm}$ is given by:
\bea
16\pi G_{0}\, S_{ren} &=& \int dzdudx\, \sqrt{G} \( {d(d+1) \over \alpha^{2}\ello^{2}} + R[G] \) + 2 \int\limits_{z=\epsilon} dudx \sqrt{q}\, Q - 2 \int\limits^{u=u_{+}}_{u=u_{-}} dz dx\, \beta \sqrt{\g}\ \na_{\mu}n_{+}^{\mu} \nn\\[5pt]
&+& {2 \over \a\,\ello} \int\limits_{z = \epsilon}dudx\, \sqrt{q} + 6 \int\limits^{u=u_{+}}_{\substack{u=u_{-}\\ \hspace{-7pt} z=\e}} dx\, \sqrt{\g} - 2\,\a\ello \int\limits^{u=u_{+}}_{\substack{u=u_{-}\\ \hspace{-7pt} z=\e}} dx\, \sqrt{\g}\, K \log\e\ .
\eea
The last corner integral is exactly the anomalous counterterm that we found in \eqref{corner}.\\

\section{Non-backreacting matter}\label{matter}

In the remainder of this work we will be interested in computing the zero $\L$ limit of the vacuum expectation value and two-point correlator of a QFT operator dual to a non-backreacting massive scalar field in AdS$_{d+2}$. The background metric we are interested in is pure AdS with the cross section of the asymptotic boundary with a spacelike hypersurface topologically $\Real^{d}$. In our coordinate system the metric reads:
\bea\label{AdS}
ds^{2} &=& G_{\mu\nu}dx^{\mu}dx^{\nu} \nn\\[5pt]
&=& {\ello^{2} \over z^{2}} \( -{1 \over \alpha^{2}}\, du^{2} - 2 du dz + d\vec{x}^{\,2}_{d} \)\ . 
\eea
In the limit $\a \to \infty$ the spacetime is a subset of Minkowski space, with $z=0$ representing future null infinity. The bulk action for the scalar field $\phi$ in this background is given by:
\bea\label{scalar_action}
S &=& \hf \int d^{d+2}x \sqrt{G} \( G^{\mu\nu}\partial_{\mu}\phi\partial_{\nu}\phi + \({m \over \a}\)^{2}\phi^{2} \)\ .
\eea
The mass of the field is defined to be $M = m/\a$. For the moment we will keep $m$ arbitrary, but as is well-known, the conformal weight of the field theory operator dual to $\phi$ will be finite in the limit $\a \to \infty$ only if $m = \op(\a^{0})$.

\subsection{Solution and asymptotics}\label{matter_solution}

The equations of motion for the scalar in our background are given by:
\bea
\({m \over \a}\)^{2} \phi &=& \square_{G}\, \phi \nn\\[5pt]
&=& {z^{\D - k + 2} \over \ello^{2}}\, \bigg[ {1 \over \a^{2}} \( \varf '' - {k-1 \over z} \varf ' \) - 2 \partial_{u} \varf ' + {k-1 \over z} \partial_{u} \varf + \vec{\na}^{2}\varf + {\D(\D-(d+1)) \over \a^{2} z^{2}}\, \varf \bigg]\ , \nn\\[5pt]
\eea
where we defined $\varf := z^{k-\D} \phi$ for $\D$ constant and $k:= 2\D - (d+1)$. As usual, $\D$ will be the conformal weight of the dual field theory operator. Also, $\varf ' := \partial_{z}\varf$ and $\vec{\na}^{2} = \delta^{ij}\partial_{i}\partial_{j}$. We will be interested in computing the correlation functions of the QFT operator in Euclidean signature, so we define the Euclidean boundary time $\bar{u} := iu$. The dynamical equation then becomes:
\bea\label{dynamical_varf}
{1 \over \a^{2}} \( \varf '' - {k-1 \over z} \varf ' \) - 2 i \dot{\varf} ' + i {k-1 \over z} \dot{\varf} + \vec{\na}^{2}\varf + {\D(\D - (d+1)) - \ello^{2}\,m^{2} \over \a^{2}z^{2}}\, \varf &=& 0\ ,
\eea
where $\dot{\varf} := \partial_{\bar{u}} \varf$. We define $\D$ as the highest root of the equation \mbox{$\D(\D - (d+1)) = \ello^{2}\,m^{2}$}. We also Fourier transform the dynamical equation in the coordinates $\bar{u}$ and $x^{i}$ and obtain:
\bea
{1 \over \a^{2}} \( \hat{\varf} '' - {k-1 \over z} \hat{\varf} ' \) + 2 \omega \hat{\varf} ' -\omega\, {k-1 \over z} \hat{\varf} - \vec{p}^{\,2}\hat{\varf} &=& 0\ ,
\eea
where:
\bea
\hat{\varf}(z,\omega,p^{i}) = \int d\bar{u}\, d^{d}x\, e^{-i\omega \bar{u}}e^{-i\vec{p}\cdot \vec{x}}\, \varf(z,\bar{u},x^{i})\ .
\eea
The solution for $\hat{\varf}$ can be written in terms of Bessel functions as:
\bea\label{sol_hat_phi}
\hat{\varf}(z,\omega,p) &=& e^{-\a^{2}\omega z}\, z^{k/2} \bigg[ A(\omega,p)\, K_{k / 2}(z\, \a \sqrt{\vec{p}^{\,2} + \a^{2}\omega^{2}}) + B(\omega,p)\, I_{k/2}(z\, \a \sqrt{\vec{p}^{\, 2} + \a^{2}\omega^{2}}) \bigg]\ , \nn\\
\eea
where the coefficients $A(\omega,p)$ and $B(\omega,p)$ are arbitrary, and where $K_{k/2}(y)$ and $I_{k/2}(y)$ are the modified Bessel functions of the first and second kind. These admit the following asymptotics as $y \to 0$:
\bea
K_{k/2}(y) &=& 2^{k/2-1}\, \Gamma(k/2)\, y^{-k/2} \bigg( 1 + {(iy)^{2} \over 2(k-2)} + {(iy)^{4} \over 2(k-2) 4(k-4)}\nn\\[5pt] 
&&\hspace{9em} +\, ... + a_{k}\, y^{k} + \tilde{a}_{k}\, y^{k} \log{y^{2}} + \op(y^{>k}) \bigg) \ , \label{Bessel_K}\\[5pt]
I_{k/2}(y) &=&  {2^{-k/2} \over \Gamma(k/2+1)}\, y^{-k/2} \( y^{k} + \op(y^{>k}) \)\ ,
\eea
with $\Gamma(a)$ the gamma function and $a_{k}$ a $k$-dependent constant. The coefficient $\tilde{a}_{k}$ is non-vanishing only if $k/2$ is an integer and in such case is given by:
\bea
\tilde{a}_{k} &=& -{(-1)^{k/2}\ 2^{-k} \over \Gamma(1+k/2)\Gamma(k/2)} \qquad :\quad k/2 \in \Integer\ .
\eea
The solution for $\hat{\varf}$ therefore admits the expansion:
\bea
\hat{\varf}(z,\omega,p) &=& e^{-\a^{2}\omega z}\, \bigg[ \( 1 - {\a^{2}(\vec{p}^{\, 2} + \a^{2}\omega^{2}) \over 2(k-2)}\, z^{2} + {\a^{4}(\vec{p}^{\, 2} + \a^{2}\omega^{2})^{2} \over 8(k-2)(k-4)}\, z^{4} + ... \) \hat{\varf}_{(0)}(\omega,p) \nn\\[5pt]
&& \hspace{3em} +\, b(\omega,p)\, z^{k} + \tilde{\hat{\varf}}_{(k)}(\omega,p)\, z^{k} \log z + \op(z^{>k}) \bigg] \nn\\[5pt]
&=& \hat{\varf}_{(0)} + z\, \hat{\varf}_{(1)} + z^{2}\, \hat{\varf}_{(2)} + z^{3}\, \hat{\varf}_{(3)} + ... + z^{k}\, \hat{\varf}_{(k)} + z^{k}\log z\, \tilde{\hat{\varf}}_{(k)} + \op(z^{>k})\ , \nn\\
\eea
where we wrote the function $A(\omega,p)$ as: 
\bea\label{coeff_A}
A(\omega,p) = \dfrac{2^{1-k/2}}{\Gamma(k/2)} \(\a\sqrt{\vec{p}^{\, 2} + \a^{2}\omega^{2}}\)^{k/2} \hat{\varf}_{(0)}(\omega,p)\ .
\eea
The coefficients $\hat{\varf}_{(0)}(\omega,p)$ and $\hat{\varf}_{(k)}(\omega,p)$ are arbitrary functions in $\omega$ and $\vec{p}^{\, 2}$ and the coefficients $\hat{\varf}_{(n<k)}$ are given up to $n=3$ by:
\bea
\hat{\varf}_{(1)} &=& -\a^{2}\omega\, \hat{\varf}_{(0)}\ ,\\
\hat{\varf}_{(2)} &=& \( \hf\, \a^{4}\omega^{2} - {\a^{2}(\vec{p}^{\, 2} + \a^{2}\omega^{2}) \over 2(k-2)} \) \hat{\varf}_{(0)}\ ,\\
\hat{\varf}_{(3)} &=& \(-{1 \over 6}\, \a^{6}\omega^{3} + {\a^{4}\omega (\vec{p}^{\, 2} + \a^{2}\omega^{2}) \over 2(k-2)} \) \hat{\varf}_{(0)}\ .
\eea
The coefficient $\tilde{\hat{\varf}}_{(k)}$ of the inhomogeneous term is given by:
\bea
\tilde{\hat{\varf}}_{(k)} &=& 2\, \tilde{a}_{k} \(\a \sqrt{\vec{p}^{\,2} + \a^{2}\omega^{2}}\)^{k} \hat{\varf}_{(0)}\ .
\eea
The full solution $\phi(z,\bar{u},x^{i})$ for the scalar field is then given by:
\bea\label{scalar_asymptotics}
\phi(z,\bar{u},x) &=& z^{\Delta - k} \int d\omega d^{d}p\, e^{i\omega\bar{u}}\, e^{i\vec{p}\cdot \vec{x}}\, \hat{\varf}(z,\omega,p) \nn\\[5pt]
&=& z^{\Delta-k} \( \varf_{(0)} + z\, \varf_{(1)} + z^{2}\, \varf_{(2)} + z^{3}\, \varf_{(3)} + ... + z^{k}\, \varf_{(k)} + z^{k}\log (\mu z)\, \tilde{\varf}_{(k)} + \op(z^{>k}) \)\ , \nn\\
\eea
where we introduced a scale $\mu$ of dimension L$^{-1}$ so that the argument of the logarithm is dimensionless. The coefficients $\varf_{(0)}=\varf_{(0)}(\bar{u},x)$ and $\varf_{(k)}=\varf_{(k)}(\bar{u},x)$ are arbitrary functions and represent the standard non-normalisable and normalisable modes in the AdS/CFT correspondence. The boundary configuration $\varf_{(0)}$ is the source for the scalar operator $\op$ in the dual QFT and $\varf_{(k)}$ will be mapped to the vacuum expectation value of $\op$. The coefficients $\varf_{(n<k)}$ together with the inhomogeneous term $\tilde{\varf}_{(k)}$ are local functionals of the source for the case of $\a$ finite. Their expressions are given by:
\bea
{1 \over \a^{2}}\, \varf_{(n)} &=& {1 \over n(k-n)} \( i ( k+1-2n)\, \dot{\varf}_{(n-1)} + \vec{\na}^{2} \varf_{(n-2)} \)\quad :\ 0<n<k\ ,\label{phi_n}\\[10pt]
{1 \over \a^{2}}\, \tilde{\varf}_{(k)} &=&
\begin{cases}\label{phi_tilde_k}
{1 \over k} \( i(k-1) \dot{\varf}_{(k-1)} - \vec{\na}^{2} \varf_{(k-2)} \)\quad :\ k/2 \in \Integer\ , \\[5pt]
0\quad \rm{otherwise}\ ,
\end{cases}
\eea
where $\varf_{(-1)} := 0$. The above is exactly the asymptotic solution one would obtain by solving the dynamical equation \eqref{dynamical_varf} in powers of $z$ in a neighbourhood of $z=0$. In the case $\a^{-1} = 0$, the coefficients are non-local functionals of the sources in the same fashion as the coefficients $g_{(n)ij}$ in the asymptotic expansion \eqref{gij} of the metric that we found in section \ref{asympt_sol}. For the case of $\a$ finite, the source $\varf_{(0)}$ and the mode $\varf_{(k)}$ are arbitrary, so there will be solutions for the scalar field in AdS that diverge in the limit $\a \to \infty$. We are interested in those configurations for the field that result in well-defined solutions of the equations of motion in Minkowski space in this limit, so we henceforth restrict our space of solutions in AdS to the subspace of those that admit the limit. This discussion mimics that in section \ref{asympt_sol} for the spacetime metric. This is enforced by requiring that the coefficients in the asymptotics \eqref{scalar_asymptotics} be non-divergent as $\a \to \infty$. Since the modes $\varf_{(n<k)}$ and $\tilde{\varf}_{(k)}$ are functionals of $\varf_{(0)}$, this requirement imposes constraints on the behaviour in $\a$ of the derivatives of the source. For $k$ non-odd, these will be constraints on the time derivatives. As an example, from $n=1, 2,3$ it follows that:
\bea
\dot{\varf}_{(0)} &=& \op(\a^{-2}) \ ,\label{constraint-alpha-1}\\[5pt]
\ddot{\varf}_{(0)} &=& {1 \over \a^{2}} \({1 \over k-3}\, \vec{\na}^{2}\varf_{(0)}\) + \op(\a^{-4})\ , \label{constraint-alpha-2} \\[5pt]
\dddot{\varf}_{(0)} &=& {1 \over \a^{4}} \({3 \over k-5}\, \vec{\na}^{2}\big(\a^{2}\dot{\varf}_{(0)}\big) \) + \op(\a^{-6})\ . \label{constraint-alpha-3}
\eea
On the other hand, for odd values of $k$ there will be a further constraint, this time on the spatial derivatives: $\vec{\na}^{k-1} \varf_{(0)} = \op(\a^{-(k-1)})$. As in section \ref{asympt_sol}, we find again that the well-definedness of the bulk solutions in the zero $\L$ limit translates into a statement about the sources and states on the dual QFT and, in particular, that the existence of the limit is connected with the behaviour in $\a$ of the time and spatial derivatives of the source. The specific dependence in powers of $\alpha$ found in \eqref{constraint-alpha-1}--\eqref{constraint-alpha-3} of the different time derivatives of the source follow directly from the recursion relations \eqref{phi_n}--\eqref{phi_tilde_k}. For a given even value of $k$, for example, there will be $k$ constraints on the time derivatives of $\varf_{(0)}$. If at least one of these constraints is not satisfied by the source, then at least one of the coefficients $\varf_{(n)}$ in the asymptotic expansion \eqref{scalar_asymptotics} will be divergent and the solution will not be well-defined in the limit $\alpha \to \infty$. The same statement holds for $k$ odd with an additional constraint on the spatial derivatives as above. As an exercise, from equation \eqref{phi_n} for $n=1$ we have that: 
\bea
\dot{\varf}_{(0)} = -i \alpha^{-2} \varf_{(1)}\ . \label{phi-0-phi-1}
\eea 
If the constraint \eqref{constraint-alpha-1} is not satisfied it follows that $\varf_{(1)}$ necessarily diverges as $\a \to \infty$. We can also differentiate equation \eqref{phi-0-phi-1} with respect to time and use the recursion relation \eqref{phi_n} for $n=2$ to obtain:
\bea
\ddot{\varf}_{(0)} = {1 \over \a^{2}} \({1 \over k-3}\, \vec{\na}^{2}\varf_{(0)}\) - {1 \over \a^{4}} \({k-2 \over k-3}\, \varf_{(2)}\)\ .
\eea
If the constraint \eqref{constraint-alpha-1} is satisfied but not \eqref{constraint-alpha-2}, then it is the coefficient $\varf_{(2)}$ that necessarily diverges as $\a \to \infty$. The same reasoning can be repeated for the remaining constraints.

Equation \eqref{constraint-alpha-1} is particularly relevant and it implies that the source is time-independent in the zero $\L$ limit. This is not an issue for the variational problem that we discuss in the next sections because the zero $\L$ limit is taken after varying the gravitational action with respect to the source, so the latter remains arbitrary until the vacuum expectation values and correlators are computed. In section \ref{asympt_sol} we have also discussed the behaviour in $\alpha$ of the time derivatives of the source $g_{(0)ij}$ for the boundary stress-tensor and we have found in particular from equation \eqref{g0-alpha} that one can choose a time coordinate $u$ such that $g_{(0)ij}$ is time-independent in the limit $\alpha \to \infty$ in the same fashion as the source $\varf_{(0)}$ (recall that the boundary shift $\sigma_{(0)}^i$ can be fix to zero by the boundary diffeomorphism $x^{i} \to x^{i} - \int du \sigma_{(0)}^i$). Furthermore, we will find in section \ref{correlator} that the two-point correlators of scalar operators are also independent of the time coordinate in the limit $\a \to \infty$. These results are compatible with those discussed in \cite{Bagchi:2012xr,Bagchi:2012cy} and suggest that the zero $\Lambda$ limit induces an ultra-relativistic contraction $(u,\vec{x}) \to (\a^{-1} u, \vec{x})$ on the boundary field theory.

\subsection{Renormalization and vacuum expectation values}

In this section we will renormalize holographically the bulk action for the scalar field in the AdS background \eqref{AdS}, analyse the limit $\a \to \infty$ and compute the vev of the dual operator. Under this limit the spacetime becomes Minkowski space and the solution in AdS is mapped to a solution of the scalar field equations in Minkowski. As in section \ref{onshell_action}, we proceed by replacing the asymptotic boundary of the spacetime by a regulating surface $z=\e$ and evaluate \eqref{scalar_action} on-shell:
\bea
i S^{on-shell} &=& {\ello^{d} \over 2\a^{2}} \int\limits_{z=\e} d\bar{u} d^{d}x\, \e^{-k} \Big( (\D-k)\, \varf^{2} + \e\, \varf\,\varf ' \Big) - {\ello^{d} \over 4} \int\limits_{z=\e} d\bar{u} d^{d}x\, \e^{-k+1} \partial_{\bar{u}} \varf^{2} .\quad
\eea
The integrand in the last integral is a total derivative and therefore can be removed from the on-shell action in the absence of null boundaries $\{u=constant\}$ for the spacetime. We then use the asymptotic solution \eqref{scalar_asymptotics} to replace for $\varf$ and find those terms that diverge if we take the limit $\e \to 0$. For finite $\a$ these will be local functionals of the source $\varf_{(0)}$ and therefore, up to anomalies, can be rewritten covariantly as described in section \ref{onshell_action}. The resulting divergent terms can then be subtracted by a covariant counterterm action $S_{ct}$ consisting of minus such terms. The renormalized action $S_{ren}$ will then be given by $S_{ren} = S + S_{ct}$. The number of counterterms increases with $k$, so we will focus separately on the cases $k=2$ and $k=4$.

\subsubsection{k=2}

In this case the procedure described above results in the following renormalized action:
\bea\label{scalar_S_ren}
i S_{ren} &=& \hf \int d^{d+2}x \sqrt{G} \( G^{\mu\nu}\partial_{\mu}\phi\partial_{\nu}\phi + \({m \over \a}\)^{2}\phi^{2} \) \nn\\[5pt]
&+& \hf \int\limits_{z=\e} d^{d+1}x\, \sqrt{q} \( -{\D - k \over \a\ello}\, \phi^{2} + (\a\ello)\, \phi \square_{q}\phi\, \log\e \)\ ,
\eea
where $q_{ab}$ is the induced metric on the regulating surface:
\bea
q_{ab}dx^{a}dx^{b} &=& {\ello^{2} \over \e^{2}} \( {1 \over \a^{2}}\, d\bar{u}^{2} + d\vec{x}_{d}^{2} \) \nn\\[5pt]
&=& {\ello^{2} \over \e^{2}}\, q_{(0)ab} dx^{a}dx^{b}\ ,
\eea
and where $\square_{q}$ is the Laplacian with respect to $q_{ab}$ and $q_{(0)ab}$ is the QFT metric. The resulting couterterms are precisely the canonical ones from standard holographic renormalization in the AdS/CFT correspondence (see \eg \cite{Skenderis:2002wp}). This is expected because the canonical counterterm action is covariant up to the anomaly in $\log \e$. The latter breaks invariance of the action under specific bulk diffeomorphisms involving the radial coordinate $z$, but our background \eqref{AdS} is mapped to the Poincar\'e patch of AdS by the boundary diffeomorphism $u \to \a^{2}(u-z),\, x^{i} \to \a\, x^{i}$. The surfaces of constant $z$ are therefore preserved by the diffeomorphism and hence the canonical counterterm action is not affected by the transformation.

The next step is to determine whether the counterterms spoil the zero $\L$ limit of the renormalized on-shell action. For that purpose we evaluate $S_{ren}$ on-shell, take the limit $\e \to 0$ and look for those terms proportional to positive powers of $\a$ as described in section \ref{onshell_action}. In the simple case of $k=2$ no such terms survive once the regulator is removed and therefore the couterterm action does not spoil the zero $\L$ limit. As we increase the value of $k$ we will see that further counterterms are needed apart from the canonical ones in order to restore the well-behaved-ness of the action in the limit $\a \to \infty$.\\

\paragraph{Vacuum expectation value\\}\n

\n The variation of the renormalized on-shell action is given by:
\bea
i\, \delta S_{ren}^{on-shell} &=& \int\limits_{z=\e} d^{d+1}x\, \sqrt{q} \( {\e \over \a\ello}\( \phi ' - i\a^{2} \dot{\phi} \) - {\D - k \over \a\ello} \phi + (\a\ello)\, \square_{q}\phi\, \log\e \) \delta \phi\ . \qquad
\eea
Using the AdS/CFT prescription, the one-point function of the dual operator $\op$ is then given by:\footnote{
Recall from section \ref{vevs_3} that the well-defined observables are always the tensor densities, in this case $\sqrt{q_{(0)}}\, \< \op \>$. By construction, the n-point functions themselves are divergent in the zero $\L$ limit because the boundary lapse vanishes in this limit. In particular for the 1-point function: $(1/\sqrt{q_{(0)}})\ i\delta S_{ren}^{on-shell}/\delta \varf_{(0)} = \a \( 1/(N_{(0)}\sqrt{g_{(0)}})\ i\delta S_{ren}^{on-shell}/\delta \varf_{(0)} \)$, which diverges as $\a \to \infty$, where in this case $N_{(0)} = 1$ and $g_{(0)ij} = \delta_{ij}$.
}
\bea\label{O_vev}
\sqrt{q_{(0)}}\, \< \op \> &=& {i \delta S_{ren}^{on-shell} \over \delta \varf_{(0)}}\ =\ \lim_{\e \to 0} \( \e^{\D-k}\, {i\delta S_{ren}^{on-shell} \over \delta \phi} \) \nn\\[5pt]
&=& {\ello^{d} \over \a^{2}}\, \Big( 2 \varf_{(2)} - \tilde{\varf}_{(2)} \Big) - \ello^{d}\, \vec{\na}^{2} \varf_{(0)}\ .
\eea
We therefore find that the vev is mapped to the normalisable mode $\varf_{(2)}$ for finite $\a$ as expected. The term proportional to $\tilde{\varf}_{(2)}$ is unphysical in the sense that it can be subtracted from the expectation value by adding to the renormalized action the finite covariant counterterm (finite both in $\e$ and $\a$):
\bea\label{scheme_ct}
-{\a\ello \over 4} \int\limits_{z=\e} d^{d+1}x\, \sqrt{q}\, \phi \square_{q}\phi\ .
\eea
The variation of this term is then proportional to $\tilde{\varf}_{(2)}$:
\bea
{i \delta \over \delta \varf_{(0)}} \( -{\a\ello \over 4} \int\limits_{z=\e} d^{d+1}x\, \sqrt{q}\, \phi \square_{q}\phi \) &=& \lim_{\e \to 0} \( -{\a\ello \over 4}\, \e^{\Delta - k}  {i\delta \over \delta \phi}\, \int\limits_{z=\e} d^{d+1}x\, \sqrt{q}\, \phi \square_{q}\phi \) \nn\\
&=& {\ello^{d} \over \a^{2}}\, \tilde{\varf}_{(2)}\ .
\eea
The term proportional to the spatial Laplacian of the source cannot be subtracted without partially breaking diffeomorphism invariance of the bulk action. The finite counterterm that subtracts this term is given by:
\bea
-{\a\ello \over 4} \int\limits_{z=\e} d^{d+1}x\, \sqrt{q}\, \phi \vec{\na}^{2}_{\g}\phi\ ,
\eea
where $\vec{\na}^{2}_{\g}$ is the Laplacian with respect to the spatial metric $\g_{ij}dx^{i}dx^{j} = {\ello^{2}/\e^{2}}\, d\vec{x}^{2}_{d}$ and, therefore, breaks invariance under diffeomorphisms that are not foliation preserving. This is the same type of anomalous counterterm that we found in section \ref{four_dims}. However, there is no need for a counterterm of this type in the present case. It may seem that the spatial Laplacian of the source in the vev \eqref{O_vev} will give rise to contact terms proportional to the spatial Laplacian of delta functions and, therefore, that partially break diffeomorphism invariance of the two-point correlator computed by taking the variation of the vev. However, this will not be the case because the variation of the normalisable mode $\varf_{(2)}$ will provide a contribution that precisely cancels these so that the two-point function is completely covariant for finite $\a$. We will see that this is indeed the case in section \ref{correlator}.

Finally, note that the vev admits a well-behaved zero $\L$ limit. If we switch off the source and take the limit $\a \to \infty$, the expectation value of the operator vanishes identically. In other words, any scalar operator of conformal dimension $\D = 1 + (d+1)/2$ evaluated on QFT states dual to gravity solutions with $\L = 0$ necessarily has a vanishing expectation value in the absence of the source.

\subsubsection{k=4}

In this case the renormalized action is given by:
\bea\label{scalar_S_ren_4}
i S_{ren} &=& \hf \int d^{d+2}x \sqrt{G} \( G^{\mu\nu}\partial_{\mu}\phi\partial_{\nu}\phi + \({m \over \a}\)^{2}\phi^{2} \) \nn\\[5pt]
&+& \hf \int\limits_{z=\e} d^{d+1}x\, \sqrt{q} \( -{\D - k \over \a\ello}\, \phi^{2} - {\a\ello \over k-2}\, \phi \square_{q}\phi +  {(\a\ello)^{3} \over 4}\, \phi (\square_{q})^{2}\phi\, \log \e \)\ ,
\eea
where the counterterm action again coincides with the canonical one. Let us now verify whether the counterterms spoil the zero $\L$ limit of the action. If we evaluate $S_{ren}$ on-shell, take the limit as the regulator $\e \to 0$ and look for those terms proportional to positive powers of $\a$, we find:
\bea
\lim_{\e \to 0}\ i S_{ren}^{on-shell} &=& -{\ello^{d} \over 4}\, \int\limits_{z=0} d\bar{u}\,d^{d}x\, \Big( \a^{2} \varf_{(0)}\,\ddot{\varf}_{(2)} + \a^{2} \varf_{(1)}\, \ddot{\varf}_{(1)} + \a^{2} \varf_{(2)}\, \ddot{\varf}_{(0)} \Big) + \op(\a^{0})\ .\qquad
\eea
The second and third terms are of order $\op(\a^{0})$. This is so because from equation \eqref{phi_n} for $n=1, 2$ we have:
\bea
\dot{\varf}_{(0)} &=& \op(\a^{-2})\ \Rightarrow\ \varf_{(2)}\,\ddot{\varf}_{(0)} = \op(\a^{-2})\ ,\\
\dot{\varf}_{(1)} &=& -i \( {4 \over \a^{2}}\, \varf_{(2)} - \vec{\na}^{2}\varf_{(0)} \)\ \Rightarrow\ \ddot{\varf}_{(1)}\, =\, -i \( {4 \over \a^{2}}\, \dot{\varf}_{(2)} + {i \over \a^{2}}\, \vec{\na}^{2}\varf_{(1)} \)\ \Rightarrow\ \varf_{(1)}\, \ddot{\varf}_{(1)} = \op(\a^{-2})\ . \nn\\
\eea
On the other hand, the first term is of order $\a^{2}$. If we use again equation \eqref{phi_n} but for $n=3$, we find:
\bea
&&\dot{\varf}_{(2)}\, =\, i \( {3 \over \a^{2}}\, \varf_{(3)} - \vec{\na}^{2} \varf_{(1)} \)\ \Rightarrow\ \ddot{\varf}_{(2)}\, =\,  i \( {3 \over \a^{2}}\, \dot{\varf}_{(3)} + i \( {4 \over \a^{2}}\, \vec{\na}^{2} \varf_{(2)} - \vec{\na}^{4} \varf_{(0)} \) \)\ \nn\\[5pt]
&&\Rightarrow\ \varf_{(0)}\, \ddot{\varf}_{(2)} = \vec{\na}^{4}\varf_{(0)} + \op(\a^{-2})\ .
\eea
In this way we find that the zero $\L$ limit of the action is spoiled by the counterterm action:
\bea
\lim_{\e \to 0}\ i S_{ren}^{on-shell} = -\a^{2}\, {\ello^{d} \over 4} \int\limits_{z=0} d\bar{u}\,d^{d}x\, \varf_{(0)} ( \vec{\na}^{2} )^{2} \varf_{(0)} + \op(\a^{0})\ .
\eea
This divergence in $\a$ is subtracted by the finite counterterm (finite in $\e$):
\bea\label{anomalous_ct_4}
{(\a\ello)^{3} \over 4} \int\limits_{z=\e} d\bar{u}d^{d}x\, \sqrt{q}\ \phi\, ( \vec{\na}_{\g}^{2} )^{2} \phi\ ,
\eea
where $\vec{\na}^{2}$ is the Laplacian with respect to the spatial metric $\g_{ij}dx^{i}dx^{j} = \ello^{2}/\e^{2} d\vec{x}_{d}^{2}$ as before. Unlike the case of $k=2$, this new counterterm is needed in order to restore the well-behaved-ness of the action in the zero $\L$ limit. This is done, however, at the expense of breaking invariance of the renormalized action under diffeomorphisms that are not foliation preserving. Since this counterterm is finite with respect to the regulator, it is associated to a choice of scheme on the QFT side. This means that a renormalization scheme that breaks invariance of the QFT under transformations that do not preserve the spacelike foliation of the boundary is a necessary requirement, so that the QFT states result in finite expectation values and correlators once the QFT limit associated to the zero $\L$ limit is taken. The final renormalized action is then given by:
\bea\label{scalar_S_ren_4_final}
i S_{ren} &=& \hf \int d^{d+2}x \sqrt{G} \( G^{\mu\nu}\partial_{\mu}\phi\partial_{\nu}\phi + \({m \over \a}\)^{2}\phi^{2} \) \nn\\[5pt]
&+& \hf \int\limits_{z=\e} d^{d+1}x\, \sqrt{q} \( -{\D - k \over \a\ello}\, \phi^{2} - {\a\ello \over k-2}\, \phi \square_{q}\phi +  {(\a\ello)^{3} \over 4}\, \phi\, (\square_{q})^{2}\phi\, \log \e \) \nn\\[5pt]
&+& \hf \int\limits_{z=\e} d^{d+1}x\, \sqrt{q}\ \( {(\a\ello)^{3} \over 2}\, \phi\, ( \vec{\na}_{\g}^{2} )^{2} \phi \)\ . \\ \nn
\eea

\paragraph{Vacuum expectation value\\}\n

\n The variation of the on-shell action is given by:
\bea
i\, \delta S_{ren}^{on-shell} = \int\limits_{z=\e} d^{d+1}x\, \sqrt{q} &\bigg(& {\e \over \a\ello}\( \phi ' - i\a^{2} \dot{\phi} \) - {\D - k \over \a\ello} \phi - {\a\ello \over k-2}\, \square_{q}\phi \nn\\[5pt]
&+& {(\a\ello)^{3} \over 2}\, (\vec{\na}_{\g}^{2})^{2}\phi + {(\a\ello)^{3} \over 4}\, (\square_{q})^{2}\phi\, \log\e \bigg) \delta\phi\ .
\eea
The vacuum expectation value of the dual QFT operator is then given by:
\bea\label{O_vev_4}
\sqrt{q_{(0)}}\, \< \op \> &=& {i \delta S_{ren}^{on-shell} \over \delta \varf_{(0)}}\ =\ \lim_{\e \to 0} \( \e^{\D-k}\, {i\delta S_{ren}^{on-shell} \over \delta \phi} \) \nn\\[5pt]
&=& {\ello^{d} \over \a^{2}}\, \Big( 4 \varf_{(4)} - {7 \over 3}\,\tilde{\varf}_{(4)} \Big) + {2\ello^{d} \over 3}\, \vec{\na}^{2} \varf_{(2)}\ .
\eea
For finite $\a$, the vev is again mapped to the normalisable mode $\varf_{(4)}$. The term proportional to $\tilde{\varf}_{(4)}$ can be subtracted by adding the finite covariant counterterm to the action (finite both in $\e$ and $\a$):\footnote{
As a technical point, the fact that the integrand is finite in $\a$ follows from the discussion at the end of section \ref{matter_solution}. From equation \eqref{phi_tilde_k} with $k=4$ it follows that: $\square_{q_{(0)}}^{2} \varf_{(0)} = -(16/\a^{4})\, \tilde{\varf}_{(4)}$. The coefficient $\tilde{\varf}_{(4)}$ is non-divergent in $\a$ by definition (recall that we resctricted the space of solutions in AdS to the subspace where the coefficients are well-behaved as $\a \to \infty$, \ie we focus only on those solutions in AdS that result in solutions in Minkowski space in this limit). This implies that $\square_{q_{(0)}}^{2} \varf_{(0)} = \op(\a^{-4})$.}
\bea\label{scheme_ct_4}
-{7 \over 96}\, (\a\ello)^{3} \int\limits_{z=\e} d^{d+1}x\, \sqrt{q}\, \phi (\square_{q})^{2} \phi\ .
\eea
The term proportional to the spatial Laplacian of $\varf_{(2)}$, however, remains. Note then that the expectation value admits a well-behaved zero $\L$ limit. For finite $\a$, the coefficient $\varf_{(2)}$ is a functional of $\varf_{(0)}$, so setting the source to zero and then taking the limit $\a \to \infty$ results in a vanishing vev for the operator. On the other hand, in the case $\a^{-1} = 0$ the coefficient $\varf_{(2)}$ is a non-local functional of $\varf_{(0)}$. From equation \eqref{phi_n} for $n=2,3$ with $\a^{-1}=0$ we find that $\varf_{(2)}$ is defined by the differential equation: $\ddot{\varf}_{(2)} = (\vec{\na}^{2})^{2}\varf_{(0)}$. In this way, setting first $\a^{-1}=0$ in the vev and then switching off the source results in a non-trivial expectation value for the operator: $\sqrt{q_{(0)}}\, \< \op \> \sim \vec{\na}^{2}\varf_{(2)}$, where $\ddot{\varf}_{(2)} = 0$. We expect this type of behaviour to be reproduced for generic values of $k \geq 4$.
 
\subsection{Two-point correlator}\label{correlator}

In this last section we will compute the 2-point function for the scalar operator with $k=2, 4$ and analyse its zero $\L$ limit. This is done by choosing a full solution of the equations of motion that is well-behaved in the bulk interior and then taking a first-order variation of the vevs \eqref{O_vev} and \eqref{O_vev_4} in the presence of the source. If we return to equation \eqref{sol_hat_phi} for the Fourier transform $\hat{\varf}$ of the scalar field and look at the behaviour of the Bessel functions as $z \to \infty$, we find that $\hat{\varf}$ diverges as $z \to \infty$ unless we set the coefficient $B(\omega,p) = 0$. In this way, the solution that is well-behaved in the interior is given by:
\bea\label{phi_K_sol}
\phi(z,\bar{u},\vec{x}) &=& \dfrac{2^{1-k/2}}{\Gamma(k/2)}\, z^{\Delta - k/2} \int d\omega d^{d}p\, e^{i\omega\bar{u}}\, e^{i\vec{p}\cdot \vec{x}}\, e^{-\a^{2}\omega z}\, \hat{\varf}_{(0)}(\omega,\vec{p}) \(\a|p|\)^{k/2}\, K_{k / 2}(\a z |p|) \ ,\hspace{3em}
%&=& z^{\Delta-k} \( \varf_{(0)} + z\, \varf_{(1)} + z^{2}\, \varf_{(2)} + z^{3}\, \varf_{(3)} + ... + z^{k}\, \varf_{(k)} + z^{k}\log z\, \tilde{\varf}_{(k)} + \op(z^{>k}) \)\ , \nn\\
\eea
where we used the expression \eqref{coeff_A} for the coefficient $A(\omega,p)$, and where $|p| := \sqrt{\vec{p}^{\, 2} + \a^{2}\omega^{2}}$. %We have also redefined the source as:
%\bea
%\hat{\varf}_{(0)}(\omega,\vec{p})\ \to\ c\,\hat{\varf}_{(0)}(\omega c,\vec{p})\ ,
%\eea
%with $c$ a constant that we will fix below. 
The solution can be rewritten as an integration in position space by defining:
%\bea
%\varf_{(0)}(\bar{v}/c,\vec{y}) &=& \int d\omega' d^{d}p\, e^{i\omega' \bar{v}/c}\, e^{i \vec{p}\cdot\vec{y}}\, \hat{\varf}_{(0)}(\omega ',\vec{p})\, =\, c\int d\omega d^{d}p\, e^{i\omega \bar{v}}\, e^{i \vec{p}\cdot\vec{y}}\, \hat{\varf}_{(0)}(\omega c,\vec{p})\ , \qquad
%\eea
\bea
\varf_{(0)}(\bar{v},\vec{y}) &=& \int d\omega d^{d}p\, e^{i\omega \bar{v}}\, e^{i \vec{p}\cdot\vec{y}}\, \hat{\varf}_{(0)}(\omega,\vec{p})\ , \qquad
\eea
and using the identity:
\bea
\int d^{d+1}X\, {e^{-i p\cdot X} \over \(\epsilon^{2} + |X|^{2} \)^{\Delta}} &=& a(k)\, \e^{-k/2}\, |p|^{k/2} K_{k/2}(\e |p|)\ ,
\eea
where $k=2\Delta - (d+1)$, $|X|^{2} = X_{0}^{2} + X^{i}X^{i}$, $|p| = \sqrt{\vec{p}^{\,2} + \a^{2}\omega^{2}}$, and $a(k)$ is a proportionality constant that depends only on $k$. The solution \eqref{phi_K_sol} can then be rewritten as:
\bea
\phi(z,\bar{u},\vec{x}) &=& {2^{1-k/2} \over \Gamma(k/2)}\, {\a^{k-\Delta-1} \over a(k)} \int d\bar{v}d^{d}y\, \varf_{(0)}(\bar{v},\vec{y})\, {(\a z)^{\Delta} \over \( \(\a z\)^{2} + \( {\bar{u}-\bar{v} \over \a} + i \a z \)^{2} + |\vec{x} - \vec{y}|^{2} \)^{\Delta} }\ . \nn\\[5pt]
\eea
This is precisely the expression one would obtain by solving the scalar field equation in Euclidean AdS$_{d+2}$ in Poincar\'e coordinates (see \eg \cite{Witten:1998qj}), requiring that the solution be well-behaved in the bulk interior and finally transforming the scalar field to the coordinate system \eqref{AdS}. From this representation we can immediately read the bulk-to-boundary propagator and obtain the expression for the unrenormalized two-point function. If we use the identity \cite{Witten:1998qj}:
\bea
\lim_{z \to 0}\, {(\a z)^{\Delta} \over \( \(\a z\)^{2} + \( {\bar{u}-\bar{v} \over \a} + i \a z \)^{2} + |\vec{x} - \vec{y}|^{2} \)^{\Delta} }\ \sim\ \a\, b(k)\, (\a z)^{\Delta-k} \delta(\bar{u}-\bar{v}) \delta^{d}(\vec{x}-\vec{y})\ ,
\eea
with $b(k)$ a constant that depends only on $k$, then the on-shell bare action \eqref{scalar_action} is given by:
\bea
S^{on-shell} &=& \hf \int\limits_{z=\e} d^{d+1}x\, \sqrt{G}\, \phi\, G^{z\mu}\partial_{\mu}\phi \nn\\[5pt]
&=& {\a^{k-3}\ello^{d} \over \tilde{b}(k)} \int\limits_{z=\e} d\bar{u}d^{d}x \int d\bar{v}d^{d}y\ {\varf_{(0)}(\bar{u},\vec{x})\, \varf_{(0)}(\bar{v},\vec{y}) \over \( \({\bar{u}-\bar{v} \over \a } \)^{2} + | \vec{x} - \vec{y} |^{2} \)^{\Delta} } \( 1 + \op(z) \)\ ,
\eea
with $\tilde{b}(k)$ a dimensionless constant. Taking the variations of the on-shell action with respect to the source and absorbing the overall proportionality constant in the normalisation of the operator results in the unrenormalized two-point correlator:
\bea\label{correlator_unren}
\sqrt{q_{(0)}}^{\,2} \< \op(\bar{v},\vec{y}) \op(\bar{u},\vec{x}) \> &=& {i\delta^{2}S^{on-shell} \over \delta\varf_{(0)}(\bar{v},\vec{y})\delta\varf_{(0)}(\bar{u},\vec{x}) } \nn\\[5pt]
&=& {1 \over \( \({\bar{u}-\bar{v} \over \a } \)^{2} + | \vec{x} - \vec{y} |^{2} \)^{\Delta} }\ .
\eea
In the zero $\L$ limit and away from coincident points, this results in the correct expression for the two-point function of a scalar operator of weight $\D$ but in $d$ dimensions.\\

In order to compute the renormalized correlator, we return to our original representation \eqref{phi_K_sol} for the physical solution and use the expansion \eqref{Bessel_K} around $z=0$ for the Bessel function with $k=2,4$ to find:
\bea
\phi(z,\bar{u},\vec{x}) &=& z^{\D-k} \( \varf_{(0)} + ... + z^{k}\varf_{(k)} + z^{k}\log(\mu z)\, \tilde{\varf}_{(k)} + ... \)\ ,
\eea
where the normalisable mode $\varf_{(k)}$ for $k=2,4$ is given in terms of the source by:
\bea
\begin{cases}
\varf_{(k=2)} = \dfrac{\a^{2}}{4} \( \vec{\na}^{2} - \a^{2}\partial_{\bar{u}}^{2} - \(2\g_{E} - 2\log 2 + \log \Big( - \dfrac{\a^{2}}{\mu^{2}}\, \square_{q_{(0)}}\Big) \) \square_{q_{(0)}} \) \varf_{(0)}\ , \label{normalisable_k-2}\\[10pt]
\varf_{(k=4)} = \dfrac{\a^{4}}{24} \( -3\a^{2}\partial_{\bar{u}}^{2}\vec{\na}^{2} - 2\a^{4}\partial_{\bar{u}}^{4} - {3 \over 4} \( 2\g_{E} - {3 \over 2} - 2\log 2 + \log \Big(-\dfrac{\a^{2}}{\mu^{2}}\, \square_{q_{(0)}} \Big) \) \square_{q_{(0)}}^{2} \) \varf_{(0)}\ , \label{normalisable_k-4}
\end{cases}
\eea
with $\g_{E}$ the Euler constant and $\square_{q_{(0)}} = \a^{2}\partial_{\bar{u}} + \vec{\na}^{2}$ the Laplacian with respect to the QFT metric. At the end of section \ref{matter_solution} we found that the requirement that the coefficients $\varf_{(n<k)}$ and $\tilde{\varf}_{(k)}$ in the asymptotics be well-defined in the limit $\a \to \infty$ results in constraints on the behaviour in $\a$ of the time derivatives of the source. Since the normalisable mode for each $k$ is also well-defined in the limit $\a \to \infty$ by definition, and from the above expressions \eqref{normalisable_k-2} for the physical solution we have that $\varf_{(k)}$ is now a functional of the source, we find that the requirement that the solution be well-behaved in the interior results in a further constraint on the source for each value of $k$. The constraint will be on the behaviour in $\a$ of the spatial derivatives. From equations \eqref{phi_tilde_k} and \eqref{phi_n} for $k=2,4$ we have in particular that:
\bea
\square_{q_{(0)}}^{k/2}\, \varf_{(0)} &=& - {k^{k/2} \over \a^{k}}\, \tilde{\varf}_{(k)}\, =\, \op(\a^{-k})\ .
\eea
It then follows from equation \eqref{normalisable_k-2} that the non-normalisable mode of the physical solution for $k=2,4$ needs to satisfy:
\bea
\vec{\na}^{k} \varf_{(0)} &=& \op(\a^{-2})\ .
\eea
For $k=2$ this implies that the vev \eqref{O_vev} evaluated on such a solution is identically zero in the zero $\L$ limit. For $k=4$ it implies that the bulk action \eqref{scalar_S_ren_4} evaluated on such a solution is well-defined in the zero $\L$ limit, as well as the vev for the dual QFT operator, without the need for the anomalous counterterm. Nonetheless, the renormalization should hold for any solution of the bulk equations of motion, so in general the anomalous conterterm is needed to restore the well-behaved-ness of the zero $\L$ limit of the bulk action.\\

\paragraph{Case k=2\\}\n 

\n If we take the variation of the one-point function \eqref{O_vev} (with $\tilde{\varf}_{(2)}$ subtracted) with respect to the source $\varf_{(0)}$ and use the expression \eqref{normalisable_k-2} for the coefficient $\varf_{(2)}$, we obtain:
\bea
\sqrt{q_{(0)}}^{\,2} \< \op(\bar{v},\vec{y}) \op(\bar{u},\vec{x}) \> &=& {\delta \over \delta \varf_{(0)}(\bar{v},\vec{y})} \( \sqrt{q_{(0)}}\, \< \op(\bar{u},\vec{x}) \> \) \nn\\[5pt]
&=& - {\ello^{d} \over 2}\, \Big( 1+2\g_{E}-2\log2+2\log\a \Big)\, \square_{q_{(0)}} \delta(\bar{u}-\bar{v}) \delta^{d}(\vec{x} - \vec{y}) \nn\\[5pt] 
&& - {\ello^{d} \over 2}\, \log\(- \mu^{-2}\, \square_{q_{(0)}}\) \square_{q_{(0)}}\, \delta(\bar{u}-\bar{v}) \delta^{d}(\vec{x} - \vec{y})\ .\nn\\
\eea
The first term proportional to the Laplacian on the delta functions is scheme dependent and it can be removed by adding a finite and {\it local} counterterm to the action proportional to \eqref{scheme_ct}. The scheme-independent piece is then:
\bea\label{correlator_1}
\sqrt{q_{(0)}}^{\,2} \< \op(\bar{v},\vec{y}) \op(\bar{u},\vec{x}) \> &=& - {\ello^{d} \over 2}\, \log\(- \mu^{-2}\, \square_{q_{(0)}}\) \square_{q_{(0)}}\, \delta(\bar{u}-\bar{v}) \delta^{d}(\vec{x} - \vec{y})\ .
\eea
If we use the identity \cite{Freedman:1991tk}:
\bea
\int d^{d+1}X {e^{i p\cdot X} \over |X|^{d-1}}\, \log \big(\tilde{\mu}^{2}|X|^{2}\big) &=& -{c \over |p|^{2}}\, \log \big( \mu^{-2} |p|^{2} \big)\ ,
\eea
with $\tilde{\mu} = \g_{E}\, \mu/2$ and $c$ a proportionality constant that depends only on $d$, and Fourier transform it, we find:
\bea\label{identity_1}
\square^{n+1} {\log \big( \tilde{\mu}^{2}|X|^{2} \big) \over |X|^{d-1}} &=& c\, \log \big( -\mu^{-2} \square \big) \square^{n}\, \delta^{d+1}(X)\ .
\eea
If we apply this identity to the right-hand side of \eqref{correlator_1} we obtain:
\bea
\sqrt{q_{(0)}}^{\,2} \< \op(\bar{v},\vec{y}) \op(\bar{u},\vec{x}) \> &=& -{\ello^{d} \over 2\a c}\, \square_{q_{(0)}}^{2} {\log \( \tilde{\mu}^{2} \Big[ \( { \bar{u} - \bar{v} \over \a } \)^{2} + |\vec{x} - \vec{y}|^{2} \Big] \) \over \Big| \( { \bar{u} - \bar{v} \over \a } \)^{2} + | \vec{x} - \vec{y} |^{2}\, \Big|^{(d-1)/2} } \nn\\[5pt]
&=& \tilde{c}\, \mathcal{R} {1 \over \big| \( { \bar{u} - \bar{v} \over \a } \)^{2} + | \vec{x} - \vec{y} |^{2}\, \big|^{\D}}\ ,
\eea
where $\D = 1+(d+1)/2$. The proportionality constant $\tilde{c}$ can be absorbed in a normalisation of $\op$. The term $\mathcal{R} \(1/|X|^{2\D}\)$ on the right-hand side is the renormalized version of the correlator $1/|X|^{2\D}$ and it coincides with the latter away from coincident points \cite{Osborn:1993cr}. In the zero $\L$ limit we find:
\bea
\lim_{\a \to \infty}\, \sqrt{q_{(0)}}^{\,2} \< \op(\bar{v},\vec{y}) \op(\bar{u},\vec{x}) \> &=& \mathcal{R} {1 \over | \vec{x} - \vec{y} |^{2\D}}\ ,
\eea
which is the renormalized version of the correlator that we found in \eqref{correlator_unren} in the zero $\L$ limit.\\

\paragraph{Case k=4\\}\n 

\n In this case the one-point function for the QFT operator receives a contribution from the anomalous counterterm \eqref{anomalous_ct_4}. This term renders the vacuum expectation value finite in the zero $\L$ limit, but it introduces contact terms in the two point function. In order to verify this more explicitly, we isolate the contribution from this term in the vev:
\bea\label{O_vev_4}
\sqrt{q_{(0)}}\, \< \op \> &=& \( {4\ello^{d} \over \a^{2}}\, \varf_{(4)} + {2\ello^{d} \over 3}\, \vec{\na}^{2} \varf_{(2)} - {\a^{2}\ello^{d} \over 2}\, \vec{\na}^{4} \varf_{(0)}\) + {\a^{2}\ello^{d} \over 2}\, \vec{\na}^{4} \varf_{(0)}\ , \qquad
\eea
where the last term represents the contribution from the anomalous counterterm. We have also subtracted the term proportional to $\tilde{\varf}_{(4)}$ which is scheme dependent. If we use the expression \eqref{normalisable_k-4} for the normalisable mode $\varf_{(4)}$ and take the variation of the one-point function with respect to the source, we obtain:
\bea
\sqrt{q_{(0)}}^{\,2} \< \op(\bar{v},\vec{y}) \op(\bar{u},\vec{x}) \> &=& {\delta \over \delta \varf_{(0)}(\bar{v},\vec{y})} \( \sqrt{q_{(0)}}\, \< \op(\bar{u},\vec{x}) \> \) \nn\\[5pt]
&=& -{\a^{2}\ello^{d} \over 6} \( 2 + {3 \over 4}\( 2\g_{E} - 2\log2 - {3 \over 2} + 2\log\a \) \) \square_{q_{(0)}}^{2}\delta(\bar{u}-\bar{v})\delta^{d}(\vec{x}-\vec{y}) \nn\\[5pt]
&& - {\a^{2}\ello^{d} \over 8}\, \log\(- \mu^{-2}\, \square_{q_{(0)}}\) \square_{q_{(0)}}^{2}\, \delta(\bar{u}-\bar{v}) \delta^{d}(\vec{x} - \vec{y}) \nn\\[5pt]
&& + {\a^{2}\ello^{d} \over 2}\, \delta(\bar{u}-\bar{v})\, \vec{\na}^{4}\delta^{d}(\vec{x}-\vec{y})\ .
\eea
The first term proportional to the square of the Laplacian can be removed by adding a finite and local counterterm to the action proportional to \eqref{scheme_ct_4}. The last term arising from the anomalous counterterm is a contact term that diverges when the operators are defined at equal time $\bar{u}=\bar{v}$. This piece cannot be removed from the correlator by a counterterm without spoiling the zero $\L$ limit of the bulk action. This type of contact terms spoils the behaviour of the correlator at coincident points in time and will always appear in the two-point functions for values of $k \geq 4$ if we simultaneously require that the bulk action be well-defined in the zero $\L$ limit. At non-coincident points, if we subtract the scheme-dependent term and use the identity \eqref{identity_1}, we find:
\bea
\sqrt{q_{(0)}}^{\,2} \< \op(\bar{v},\vec{y}) \op(\bar{u},\vec{x}) \> &=& -{\a \ello^{d} \over 8 c}\, \square_{q_{(0)}}^{3}\, {\log \( \tilde{\mu}^{2} \Big[ \( { \bar{u} - \bar{v} \over \a } \)^{2} + |\vec{x} - \vec{y}|^{2} \Big] \) \over \Big| \( { \bar{u} - \bar{v} \over \a } \)^{2} + | \vec{x} - \vec{y} |^{2}\, \Big|^{(d-1)/2} } \nn\\[5pt]
&=& \tilde{c}\, \mathcal{R} {1 \over \big| \( { \bar{u} - \bar{v} \over \a } \)^{2} + | \vec{x} - \vec{y} |^{2}\, \big|^{\D}}\ \qquad (\bar{u} \neq \bar{v})\ .
\eea
where $\Delta = 2+(d+1)/2$. If we absorb the constant $\tilde{c}$ in the normalisation of the operator and take the limit $\a \to \infty$, we again find the renormalized version of the correlator that we obtained in \eqref{correlator_unren} in this limit.\\

\section{Conclusions}

\qquad In this article we discussed the zero $\L$ limit of vacuum expectation values and correlation functions in AdS/CFT at a formal level, the associated issues and attempted to address them. We found that the analysis requires a suitable foliation of the spacetime and we derived the mapping between bulk and boundary data in the associated coordinate system. We focused specifically on the case of the bulk spacetime metric and a non-backreacting scalar field, determined their unique asymptotics, computed the one-point function of the dual operators and discussed the necessary conditions for the correspondence between the near-boundary asymptotics and the vevs to admit a well-behaved zero $\L$ limit. We found that the existence of the limit essentially translates into a statement about the sources and states of the boundary theory. We discussed the holographic Ward identities in three and four bulk dimensions in the case of pure gravity, and reproduced the central charge that arises in the central extension of the asymptotic symmetry algebra of three-dimensional flat space via the zero $\L$ limit of the holographic Weyl anomaly. We also found that the energy and momentum of the QFT states dual to three-dimensional flat cosmological spaces and to the Kerr spacetime match with those of the bulk solutions. In the context of holographic renormalization, we analysed the behaviour of the holographic counterterms in the zero $\L$ limit and showed that the well-behaved-ness of the gravity action in this limit can only be preserved by means of anomalous counterterms. Based on the AdS/CFT dictionary, we then argued that the renormalization of QFTs with states dual to asymptotically flat solutions generically requires renormalization schemes that break invariance of the QFT under transformations that do not preserve the spacelike foliation at the boundary. Finally, for the case of the non-backreacting bulk scalar, we computed holographically the renormalized two-point function of the dual operator in the zero $\L$ limit and found it to be consistent with that of a conformal operator in two dimensions less. In this case, however, we found that the anomalous counterterms introduce contact terms in the correlator that spoil the behaviour of the latter at coincident points.\\

In the context of the metric asymptotics at null infinity, we emphasized the differences between the asymptotics obtained in the zero $\L$ limit and the standard definitions of asymptotic flatness in the literature in the case of radiating spacetimes in odd dimensions. It would be interesting to investigate more precisely to which extent perturbations of the asymptotically flat metric do not preserve the asymptotics in odd dimensions when the spacetime contains gravitational radiation. We also did not address the question of how to compute flat space S-matrix elements in general from the zero $\L$ limit of boundary correlators. An interesting direction would be to verify whether correlation functions obtained by taking variations of the bulk action with respect to those boundary configurations at past and future temporal infinity (the `corners' discussed in section \ref{corner_section}) can be used in a holographic computation of the bulk S-matrix elements.\\

\section*{Acknowledgements}

I would like to thank Marika Taylor and Kostas Skenderis and the University of Southampton, School of Mathematical Sciences, for hospitality during the course of this work.

\appendix
%\phantomsection
\section*{Appendices}
\addcontentsline{toc}{section}{Appendices:}

\vspace{4pt}

\section{Conformal compactness}\label{appA}

A manifold $(\mM,G)$ is defined to be $C^{n\geq 0}$ conformally compact if there exists an {\it asymptote} $(\tilde{\mM},\tilde{G},\r)$ consisting of a manifold-with-boundary $(\tilde{\mM},\tilde{G})$ with boundary $\del \tilde{\mM}$ and a defining function $\r(x) : \tilde{\mM} \to \Real^{+}$ satisfying the following properties \cite{penrose,geroch,Anderson:2004yi}:

\begin{itemize}
\item[1)] $\mM= \text{int } \tilde{\mM} = \{p \in \tilde{\mM} : \exists\ \text{open set}\ p \ni U \subset \tilde{\mM} \} \ , $
\item[2)] $\tilde{G}_{\mu\nu} = \r^{2}(x)\, G_{\mu\nu}\ $ : $\ \tilde{\mM} = \{\r \geq 0\}$ , $\del \tilde{\mM} = \{\r = 0\}\ ,$
\item[3)] $d\r \neq 0$ on $\del \tilde{\mM}$ ,
\end{itemize}

\n with $\r(x)$ of class $C^{\infty}$ and $\tilde{G}$ non-degenerate and of class $C^{n \geq 0}$ in $\tilde{\mM}$. The region $\{\r=0\}$ of $\tilde{\mM}$ is referred to as the conformal boundary of $\mM$ and $\tilde{\mM}$ as the conformal embedding.\\

\section{Gaussian null coordinates}\label{appAB}

In this section we will derive our coordinate system by performing a brief ADM analysis of the spacetime metric $G_{\mu\nu}$. For a thorough treatment see the original works in \cite{Moncrief:1983,Friedrich:1998wq,Reall:2002bh}. We introduce coordinates $x^{\mu} = (u,x^{A}) = (u,r,x^{i}) = (r,x^{a})$ and define the surfaces of constant $u$ to be null. We then do an ADM decomposition of $G_{\mu\nu}$ with respect to these surfaces as:
\bea
ds^{2} &=& - \a^{2} du^{2} + h_{AB} \( dx^{A} + \a^{A}du \) \( dx^{B} + \a^{B}du \)\ .
\eea
We also decompose the induced metric $h_{AB}$ with respect to the surfaces of constant $r$ as:
\bea
h_{AB}dx^{A}dx^{B} &=& \b^{2} dr^{2} + \g_{ij}\( dx^{i} + \b^{i}dr \) \( dx^{j} + \b^{j}dr \)\ ,
\eea
and define the spatial metric $\g_{ij}$ to be positive-definite. Since the surfaces of constant $u$ are null by definition, the induced metric $h_{AB}$ must be degenerate. Since the determinant $\sqrt{h} = \b \sqrt{\g}$ and $\g_{ij} > 0$, the degeneracy of $h_{AB}$ implies that $\b = 0$ everywhere. With this condition, we rewrite $G_{\mu\nu}$ without loss of generality as:
\bea
ds^{2} &=& - \phi\, du^{2} + 2 M du dr + \g_{ij} \( dx^{i} + \s^{i}du + \b^{i}dr \) \( dx^{j} + \s^{j}du + \b^{j}dr \) \label{G_mn} \\[5pt]
&=& N^{2}dr^{2} + q_{ab} \Big( dx^{a} + N^{a}dr \Big) \( dx^{b} + N^{b}dr \)\ , \label{radial_foliation}
\eea
where $(N,N^{a})$ are the lapse and shift of the radial foliation in $r$ and where the induced metric $q_{ab}$ is given by:
\be\label{q_ab}
q_{ab}dx^{a}dx^{b} = -\phi\, du^{2} + \g_{ij} \( dx^{i} + \s^{i}du \) \( dx^{j} + \s^{j}du \)\ .
\ee
Let us then perform an ADM decomposition of the Einstein-Hilbert Lagrangian with respect to the radial foliation \eqref{radial_foliation}:
\be\label{lagrangian}
\mL = \sqrt{G}\, R[G] = N\sqrt{q}\, \Big( R[q] + Q^{2} - Q \cdot Q - 2\na_{\mu}v^{\mu} \Big)\ ,
\ee
where $Q_{ab} = 1/(2N) \( \partial_{r} - \Lie_{N} \) q_{ab}$ is the extrinsic curvature of the surfaces of constant $r$ and: $v^{\mu} = Q n^{\mu} - a^{\mu}$, with $n^{\mu}$ and $a^{\mu}$ the unit normal and acceleration of these surfaces, respectively. The last term in the Lagrangian is a total derivative and thus will be discarded. The decomposed Lagrangian is now a functional of the lapse, shift and induced metric $N$, $N^{a}$ and $q_{ab}$. A quick inspection of $\mL$ then reveals that only $q_{ab}$ contains radial derivatives and therefore the equations of motion for the metric will be second order differential equations in $r$ for $q_{ab}$ only. This indicates as usual that $N$ and $N^{a}$ do not represent true degrees of freedom and therefore can be gauge-fixed, \ie can be brought to any configuration by diffeomorphisms near a surface of constant $r$. If we then return to \eqref{q_ab} we find that $q_{ab}$ depends only on $\phi, \s^{i}$ and $\g_{ij}$. This means that the Lagrangian does not contain radial derivatives of the functions $M$ and $\b^{i}$ that appear in \eqref{G_mn} and therefore these can be gauge-fixed by diffeomorphisms. The simplest gauge we can choose is the Gaussian gauge $\(M = 1, \b^{i} = 0\)$ in which the spacetime metric assumes the final form:
\be\label{g_null-old}
ds^{2} = - \phi\, du^{2} + 2du dr + \g_{ij} \( dx^{i} + \s^{i}du\) \( dx^{j} + \s^{j}du \)\ ,
\ee
with determinant $\sqrt{G} = \sqrt{\g}$. In the particular case of black hole spacetimes in gaussian null coordinates, the horizon is defined to consist of the surface $r=0$. Then note that it is still possible to use a further diffeomorphism of the form $x^{i} \to x^{i} + f^{i}(x,u)$ in \eqref{g_null-old} and choose the set of functions $f^{i}$ such that:
\bea
\s^{i} &\to& r^{\a}\tilde{\s}^{i}(r,u,x)\quad :\quad \a > 0\ ,\ \tilde{\s}^{i} = \op(r^{\geq 0})\ .
\eea
Also, since the horizon is a null surface, we find that the function $\phi$ must behave near $r=0$ at least as: 
\bea
\phi &=& r^{\b}\varphi(r,u,x)\quad :\quad \b > 0\ ,\ \varphi = \op(r^{\geq 0})\ . \label{phi_r}
\eea
In most cases the equations of motion near the horizon then fix the exponents $\a, \b = 1$ for a non-degenerate horizon, and $\a=1, \b = 2$ for a degenerate one.

\section{Ricci tensor}\label{appB}

In this section we provide the decomposition of the Ricci tensor of our gauge-fixed metric:
\bea
ds^{2}_{d+2} &=& G_{\mu\nu}dx^{\mu}dx^{\nu} \nn\\
&=& {\ello^{2} \over z^{2}}\, \Big( -\varf N_{(0)} du^{2} - 2N_{(0)}dudz + g_{ij} \( dx^{i} + \s^{i}du \) \( dx^{j} + \s^{j}du \) \Big)
\eea
where $N_{(0)} = N_{(0)}(u,x^{i})$ and the remaining components of the metric depend on all coordinates. The inverse and determinant of the metric are given by:
\bea
G^{\mu\nu} &=& \({z \over \ello}\)^{2}N_{(0)}^{-1}
\begin{pmatrix} 
0 & -1 & 0\\
-1 & \varf & \s^{i}\\
0 & \s^{i} & N_{(0)}g^{ij}
\end{pmatrix}\ ,\\[5pt]
\sqrt{G} &=& (\ello/z)^{d+2}N_{(0)} \sqrt{g}\ .
\eea
Define:
\bea
k_{ij} &:=& {1 \over 2N_{(0)}} \(\partial_{u} - \Lie_{\s} \) g_{ij}\ .
\eea
The decomposition of the Ricci tensor $R_{\mu\nu}[G]$ is then given by \cite{Moncrief:1983,Tanabe:2011es}:
\begin{align}
&2 R_{zi}[G]\, =\, {1 \over N_{(0)}} \( -  (g\cdot\s')_{i}' + {d \over z}\, \Big( (g\cdot\s')_{i} - \partial_{i}N_{(0)} \Big) - \hf\, \Tr[g^{-1}g']\, \Big( (g\cdot\s')_{i} - \partial_{i}N_{(0)} \Big) \) \nn\\[5pt]
&\hspace{4em} + \na_{j} ( g^{-1}g' )^{j}_{~i} - \partial_{i}\Tr[g^{-1}g']\ , \label{Rzi}\\[15pt]
&2 \Big( R_{zu}[G] - \s^{i}R_{zi}[G] \Big)\, =\, \varf '' - {d+2 \over z}\, \varf ' + {2(d+1) \over z^{2}}\, \varf + \Tr[g^{-1}g'] \( \hf\, \varf ' - {1 \over z}\, \varf \) \nn\\[5pt]
&\hspace{4em} - \na_{i} \Big( \s'^{\,i} - g^{ij}\partial_{j}N_{(0)} \Big) - {1 \over N_{(0)}}\, \s'^{\,i}\, \Big( (g\cdot\s')_{i} - \partial_{i}N_{(0)}\Big)\nn\\[5pt] 
&\hspace{4em} - N_{(0)} \( 2\, \Tr[g^{-1}k]' - {2 \over z}\, \Tr[g^{-1}k] + (k \cdot g') \)\ , \label{Rzu}
\end{align}
\begin{align}
&2 R_{ij}[G]\, =\, 2 R_{ij}[g] + {1 \over N_{(0)}}\, \Bigg[ - (\varf\, g_{ij}')' + {d \over z}\, \varf\, g_{ij}' + {2 \over z}\, \varf ' g_{ij} - {2(d+1) \over z^{2}}\, \varf\, g_{ij} \nn\\[5pt] 
&\hspace{4em} + \varf \( {1 \over z}\, g_{ij} - \hf\, g_{ij}' \) \Tr[g^{-1}g'] + \varf\, (g' \cdot g')_{ij} \nn\\[5pt] 
&\hspace{4em} + 2 \na_{(i} \Big( (g\cdot\s')_{j)} - \partial_{j)}N_{(0)} \Big) - N_{(0)}^{-1} (g\cdot\s')_{i}\,(g\cdot\s')_{j} \Bigg] + \partial_{i}\log N_{(0)}\partial_{j}\log N_{(0)} \nn\\[5pt] 
&\hspace{4em} + 4\, k_{ij}' - {2d \over z}\, k_{ij} + \Tr[g^{-1}g'] k_{ij} + \( g_{ij}' - {2 \over z}\, g_{ij} \) \Tr[g^{-1}k] - 4 (k\cdot g')_{(ij)}\ , \label{Rij}\\[15pt]
&2 R_{zz}[G]\, =\, - \Tr[g^{-1}g''] + \hf\, (g'\cdot g')\ , \label{Rzz}\\[15pt]
&2 \Big( R_{ui}[G] - \s^{j}R_{ij}[G] - \varf\, R_{zi}[G] \Big) = \nn\\[5pt]
&\hspace{4em} \(\partial_{u} - \Lie_{\s}\)\, \Big[ {1 \over N_{(0)}}\, \Big( (g\cdot\s')_{i} - \partial_{i}N_{(0)} \Big) \Big] + \Tr[g^{-1}k]\, \Big( (g\cdot \s')_{i} - \partial_{i}N_{(0)} \Big) \nn\\[5pt] 
&\hspace{4em} + 2\, (g^{-1}k)^{j}_{~i}\,\partial_{j}N_{(0)} - (g^{-1}g')^{j}_{~i}\, \partial_{j}\varf + \partial_{i}\varf ' + N_{(0)} \( -{d \over z} + \hf\, \Tr[g^{-1}g'] \) \partial_{i} \( \varf / N_{(0)} \) \nn\\[5pt]
&\hspace{4em} + 2N_{(0)} \( \na_{j}(g^{-1}k)^{j}_{~i} - \partial_{i}\Tr[g^{-1}k] \) - \varf \( \na_{j} (g^{-1}g')^{j}_{~i} - \partial_{i} \Tr[g^{-1}g'] \)  \ , \label{Rui}\\[15pt]
&{2 \over N_{(0)}}\, \Big[ R_{uu}[G] - 2\, \s^{i}R_{ui}[G] + \s^{i}\s^{j}R_{ij}[G] - \varf \( R_{zu}[G] - \s^{i}R_{zi}[G]\) \Big] = \nn\\[5pt]
&\hspace{4em} \( -{d \over z} + \hf\, \Tr[g^{-1}g'] \) \( \partial_{u} - \Lie_{\s} \) (\varf / N_{(0)}) - 2 \( \partial_{u} - \Lie_{\s} \) \Tr[g^{-1}k] \nn\\[5pt]
&\hspace{4em} + \varf\, \Big( 2\, \Tr[g^{-1}k]' + (k \cdot g') \Big) - \varf '\, \Tr[g^{-1}k] - 2N_{(0)} \big( k \cdot k \big) + \na^{i}\na_ {i}\varf \nn\\[5pt]
&\hspace{4em} + g^{ij}\, \partial_{i}\varf\, \partial_{j}\log N_{(0)} + {1 \over N_{(0)}} \( \varf \na_{i} \s'^{\, i} - \s'^{\, i}\partial_{i}\varf \) \ , \label{Ruu}
\end{align}
where the prime denotes differentiation with respect to $z$, the trace and inner product are taken with respect to $g_{ij}$, and where $\na_{i}g_{jk}:=0$. When replaced by the Einstein equations: 
\bea\label{Einstein_eqs}
R_{\mu\nu}[G] &=& -{d+1 \over \a^{2} \ello^{2}}\, G_{\mu\nu}\ , \label{E-eq}
\eea 
we find that equations \eqref{Rzi}--\eqref{Rij} represent the dynamical equations for the metric components $\s^{i}, \varf$ and $g_{ij}$, respectively, whereas equations \eqref{Rui} and \eqref{Ruu} are constraint equations since they do not contain second order derivatives in $z$. After \eqref{Rij} is solved, equation \eqref{Rzz} can also be seen as a constraint equation because it can be replaced by an equation without second order derivatives in $z$ if we use the trace of \eqref{Rij}.

\section{Terms $X_{ij}$ and $X_{i}$}\label{Xij}

The algebraic expressions for the terms $X_{ij}$ and $X_{i}$ that appear in equations \eqref{sij_4} and \eqref{ji_4} depend on the coefficient $g_{(1)ij}$ and vanish if the boundary metric is static. In general, the expressions are given by:
\bea
X_{ij} &=& {1 \over 4\a^{2}}\, g_{(1)ij} \( \Tr^{2}[g_{(0)}^{-1}g_{(1)}] + \( g_{(1)}\cdot g_{(1)} \) \) - {3 \over 4}\, \Tr[g_{(0)}^{-1}g_{(1)}]\, k_{(1)ij} - {5 \over 4}\, \Tr[g_{(0)}^{-1}k_{(1)}]\, g_{(1)ij} \nn\\[5pt] 
&+& \hf\, R_{(0)}\, g_{(1)ij} - {3 \over 2}\, g_{(1)(i}^{~~~k} \cov_{j)} \partial_{k} \log N_{(0)} + \qt \( \cov_{i}\partial_{j} \Tr[g_{(0)}^{-1}g_{(1)}] -\, ^{(0)}\square g_{(1)ij} \) \nn\\[5pt]
&+& {1 \over 4N_{(0)}} \( \Tr[g_{(0)}^{-1}g_{(1)}]\, \cov_{i}\partial_{j} N_{(0)} + g_{(1)ij}\, ^{(0)}\square N_{(0)} - g_{(0)ij}\, \Tr[g_{(0)}^{-1}g_{(1)}]\, ^{(0)}\square N_{(0)} \)\ ,\qquad \\[10pt]
X_{i} &=& {3 \over 8}\, \big(g_{(1)}\cdot g_{(1)}\big)_{i}^{~j}\partial_{j}\log N_{(0)} + \hf\, g_{(1)ij} \cov_{k} g_{(1)}^{kj} + \hf\, \cov_{k}  \big( g_{(1)}\cdot g_{(1)} \big)^{k}_{~i} - {3 \over 4}\, g_{(1)i}^{~j}\partial_{j} \Tr[g_{(0)}^{-1}g_{(1)}] \nn\\[5pt] 
&+& {1 \over 16}\, \partial_{i} \Tr^{2}[ g_{(0)}^{-1}g_{(1)}] - {5 \over 16}\, \partial_{i} \( g_{(1)} \cdot g_{(1)} \)\ .
\eea
To obtain these expressions we made use of the matrix identity: 
\bea
(AB^{-1}A)_{ij} - \hf B_{ij} \Tr[B^{-1}AB^{-1}A] &=& \Tr[B^{-1}A] \( A_{ij} - \hf B_{ij} \Tr[B^{-1}A] \)\ ,\label{matrix_id}
\eea
for any 2x2 matrices $A$ and $B$ such that $\det{B}\neq 0$.

\vspace{10pt}

\bibliographystyle{utphys}
\bibliography{biblio}

\end{document}